\documentclass[11pt]{article}
\usepackage{amscd}
\usepackage[all,cmtip]{xy}
\usepackage{tikz}
\usepackage{pgfplots}
\usetikzlibrary{arrows.meta}
\usepackage{graphicx}
\usepackage{caption}
\usepackage{subcaption}
\usepackage{hyperref}
\usepackage{hyperendnotes}
\usepackage{authblk}
\usepackage{latexsym,calligra,amsmath,amsthm,amssymb,stmaryrd,MnSymbol}
\usepackage[parfill]{parskip} 
\usepackage{mathtools}

\usetikzlibrary{arrows.meta}
\newif\ifdviwin
\usepackage[english]{babel}
\usepackage{microtype}
\usepackage{indentfirst}
\usepackage[mathscr]{eucal}
\usepackage{amssymb,amsmath,amsfonts}
\usepackage{graphicx}
\usepackage{amsmath,amscd}
\usepackage{xxcolor}
\newtheorem{theorem}{Theorem}

\newtheorem{definition}[theorem]{Definition}

\numberwithin{equation}{section}

\newcommand{\ms}[1]{\mathscr{#1}}

\newcommand{\tn}[1]{\textnormal{#1}}

\newcommand{\comment}[1]{}

\pgfplotsset{soldot/.style={color=black,only marks,mark=*}} \pgfplotsset{holdot/.style={color=black,fill=white,only marks,mark=*}}

\begin{document}

\title{Network Horizon Dynamics I: Qualitative Aspects}
\author{B. Dribus, A. Sumner, K. Bist, N. Regmi, J. Sircar, S. Upreti} 
\affil{{\small William Carey University Department of Mathematics}\\\vspace*{.3cm} \href{mailto:bdribus@wmcarey.edu}{\color{blue}bdribus@wmcarey.edu}, \href{mailto:bdribus@gmail.com}{\color{blue}bdribus@gmail.com}}
\date{}
\maketitle

\begin{abstract} \noindent Mostly acyclic directed networks, treated mathematically as directed graphs, arise in machine learning, biology, social science, physics, and other applications. Newman \cite{NewmanNetworks18} has noted the mathematical challenges of such networks.  In this series of papers, we study their connectivity properties, focusing on three types of phase transitions that affect horizon sizes for typical nodes.  The first two types involve the familiar emergence of giant components as average local connectivity increases, while the third type involves small-world horizon growth at variable distance from a typical node.  In this first paper, we focus on qualitative behavior, simulations, and applications, leaving formal considerations for subsequent papers. We explain how such phase transitions distinguish deep neural networks from shallow machine learning architectures, and propose hybrid local/random network designs with surprising connectivity advantages. We also propose a small-world approach to the horizon problem in the cosmology of the early universe as a novel alternative to the inflationary hypothesis of Guth and Linde.
\end{abstract}

\section{Introduction}
\label{sectionintro}

\begin{figure}[h]
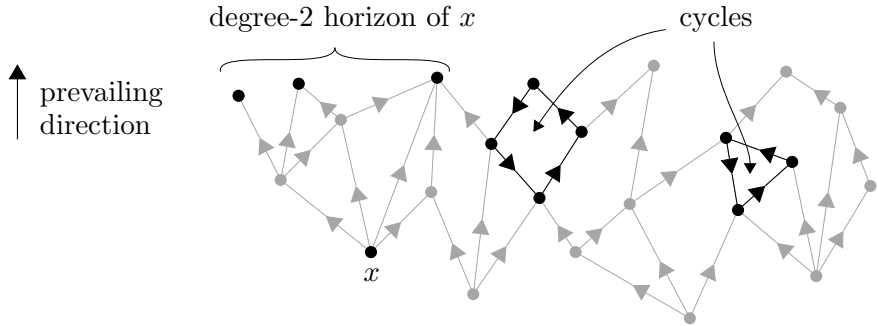

\begin{pgfpicture}{0cm}{0cm}{15cm}{4.4cm}
\begin{pgftranslate}{\pgfpoint{3.8cm}{0cm}}
\begin{pgftranslate}{\pgfpoint{1cm}{-.2cm}}
\begin{pgfmagnify}{.8}{.8}
\color{black!35}
\pgfnodecircle{Node01}[fill]{\pgfxy(-1,2.7)}{0.1cm}
\pgfnodecircle{Node02}[fill]{\pgfxy(-1.7,4.1)}{0.1cm}
\pgfnodecircle{Node03}[fill]{\pgfxy(-.7,4.3)}{0.1cm}
\pgfnodecircle{Node1}[fill]{\pgfxy(0,3.7)}{0.1cm}
\pgfnodecircle{Node2}[fill]{\pgfxy(.5,1.5)}{0.1cm}
\pgfnodecircle{Node3}[fill]{\pgfxy(1.5,2.5)}{0.1cm}
\pgfnodecircle{Node4}[fill]{\pgfxy(1.6,4.4)}{0.1cm}
\pgfnodecircle{Node5}[fill]{\pgfxy(2.2,.8)}{0.1cm}
\pgfnodecircle{Node6}[fill]{\pgfxy(2.5,3.3)}{0.1cm}
\pgfnodecircle{Node7}[fill]{\pgfxy(3.2,4.3)}{0.1cm}
\pgfnodecircle{Node8}[fill]{\pgfxy(3.3,2.4)}{0.1cm}
\pgfnodecircle{Node9}[fill]{\pgfxy(3.9,1.5)}{0.1cm}
\pgfnodecircle{Node10}[fill]{\pgfxy(4,3.5)}{0.1cm}
\pgfnodecircle{Node11}[fill]{\pgfxy(4.8,2.3)}{0.1cm}
\pgfnodecircle{Node12}[fill]{\pgfxy(5.2,4.6)}{0.1cm}
\pgfnodecircle{Node13}[fill]{\pgfxy(5.8,.4)}{0.1cm}
\pgfnodecircle{Node14}[fill]{\pgfxy(6.4,3.4)}{0.1cm}
\pgfnodecircle{Node15}[fill]{\pgfxy(6.6,2.2)}{0.1cm}
\pgfnodecircle{Node16}[fill]{\pgfxy(7.4,4.5)}{0.1cm}
\pgfnodecircle{Node17}[fill]{\pgfxy(7.5,3)}{0.1cm}
\pgfnodecircle{Node18}[fill]{\pgfxy(7.9,1.1)}{0.1cm}
\pgfnodecircle{Node19}[fill]{\pgfxy(8.3,3.9)}{0.1cm}
\pgfnodecircle{Node20}[fill]{\pgfxy(8.8,2.6)}{0.1cm}
\pgfxyline(-.85,3.5)(-.7,4.3)
\pgfxyline(-1.35,3.4)(-1.7,4.1)
\pgfxyline(-.5,3.2)(0,3.7)
\pgfxyline(-.35,4)(-.7,4.3)
\pgfxyline(-.25,2.1)(-1,2.7)
\pgfxyline(.25,2.6)(0,3.7)
\pgfxyline(1,2)(1.5,2.5)
\pgfxyline(1.05,2.95)(1.6,4.4)
\pgfxyline(.8,4.05)(1.6,4.4)
\pgfxyline(1.55,3.45)(1.6,4.4)
\pgfxyline(1.85,1.65)(1.5,2.5)
\pgfxyline(2.35,2.05)(2.5,3.3)
\pgfxyline(2.05,3.85)(1.6,4.4)
\pgfxyline(2.75,1.6)(3.3,2.4)
\pgfxyline(2.9,2.85)(3.3,2.4)
\pgfxyline(2.85,3.8)(2.5,3.3)
\pgfxyline(3.6,1.95)(3.3,2.4)
\pgfxyline(3.65,2.95)(4,3.5)
\pgfxyline(3.6,3.9)(3.2,4.3)
\pgfxyline(4.35,1.9)(4.8,2.3)
\pgfxyline(4.6,4.05)(5.2,4.6)
\pgfxyline(5,3.45)(5.2,4.6)
\pgfxyline(5.6,2.85)(6.4,3.4)
\pgfxyline(4.85,.95)(3.9,1.5)
\pgfxyline(5.3,1.35)(4.8,2.3)
\pgfxyline(6.2,1.3)(6.6,2.2)
\pgfxyline(6.5,2.8)(6.6,2.2)
\pgfxyline(6.9,3.95)(7.4,4.5)
\pgfxyline(7.05,2.6)(7.5,3)
\pgfxyline(6.95,3.2)(6.4,3.4)
\pgfxyline(7.25,1.65)(6.6,2.2)
\pgfxyline(7.7,2.05)(7.5,3)
\pgfxyline(8.1,2.5)(8.3,3.9)
\pgfxyline(7.85,4.2)(7.4,4.5)
\pgfxyline(8.35,1.85)(8.8,2.6)
\pgfxyline(8.55,3.25)(8.3,3.9)
\begin{pgfscope}
\pgfsetarrowsend{Triangle[scale=2pt]}
\pgfxyline(-1,2.7)(-.85,3.5)
\pgfxyline(-1,2.7)(-1.35,3.4)
\pgfxyline(-1,2.7)(-.5,3.2)
\pgfxyline(0,3.7)(-.35,4)
\pgfxyline(.5,1.5)(-.25,2.1)
\pgfxyline(.5,1.5)(.25,2.6)
\pgfxyline(.5,1.5)(1,2)
\pgfxyline(.5,1.5)(1.05,2.95)
\pgfxyline(0,3.7)(.8,4.05)
\pgfxyline(1.5,2.5)(1.55,3.45)
\pgfxyline(2.2,.8)(1.85,1.65)
\pgfxyline(2.2,.8)(2.35,2.05)
\pgfxyline(2.5,3.3)(2.05,3.85)
\pgfxyline(2.2,.8)(2.75,1.6)
\pgfxyline(2.5,3.3)(2.9,2.85)
\pgfxyline(3.2,4.3)(2.85,3.8)
\pgfxyline(3.9,1.5)(3.6,1.95)
\pgfxyline(3.3,2.4)(3.65,2.95)
\pgfxyline(4,3.5)(3.6,3.9)
\pgfxyline(3.9,1.5)(4.35,1.9)
\pgfxyline(4,3.5)(4.6,4.05)
\pgfxyline(4.8,2.3)(5,3.45)
\pgfxyline(4.8,2.3)(5.6,2.85)
\pgfxyline(5.8,.4)(4.85,.95)
\pgfxyline(5.8,.4)(5.3,1.35)
\pgfxyline(5.8,.4)(6.2,1.3)
\pgfxyline(6.4,3.4)(6.5,2.8)
\pgfxyline(6.4,3.4)(6.9,3.95)
\pgfxyline(6.6,2.2)(7.05,2.6)
\pgfxyline(7.5,3)(6.95,3.2)
\pgfxyline(7.9,1.1)(7.25,1.65)
\pgfxyline(7.9,1.1)(7.7,2.05)
\pgfxyline(7.9,1.1)(8.1,2.5)
\pgfxyline(8.3,3.9)(7.85,4.2)
\pgfxyline(7.9,1.1)(8.35,1.85)
\pgfxyline(8.8,2.6)(8.55,3.25)
\end{pgfscope}
\color{black}
\pgfnodecircle{Node02}[fill]{\pgfxy(-1.7,4.1)}{0.1cm}
\pgfnodecircle{Node03}[fill]{\pgfxy(-.7,4.3)}{0.1cm}
\pgfnodecircle{Node2}[fill]{\pgfxy(.5,1.5)}{0.1cm}
\pgfnodecircle{Node4}[fill]{\pgfxy(1.6,4.4)}{0.1cm}
\pgfnodecircle{Node6}[fill]{\pgfxy(2.5,3.3)}{0.1cm}
\pgfnodecircle{Node7}[fill]{\pgfxy(3.2,4.3)}{0.1cm}
\pgfnodecircle{Node8}[fill]{\pgfxy(3.3,2.4)}{0.1cm}
\pgfnodecircle{Node10}[fill]{\pgfxy(4,3.5)}{0.1cm}
\pgfnodecircle{Node14}[fill]{\pgfxy(6.4,3.4)}{0.1cm}
\pgfnodecircle{Node15}[fill]{\pgfxy(6.6,2.2)}{0.1cm}
\pgfnodecircle{Node17}[fill]{\pgfxy(7.5,3)}{0.1cm}
\pgfxyline(2.9,2.85)(3.3,2.4)
\pgfxyline(2.85,3.8)(2.5,3.3)
\pgfxyline(3.65,2.95)(4,3.5)
\pgfxyline(3.6,3.9)(3.2,4.3)
\pgfxyline(6.5,2.8)(6.6,2.2)
\pgfxyline(7.05,2.6)(7.5,3)
\pgfxyline(6.95,3.2)(6.4,3.4)
\pgfmoveto{\pgfxy(-2.0,4.5)}
\pgfcurveto{\pgfxy(-2.1,4.9)}{\pgfxy(-.1,4.5)}{\pgfxy(-.1,4.9)}
\pgfcurveto{\pgfxy(-.1,4.5)}{\pgfxy(1.9,4.9)}{\pgfxy(1.8,4.5)}
\pgfstroke
\pgfsetarrowsend{Triangle[scale=1.4pt]}
\pgfmoveto{\pgfxy(6.2,5)}
\pgfcurveto{\pgfxy(6.2,4.2)}{\pgfxy(6.8,3.4)}{\pgfxy(6.8,2.8)}
\pgfstroke
\pgfmoveto{\pgfxy(5.4,5.2)}
\pgfcurveto{\pgfxy(4.7,5.1)}{\pgfxy(4,4.7)}{\pgfxy(3.2,3.5)}
\pgfstroke
\pgfsetarrowsend{Triangle[scale=2pt]}
\pgfxyline(2.5,3.3)(2.9,2.85)
\pgfxyline(3.2,4.3)(2.85,3.8)
\pgfxyline(3.3,2.4)(3.65,2.95)
\pgfxyline(4,3.5)(3.6,3.9)
\pgfxyline(6.4,3.4)(6.5,2.8)
\pgfxyline(6.6,2.2)(7.05,2.6)
\pgfxyline(7.5,3)(6.95,3.2)
\end{pgfmagnify}
\end{pgftranslate}
\pgfsetarrowsend{Triangle[scale=1.6pt]}
\pgfxyline(-3.3,2.5)(-3.3,3.5)
\pgfputat{\pgfxy(1.3,.7)}{\pgfbox[left,center]{$x$}}
\pgfputat{\pgfxy(-.75,4.1)}{\pgfbox[left,center]{degree-$2$ horizon of $x$}}
\pgfputat{\pgfxy(5.5,4.1)}{\pgfbox[left,center]{cycles}}
\pgfputat{\pgfxy(-3,3.1)}{\pgfbox[left,center]{prevailing}}
\pgfputat{\pgfxy(-3,2.7)}{\pgfbox[left,center]{direction}}
\end{pgftranslate}
\end{pgfpicture}
\caption{Mostly acyclic directed network.}
\label{figmostlyacyclic}
\end{figure}

Mostly acyclic directed networks are networks in which information, energy, or some other quantity flows in essentially one direction, though with possible small loops back.  These are necessary to model features such as local feedback in a deep neural network \cite{Hochreiter97,Tsoi94,Campolucci99,Schmidhuber15}, eddies in a fluid flow \cite{Gustafson85,NairVortex15}, papers or patents that cite themselves or each other \cite{DribusNHDI19,DribusNHDII19,ChoePatent13,JanssenCitation02}, pairs of species that prey upon each other \cite{PolisPredation89,PalomaresPredation98,InvasivePythons11}, and closed causal curves in spacetime geometry near black holes \cite{HawkingEllis73,Wald84}.   Figure \hyperref[figmostlyacyclic]{\ref{figmostlyacyclic}} illustrates a mostly acyclic directed network.  Newman \cite{NewmanNetworks18} remarks about such networks that {\it``not that many of their properties are known--calculations for these networks appear to be harder in many ways than for other types of random graphs."}  Here, we study connectivity properties of such networks, combining results about specific simple networks with generalization strategies involving graph construction and decomposition methods.  For brevity, we work informally, leaving detailed arguments to subsequent papers.  After outlining results for strictly graded networks in Section \hyperref[sectionintro]{\ref{sectionintro}}, we discuss more general networks in Section \hyperref[sectiongeneralizations]{\ref{sectiongeneralizations}}, and applications in Section \hyperref[sectionapplications]{\ref{sectionapplications}}.

\subsection{Background and Motivation}
\label{subsectionbackground}

Our goal is to understand how {\it connection probabilities} for pairs of network nodes change as edges are added or deleted, usually randomly, or as the distance between nodes increases in some frame of reference.  This allows us to model {\it horizons,} which measure accessibility from a given node.   Connection probabilities are prominent in random graph theory \cite{MolloyCritical95,NewmanRandom01,Bollobas01,VadhanPseudo12}, network theory \cite{NewmanNetworks18, DoroNetworks01,Albert02,NewmanAcyclic09}, and percolation theory \cite{Broadbent57,Kesten82,Bollobas06,Heydenreich17,HofstadOrPerc03}.  They depend on the proportion and distribution of {\it permissible edges} actually included in a network, where permissibility depends on the choice of model.  For example, in early percolation theory, nodes often represented lattice points in a low-dimensional geometry used to model a physical system such as a porous medium.  Edges between distant nodes were forbidden, producing a {\it local} structure.  By contrast, information-theoretic networks such as social networks and artificial neural networks can be highly {\it nonlocal.}


Connection probabilities are subject to at least three different types of {\it phase transitions.}  The first two types result from randomly increasing average local connectivity, while the third type involves small-world growth of a sequence of sub-networks. First is a {\it topological phase transition,} in which a {\it giant connected component} abruptly emerges as the proportion of permissible edges increases.   In a directed network, nodes in the giant component are seldom accessible from each other immediately after this transition, since the paths connecting them are seldom directed paths. The giant component is therefore called {\it weakly connected.}  Second, for a directed network, an {\it accessibility phase transition} occurs as the proportion of permissible edges continues to increase, with a large proportion of nodes abruptly becoming accessible from a given node along directed paths.   This transition typically produces a {\it giant in-component,} from which a large proportion of nodes are accessible, a {\it giant out-component,} which is accessible from a large proportion of nodes, and their intersection, the {\it giant strongly-connected component,} which contains directed paths in both directions between any pair of its nodes.  These components may be represented schematically by the ``bow-tie" diagram of Broder et al. \cite{Broder00}.  Third, a sequence of sub-networks of a network may exhibit a {\it small-world phase transition}, in which a large proportion of the entire network is included in every member of the sequence beyond a certain critical index range.  The type of sub-network of principal interest in this paper consists of all nodes connected to a given node $x$ via directed paths of length $m$, which we call the {\it degree-$m$ horizon} of $x$, and denote by $\sigma_m(x)$.  For example, Figure \hyperref[figmostlyacyclic]{\ref{figmostlyacyclic}} shows a degree-$2$ horizon.  Negative $m$ index {\it past horizons,} defined by following directed paths backwards from $x$, while positive $m$ index {\it future horizons}.  

Despite much progress in the theory of phase transitions for directed/oriented percolation in lattices, less emphasis has been placed on mostly acyclic directed networks in general.  Instead, the literature focuses on highly cyclic networks, such as the human brain, social networks, and the Internet. Technically,  a {\it mostly acyclic directed network} is a network well-approximated by its {\it simple acyclic reduction,} given by reducing strongly-connected components to nodes.  A typical example of such a network is an interconnected family of food chains in an ecological community, called a {\it food web.}  Typical pairs of species either exhibit unambiguous predator-prey relationships or coexist amicably, but a few pairs such as alligators and pythons \cite{InvasivePythons11} prey upon each other.  {\it Locally recurrent neural networks} also offer good examples.  More profound examples are {\it discrete spacetime models,} which may exhibit cycles near black holes or at the Planck scale, but which obey an irreversible {\it arrow of time} in ordinary regimes.  While such networks cannot develop giant strongly-connected components, they still undergo accessibility phase transitions, in which a large proportion of the ``future" becomes accessible, and a large proportion of the ``past" becomes visible, from a typical node.  They also exhibit small-world phase transitions, unless limited by ambient low-dimensional geometry, as in early percolation theory.  

Among many applications of these ideas, we discuss two of particular interest.  First, {\it deep neural networks} (DNNs) have enjoyed a recent resurgence in popularity, after being temporarily eclipsed by alternative machine learning architectures \cite{Schmidhuber15,Aggarwal18,Chollet2018}.  Such networks are characterized by the use of {\it deep layers} of {\it artificial neurons} that progressively filter data flow between input and output layers.  Their functional connectivity properties are determined by {\it weights} assigned to their edges.  Networks are {\it trained} to produce desired outputs by adjusting their weights via numerical methods such as {\it gradient descent.}  Training induces functional suppression of insignificant or redundant edges, which may be supplemented by explicit {\it pruning} \cite{NVIDIAPruning17,Alford18}.  Such processes are analogous to biological {\it synaptic pruning,} an important aspect of maturation \cite{Checkik98,Checkik99}.  Phase transition theory facilitates a balance between minimizing redundancy and preserving connectivity in network design.   {\it Shallow networks} lacking deep layers can only undergo topological phase transitions, but deep networks can undergo all three types. Knowledge of this behavior may enable significant design improvements.  In particular, {\it hybrid local/random networks} constructed by augmenting {\it convolutional neural networks} \cite{FukushimaNeocog82,LeCun89,KrizhevskyCNN12} with sparse random structure exhibit superior connectivity properties at reduced computational cost.  Novel pseudorandom  designs have already eclipsed standard architectures in accuracy and efficiency \cite{Alford18,Prabhu18,Robinett18}.  Such methods may allow construction of networks capable of next-generation tasks such as recognition of individuals among a large population, while democratizing access to state-of-the art technology.


Second, something very like a small-world phase transition seems to characterize the connectivity properties of the early universe, to judge by standard astronomical data.   Evidence for this strange {\it horizon problem} was first noted by Rindler \cite{RindlerHorizon56} and later strengthened by the discovery and analysis of the {\it cosmic microwave background} \cite{Dicke65,Penzias65} around the same time that Erd\"{o}s and R\'{e}nyi popularized random graph theory \cite{ErdosRenyi59}.  The resulting explanatory issues later motivated the {\it inflationary hypothesis} of Guth \cite{GuthPaper81} and Linde \cite{LindePaper82}. While the data itself exhibits tell-tale signs of small-world behavior familiar in modern network theory, initial analysis of the problem preceded both the development of discrete spacetime models involving networks \cite{Thiemann07,RovelliQG04,Sorkinetal87,Krioukov12,Ambjorn02}, and the general theory of phase transitions for directed networks \cite{NewmanRandom01,DoroNetworks01}. Perhaps for this reason, small-world phase transitions have not been seriously investigated as a possible solution to the horizon problem until quite recently \cite{DribusDCT, DribusFQXi15, DribusAxioms13}, although other types of phase transitions are routinely studied in early cosmology \cite{MazenkoPhaseInflation85,Ellis12,AmbjornPhase12}.  General mechanisms such as Feynman's {\it path summation approach} to quantum theory \cite{FeynmanSOH48} and the phenomenon of {\it preferential attachment} in network growth processes \cite{NewmanNetworks18} tend to strengthen the case for a small-world explanation of the horizon problem.   This topic is of significant present interest, with recent data disfavoring simple inflationary models \cite{Steinhardt13,Steinhardt17,SteinhardtReply17,AdeInflation18}.


\subsection{Strictly Graded Networks and their Graphs}
\label{intro3gen}

\begin{figure}[h]
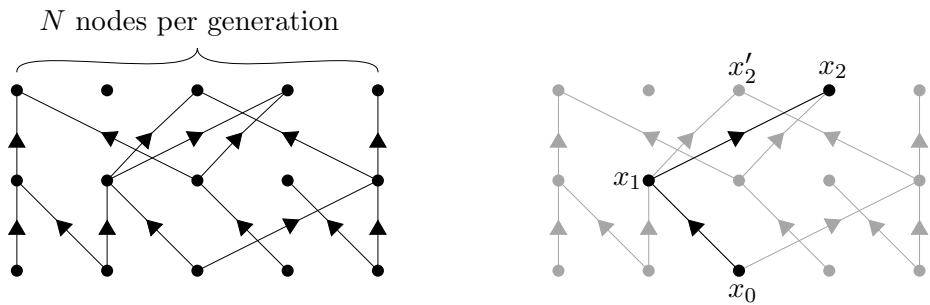

\begin{pgfpicture}{0cm}{0cm}{15cm}{4.0cm}
\begin{pgftranslate}{\pgfpoint{-2.2cm}{.4cm}}
\begin{pgfmagnify}{.8}{.8}
\pgfnodecircle{Node13}[fill]{\pgfxy(3,0)}{0.1cm}
\pgfnodecircle{Node14}[fill]{\pgfxy(4.5,0)}{0.1cm}
\pgfnodecircle{Node15}[fill]{\pgfxy(6,0)}{0.1cm}
\pgfnodecircle{Node16}[fill]{\pgfxy(7.5,0)}{0.1cm}
\pgfnodecircle{Node17}[fill]{\pgfxy(9,0)}{0.1cm}
\pgfnodecircle{Node23}[fill]{\pgfxy(3,1.5)}{0.1cm}
\pgfnodecircle{Node24}[fill]{\pgfxy(4.5,1.5)}{0.1cm}
\pgfnodecircle{Node25}[fill]{\pgfxy(6,1.5)}{0.1cm}
\pgfnodecircle{Node26}[fill]{\pgfxy(7.5,1.5)}{0.1cm}
\pgfnodecircle{Node27}[fill]{\pgfxy(9,1.5)}{0.1cm}
\pgfnodecircle{Node33}[fill]{\pgfxy(3,3)}{0.1cm}
\pgfnodecircle{Node34}[fill]{\pgfxy(4.5,3)}{0.1cm}
\pgfnodecircle{Node35}[fill]{\pgfxy(6,3)}{0.1cm}
\pgfnodecircle{Node36}[fill]{\pgfxy(7.5,3)}{0.1cm}
\pgfnodecircle{Node37}[fill]{\pgfxy(9,3)}{0.1cm}
\pgfmoveto{\pgfxy(3,3.3)}
\pgfcurveto{\pgfxy(2.95,3.8)}{\pgfxy(6,3.1)}{\pgfxy(6,3.7)}
\pgfcurveto{\pgfxy(6,3.1)}{\pgfxy(9.05,3.8)}{\pgfxy(9,3.3)}
\pgfxyline(3,.75)(3,1.5)
\pgfxyline(3.75,.75)(3,1.5)
\pgfxyline(4.5,.75)(4.5,1.5)
\pgfxyline(5.25,.75)(4.5,1.5)
\pgfxyline(7.5,.75)(9,1.5)
\pgfxyline(6.75,.75)(6,1.5)
\pgfxyline(8.25,.75)(7.5,1.5)
\pgfxyline(9,.75)(9,1.5)
\pgfxyline(3,2.2)(3,3)
\pgfxyline(4.6,2.2)(3,3)
\pgfxyline(5.15,2.2)(6,3)
\pgfxyline(5.9,2.2)(7.5,3)
\pgfxyline(6.65,2.2)(7.5,3)
\pgfxyline(7.6,2.2)(6,3)
\pgfxyline(9,2.2)(9,3)
\pgfsetarrowsend{Triangle[scale=2pt]}
\pgfxyline(3,0)(3,.85)
\pgfxyline(4.5,0)(3.65,.85)
\pgfxyline(4.5,0)(4.5,.85)
\pgfxyline(6,0)(5.15,.85)
\pgfxyline(6,0)(7.7,.85)
\pgfxyline(7.5,0)(6.65,.85)
\pgfxyline(9,0)(8.15,.85)
\pgfxyline(9,0)(9,.85)
\pgfxyline(4.5,1.5)(5.25,2.3)
\pgfxyline(4.5,1.5)(6.1,2.3)
\pgfxyline(3,1.5)(3,2.3)
\pgfxyline(6,1.5)(4.4,2.3)
\pgfxyline(6,1.5)(6.75,2.3)
\pgfxyline(9,1.5)(7.4,2.3)
\pgfxyline(9,1.5)(9,2.3)
\end{pgfmagnify}
\end{pgftranslate}
\begin{pgftranslate}{\pgfpoint{5cm}{.4cm}}
\begin{pgfmagnify}{.8}{.8}
\color{black!35}
\pgfnodecircle{Node13}[fill]{\pgfxy(3,0)}{0.1cm}
\pgfnodecircle{Node14}[fill]{\pgfxy(4.5,0)}{0.1cm}
\pgfnodecircle{Node15}[fill]{\pgfxy(6,0)}{0.1cm}
\pgfnodecircle{Node16}[fill]{\pgfxy(7.5,0)}{0.1cm}
\pgfnodecircle{Node17}[fill]{\pgfxy(9,0)}{0.1cm}
\pgfnodecircle{Node23}[fill]{\pgfxy(3,1.5)}{0.1cm}
\pgfnodecircle{Node24}[fill]{\pgfxy(4.5,1.5)}{0.1cm}
\pgfnodecircle{Node25}[fill]{\pgfxy(6,1.5)}{0.1cm}
\pgfnodecircle{Node26}[fill]{\pgfxy(7.5,1.5)}{0.1cm}
\pgfnodecircle{Node27}[fill]{\pgfxy(9,1.5)}{0.1cm}
\pgfnodecircle{Node33}[fill]{\pgfxy(3,3)}{0.1cm}
\pgfnodecircle{Node34}[fill]{\pgfxy(4.5,3)}{0.1cm}
\pgfnodecircle{Node35}[fill]{\pgfxy(6,3)}{0.1cm}
\pgfnodecircle{Node36}[fill]{\pgfxy(7.5,3)}{0.1cm}
\pgfnodecircle{Node37}[fill]{\pgfxy(9,3)}{0.1cm}
\begin{pgfscope}
\pgfxyline(3,.75)(3,1.5)
\pgfxyline(3.75,.75)(3,1.5)
\pgfxyline(4.5,.75)(4.5,1.5)
\pgfxyline(5.25,.75)(4.5,1.5)
\pgfxyline(7.5,.75)(9,1.5)
\pgfxyline(6.75,.75)(6,1.5)
\pgfxyline(8.25,.75)(7.5,1.5)
\pgfxyline(9,.75)(9,1.5)
\pgfxyline(3,2.2)(3,3)
\pgfxyline(4.6,2.2)(3,3)
\pgfxyline(5.15,2.2)(6,3)
\pgfxyline(5.9,2.2)(7.5,3)
\pgfxyline(6.65,2.2)(7.5,3)
\pgfxyline(7.6,2.2)(6,3)
\pgfxyline(9,2.2)(9,3)
\pgfsetarrowsend{Triangle[scale=2pt]}
\pgfxyline(3,0)(3,.85)
\pgfxyline(4.5,0)(3.65,.85)
\pgfxyline(4.5,0)(4.5,.85)
\pgfxyline(6,0)(5.15,.85)
\pgfxyline(6,0)(7.7,.85)
\pgfxyline(7.5,0)(6.65,.85)
\pgfxyline(9,0)(8.15,.85)
\pgfxyline(9,0)(9,.85)
\pgfxyline(4.5,1.5)(5.25,2.3)
\pgfxyline(4.5,1.5)(6.1,2.3)
\pgfxyline(3,1.5)(3,2.3)
\pgfxyline(6,1.5)(4.4,2.3)
\pgfxyline(6,1.5)(6.75,2.3)
\pgfxyline(9,1.5)(7.4,2.3)
\pgfxyline(9,1.5)(9,2.3)
\end{pgfscope}
\color{black}
\pgfnodecircle{Node15}[fill]{\pgfxy(6,0)}{0.1cm}
\pgfnodecircle{Node24}[fill]{\pgfxy(4.5,1.5)}{0.1cm}
\pgfnodecircle{Node36}[fill]{\pgfxy(7.5,3)}{0.1cm}
\pgfxyline(5.25,.75)(4.5,1.5)
\pgfxyline(5.9,2.2)(7.5,3)
\pgfsetarrowsend{Triangle[scale=2pt]}
\pgfxyline(6,0)(5.15,.85)
\pgfxyline(4.5,1.5)(6.1,2.3)
\end{pgfmagnify}
\pgfputat{\pgfxy(4.85,-.3)}{\pgfbox[center,center]{$x_0$}}
\pgfputat{\pgfxy(3.32,1.2)}{\pgfbox[center,center]{$x_1$}}
\pgfputat{\pgfxy(6.05,2.67)}{\pgfbox[center,center]{$x_2$}}
\pgfputat{\pgfxy(4.85,2.7)}{\pgfbox[center,center]{$x_2'$}}
\pgfputat{\pgfxy(-4.5,3.3)}{\pgfbox[left,center]{$N$ nodes per generation}}
\end{pgftranslate}
\end{pgfpicture}
\caption{$3$-generation graph; $2$-chain $c=x_0 x_1 x_2$.}
\label{fig3gengraph}
\end{figure}

We begin by introducing {\it strictly graded networks,} whose edges connect nodes of neighboring grade levels, called {\it generations.}  From an abstract mathematical viewpoint, such networks are  strictly graded directed graphs. Figure \hyperref[fig3gengraph]{\ref{fig3gengraph}} illustrates a $3$-generation graph with $5$ nodes per generation.  Edge directions are  ``up the page," in analogy to the {\it arrow of time} in Minkowski spacetime diagrams.

We denote a typical node of generation $m$ by $x_m$.  Permissible edges are of the form $x_m x_{m+1}$, where $x_m$ is the initial node and $x_{m+1}$ is the terminal node.  We focus first on two closely-related random graph models $G_N^M(K)$ and $\ms{G}_N^M(p)$, each with $M$ generations of size $N$.\\

\begin{definition}\label{def3gen} {\bf $M$-generation graphs} $G_N^M(K)$ and $\ms{G}_N^M(p)$ are strictly graded directed graphs with $M$ generations of $N$ nodes each.  $G_N^M(K)$ has $K$ randomly-chosen edges, and $\ms{G}_N^M(p)$ includes each edge with probability $p$.  
\end{definition}

 It is convenient to number the initial and final generations of an $M$-generation graph as $0$ and $M-1$ rather than $1$ and $M$.  The relationship between $G_N^M(K)$ and $\ms{G}_N^M(p)$ is that $K$ is the {\it expected} number of edges when $p=\frac{K}{(M-1)N^2}$.  $G_N^M(K)$ is convenient for describing graph construction processes, while $\ms{G}_N^M(p)$ is convenient for certain counting arguments.  Both may be viewed as configuration spaces equipped with appropriate probability distributions, but we work mostly with individual graphs.  We later generalize to allow generations of different sizes, generation-skipping edges, multiple edges between a given pair of nodes, and small cycles.  This provides the tools necessary to analyze connectivity properties of arbitrary mostly acyclic networks.  However, our main interest is in generic properties of specific classes of large networks arising in applications.


\subsection{Connection Probabilities and Horizons}
\label{subsecconnprob}

\begin{figure}[h]
\centering
\begin{subfigure}{.5\textwidth}
  \centering
  \includegraphics[width=1.1\linewidth]{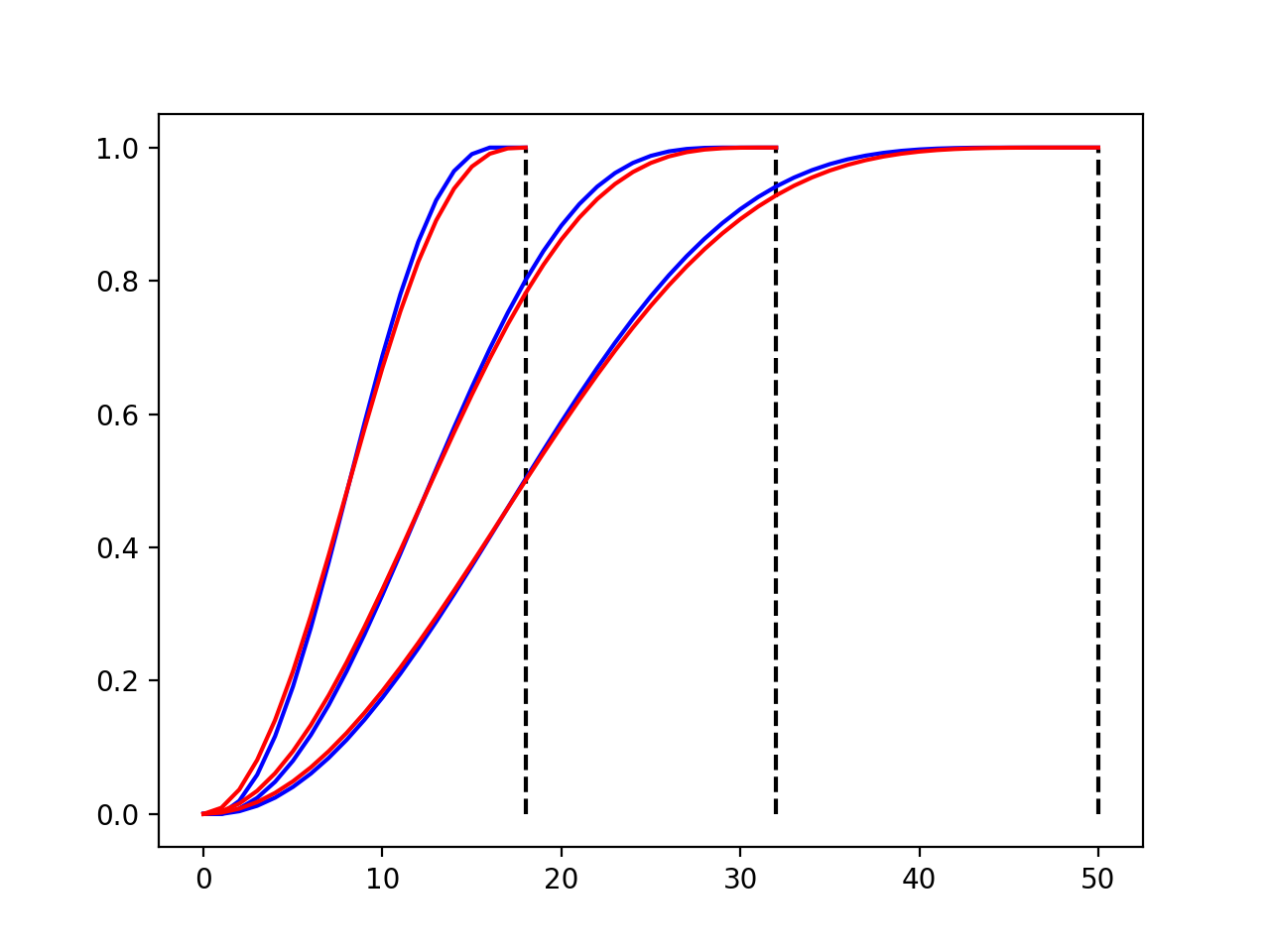}
\end{subfigure}%
\begin{subfigure}{.5\textwidth}
  \centering
\includegraphics[width=1.1\linewidth]{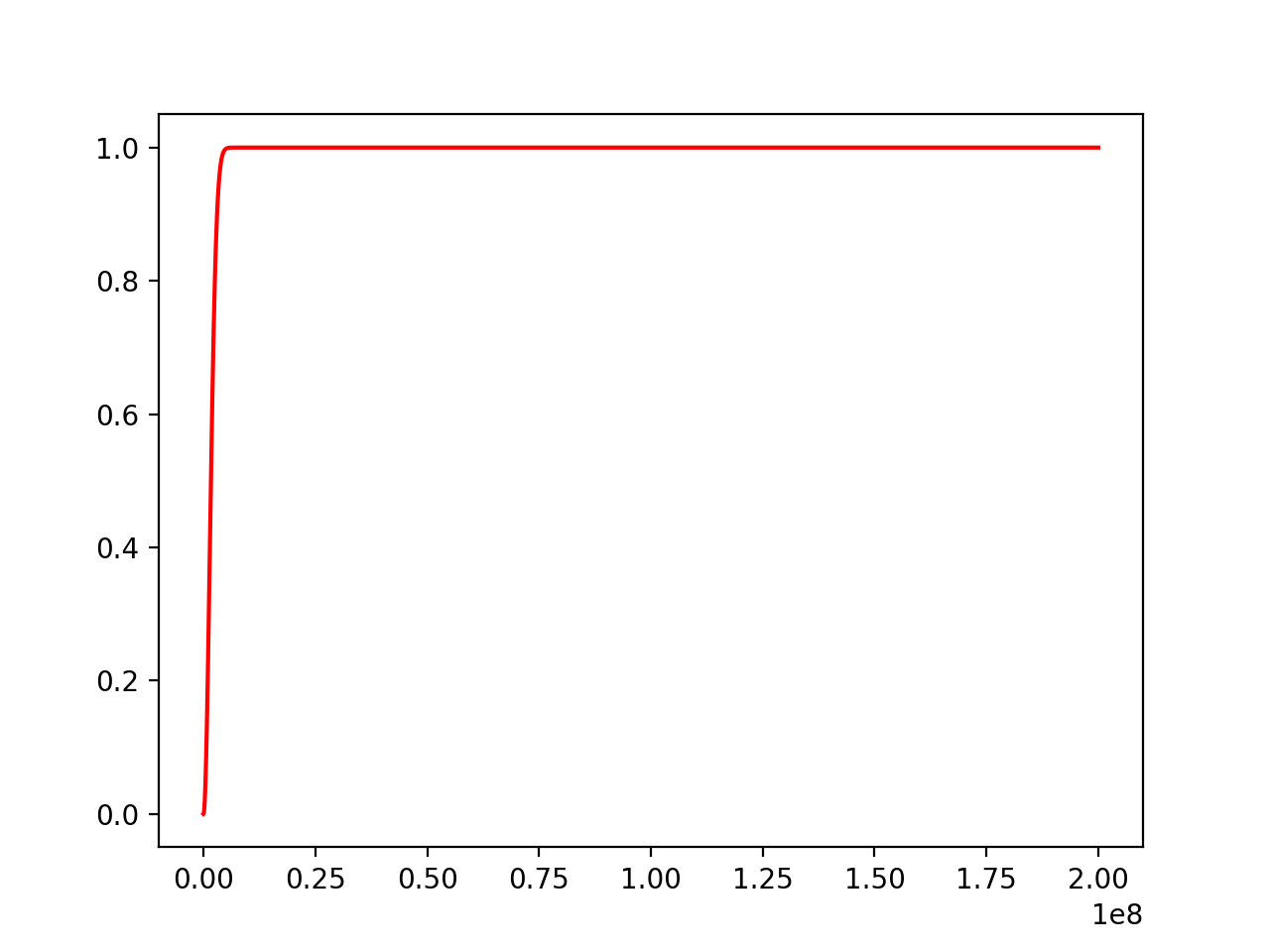}
\end{subfigure}
\caption{$P_N^3(K)$ versus $\ms{P}_N^3(p)$ for $N=3,4,5$; $\ms{P}_N^3(p)$ for $N=10^4$.}
\label{figKvp}
\end{figure}

Applications such as artificial neural networks motivate study of information flow from ``beginning to end" in a mostly acyclic directed network.  For $G_N^M(K)$ and $\ms{G}_N^M(p)$, this involves the probabilities $P_N^M(K)$ and $\ms{P}_N^M(p)$ of connection between typical initial and final-generation nodes $x_0$ and $x_{M-1}$.  A {\it chain}  $c=x_0x_1...x_{M-1}$ connecting  $x_0$ and $x_{M-1}$ is a sequence of edges $x_0x_1, x_1x_2,...,x_{M-2}x_{M-1}$, where the terminal node of each edge is the initial node of the next.  Chains are special directed paths that include each of their edges exactly once.  The {\it length} of $c$ is its number of edges $M-1$, and it is called an {\it $(M-1)$-chain.}  A $2$-chain $x_0x_1x_2$ is illustrated in the right-hand diagram in Figure \hyperref[fig3gengraph]{\ref{fig3gengraph}}.  Connection probabilities for $M=1$ are trivial, since there are no edges.  For $M=2$, the {\it bipartite} case, $P_N^2(K)=\frac{K}{N^2}$, and  $\ms{P}_N^2(p)=p$.  For $M>2$, connection probabilities quickly become more complicated, with $P_N^M(K)$ generally worse than $\ms{P}_N^M(p)$.  For example,
\begin{equation}\label{probintros}P_N^3(K)=\frac{1}{\binom{2N^2}{K}}\sum_{n=1}^N(-1)^{n+1}\binom{N}{n}\binom{2N^2-2n}{K-2n}\hspace*{.3cm}\tn{ and }\hspace*{.3cm}\ms{P}_N^3(p)=1-\left(1-p^2\right)^N.\end{equation}
For $p=\frac{K}{(M-1)N^2}$, the probabilities $P_N^M(K)$ and $\ms{P}_N^M(p)$ (rescaled) nearly coincide for large $N$.  The left-hand diagram in Figure \hyperref[figKvp]{\ref{figKvp}} shows plots of $P_N^3(K)$ (blue) versus $\ms{P}_N^3(\frac{K}{2N^2})$ (red, abscissa units $\frac{1}{2N^2}$) for $N=3,4,$ and $5$.  The dashed lines indicate the maximum number of permissible edges in each case.  As $N$ increases, the growth in the connection probabilities from near $0$ to near $1$ occurs over a progressively shorter proportion of the interval involved, due to an accessibility phase transition elaborated below.  The transition becomes arbitrarily sharp for large $N$.  The right-hand diagram shows the nearly step-function behavior of $\ms{P}_N^3(\frac{K}{2N^2})$ for $N=10^4$. 

\vspace*{-.3cm}

For any directed graph, the {\it degree-$m$ horizon} $\sigma_m(x)$ of a node $x$ is the ``cross-section" of nodes accessible from $x$ via chains of length $m$.  In Figure \hyperref[fig3gengraph]{\ref{fig3gengraph}}, $\sigma_1(x_0)=\{x_1\}$, and $\sigma_2(x_0)=\{x_2,x_2'\}$.  In relativistic physics, discussed in Section \hyperref[sectionhorizon]{\ref{sectionhorizon}},``horizons" in spacetime are defined to be boundaries rather than cross sections, but either notion suffices for studying connectivity.  Our definition is motivated by the fact that it is easier to generalize cross sections to a non-geometric setting than boundaries. We now sketch a heuristic argument that the horizon size $|\sigma_m(x)|$ for a node $x$ in $G_N^M(K)$ or $\ms{G}_N^M(p)$ behaves qualitatively like a logistic population model: it scales exponentially with increasing generation number unless it is limited by the generational size $N$, which plays the role of a ``carrying capacity."  This is a familiar type of argument for random and pseudorandom graphs \cite{NewmanNetworks18,VadhanPseudo12}.  For simplicity, we assume that $x$ belongs to generation $0$.  Typical neighboring generations are connected by roughly $k=\frac{K}{M-1}\approx pN^2$ edges each, so roughly $\frac{k}{N}\approx pN$ edges connect $x$ to generation $1$.  Working inductively, the number of chains between $x$ and generation $m$ is roughly $\frac{k}{N}\approx pN$ times the number of chains between $x$ and generation $m-1$. For small $k$ or $p$, these chains typically terminate at different nodes, yielding the estimates
\begin{equation}\label{roughsizeest}|\sigma_m(x)|\hspace*{.1cm}\approx\hspace*{.1cm}\tn{\# chains from $x$ to generation $m$}\hspace*{.1cm}\approx\hspace*{.1cm}\left(\frac{k}{N}\right)^{m}\approx \left(pN\right)^{m}.\end{equation}
Initial exponential growth, rather than decay, is expected if $k>N$ or $p>\frac{1}{N}$.  Significant horizon size is expected whenever \eqref{roughsizeest} is of order $N$, which occurs whenever $k\approx N^{\frac{m+1}{m}}$ or $p\approx N^{-\frac{m-1}{m}}$.  Once growth is halted by the generation size $N$, the average size of $\sigma_m(x)$ gradually decays unless $k\approx N^2$ or $p\approx 1$.  This is not because {\it typical} horizon sizes gradually decay, but because the na\"{i}ve constant asymptotic expected size is subject to random fluctuations, and zero is an {\it attractor} in the sense that if $|\sigma_n(x)|=0$ for some $n$, then $|\sigma_m(x)|=0$ for all $m>n$.  The half-life of this decay process can be quite long, and it is therefore irrelevant in many applications.  By contrast, the initial exponential behavior predicted by \eqref{roughsizeest} is pivotal.  It differs markedly from what occurs in geometry, where horizons grow roughly like a fixed power, namely, the dimension of the manifold.  It is also unlike percolation in a lattice, since lattices and manifolds share similar notions of locality.   Random graded graphs have no such notion; all pairs of nodes in neighboring generations are equally likely to be connected.  Hence, {\it change from random to geometric structure induces change from exponential to power-law scaling of horizons.}  This principle has significant consequences in applications such as machine learning and cosmology.


\subsection{Three Phase Transitions}
\label{subsection3pt}

We now revisit the three types of phase transitions mentioned in Section \hyperref[subsectionbackground]{\ref{subsectionbackground}}, and  discuss how these transitions affect connection probabilities and horizon sizes, focusing at first on $M$-generation graphs $G_N^M(K)$ and $\ms{G}_N^M(p)$.

\begin{figure}[h]
\centering
\begin{subfigure}{.5\textwidth}
  \centering
\includegraphics[width=1.1\linewidth]{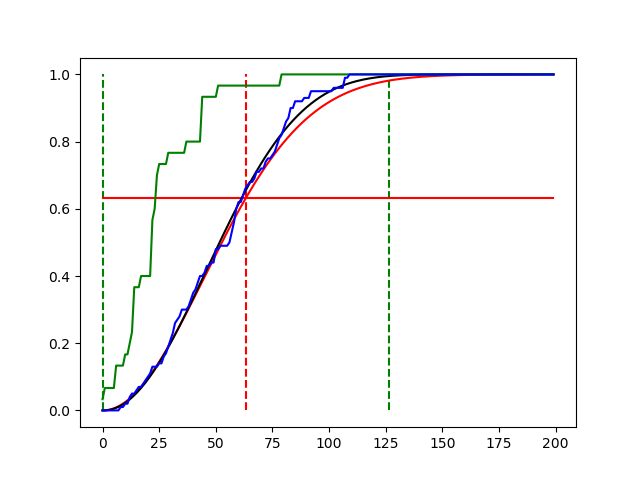}
\end{subfigure}%
\begin{subfigure}{.5\textwidth}
  \centering
\includegraphics[width=1.1\linewidth]{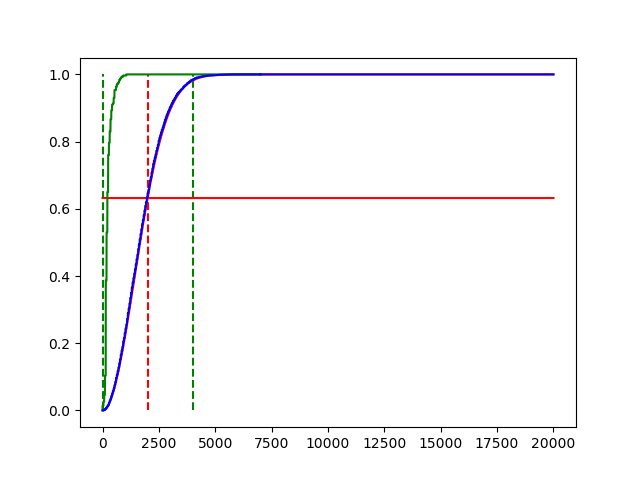}
\end{subfigure}
\caption{Topological (green) and accessibility (blue) phase transitions for $G_{10}^3(K)$ and $G_{100}^3(K)$, along with probability (black) and estimate (red).}
\label{figtenandhundred}
\end{figure}

{\bf 1. Topological phase transition.} If we randomly increase the proportion of permissible edges in an $M$-generation graph, a weakly connected giant component abruptly emerges and quickly grows to fill most of the graph.  This is the topological phase transition.  The green curves in Figure \hyperref[figtenandhundred]{\ref{figtenandhundred}} show the growth of the giant component as $K$ increases for $3$-generation graphs $G_{10}^3(K)$ and $G_{100}^3(K)$.  The connection probabilities $P_N^M(K)$ and $\ms{P}_N^M(p)$ remain small immediately after the topological phase transition, since typical connections in the giant component during its initial emergence involve undirected sequences of edges rather than chains.  Since we are principally interested in connection probabilities and horizon sizes, we do not devote much attention to the topological phase transition.  

\vspace*{-.3cm}

{\bf 2. Accessibility phase transition.} If $M>2$, then the graph $G_N^M(K)$ or $\ms{G}_N^M(p)$ has ``deep" or ``hidden" generations.  In this case, further increasing the proportion of permissible edges after the topological phase transition induces a second phase transition, during which the connection probability $P_N^M(K)$ or $\ms{P}_N^M(p)$ abruptly increases from near $0$ to near $1$.  This is the accessibility phase transition.  Equivalently, the final horizon  of a typical zeroth generation node abruptly expands to fill most of the final generation.  Powerful generating function methods \cite{NewmanRandom01} exist for predicting accessibility phase transitions in random directed graphs obeying a specified degree distribution, but mostly acyclic networks such as $G_N^M(K)$ and $\ms{G}_N^M(p)$ are not easily analyzed via such methods because their degree distributions fail to characterize their prevailing unidirectional structure.  Heuristic arguments explained in Section \hyperref[sectiongeneralizations]{\ref{sectiongeneralizations}} suggest that accessibility phase transitions for such graphs occur near the critical values $K= (M-1)N^\frac{M}{M-1}$ and $p= N^{-\frac{M-2}{M-1}}$, respectively, and that the connection probabilities are approximately 
\begin{equation}\label{Mgenintro}P_N^M(K)\approx1-\exp\left(-\frac{K^{M-1}}{(M-1)^{M-1}N^M}\right)\hspace*{.2cm} \tn{ and } \hspace*{.2cm} \ms{P}_N^M(p)\approx1-\exp\left(-p^{M-1}N^{M-2}\right).\end{equation}
Proof in the case $M=3$ will appear in \cite{DribusNHDII19}. Informally, \eqref{Mgenintro} says that for $M>2$, a very small proportion of permissible edges suffices to connect typical initial and final-generation nodes in a large $M$-generation graph. 

An advantage of $G_N^M(K)$ over $\ms{G}_N^M(p)$ is that it may be regarded as the $K$th stage of a {\it random edge-addition process,} in which permissible edges are added sequentially to a preexisting node set.  A dual process of edge removal is sometimes useful as well, for example, in {\it training} and/or {\it pruning} of artificial neural networks.  $P_N^M(K)$ may be interpreted as the expected behavior of a {\it counting function} $F_N^M(K)$, which records the proportion of connections between initial and final-generation nodes in $G_N^M(K)$ as $K$ varies. The estimate \eqref{Mgenintro} predicts a rapid increase in $F_N^M(K)$ with increasing $K$ near the critical value. Figure \hyperref[figtenandhundred]{\ref{figtenandhundred}} shows examples of $F_N^3(K)$ (blue) for $N=10$ and $N=100$.  Also shown are $P_N^3(K)$ (black), via \eqref{probintros}, the estimate \eqref{Mgenintro} (red),  and the relative size of the weakly-connected giant component (green) from the earlier topological phase transition.  For $N=100$, the blue, black, and red curves nearly coincide.  The vertical lines show the center and radius of the phase transition, and the horizontal line shows the approximation $P_N(2N^{\frac{2}{3}})\approx\frac{e-1}{e}$ at the center, to be derived in \cite{DribusNHDII19}.

In a more general setting, an accessibility phase transition may occur due to a network construction process that adds random edges to a preexisting edge structure, typically uniform, or that randomly changes a proportion of edges from a uniform to random arrangement.  For example, rather than $G_N^M(K)$ or $\ms{G}_N^M(p)$, one may begin with a local lattice-like structure in which each node is connected to a small local {\it kernel} of nodes in the previous generation.   This type of architecture characterizes {\it convolutional neural networks,} discussed further in Section \hyperref[sectionNN]{\ref{sectionNN}}.  In such a network, each initial-generation node communicates with a fixed proportion of final-generation nodes, determined by a power law derived from the size  and shape of the kernel.  This proportion may be quite small, so such networks are often supplemented by tacking on computationally-expensive {\it dense layers} that include every permissible edge.  However, adding a few random edges instead, or changing a few edges to a random arrangement, can induce an accessibility phase transition, leading to a highly connected hybrid local/random network. 

\begin{figure}[h]
\centering
\begin{subfigure}{.5\textwidth}
  \centering
  \includegraphics[width=1.1\linewidth]{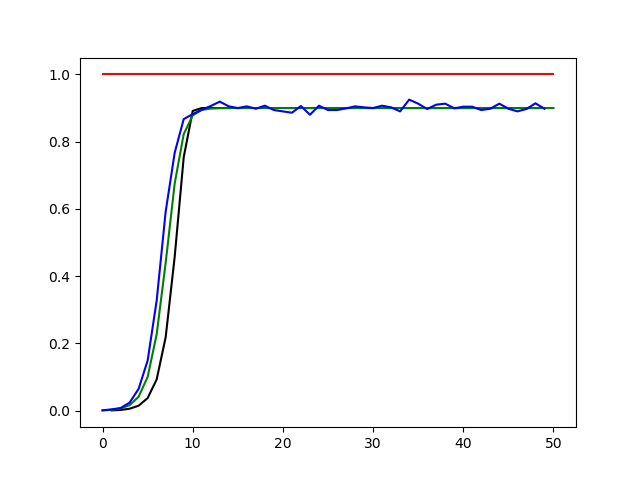}
\end{subfigure}%
\begin{subfigure}{.5\textwidth}
  \centering
\includegraphics[width=1.1\linewidth]{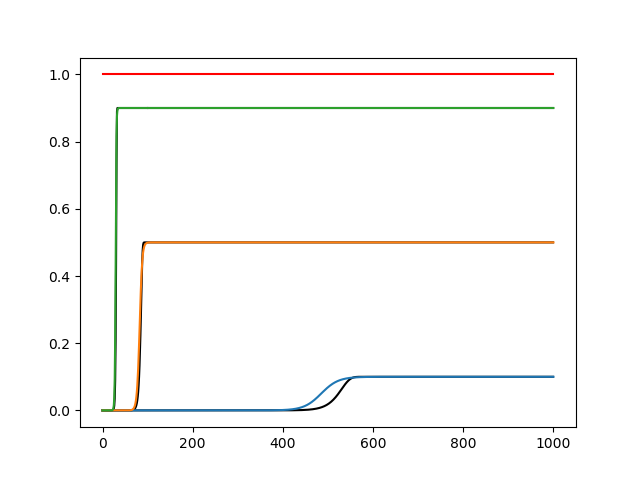}
\end{subfigure}
\caption{Relative horizon growth (blue) versus estimates for $\gamma=\frac{9}{10}$ and $N=1000$; estimates $\frac{H}{N}$ (colors) and \eqref{swcurveshape} (black, scaled by $\frac{1}{N}$) for $N=10^{12}$.}
\label{finalsize}
\end{figure}

 
{\bf 3. Small-world phase transition.} Following the accessibility phase transition, it becomes possible for the degree-$m$ horizon $\sigma_m(x)$ of a node $x$ in $G_N^M(K)$ or $\ms{G}_N^M(p)$ to approach order $N$ in size, i.e., to fill a significant proportion of its generation.  Such a generation is then a {\it small world} from the perspective of $x$.  A useful physics-motivated analogy is to view each generation as a {\it spatial section of spacetime.}  A small world then means that a large proportion of space is accessible from $x$ as one moves in the past or future direction, and a small-world phase transition occurs if the accessible proportion increases abruptly to $O(N)$ within a narrow range of generations.  To model horizon dynamics as the generation number $m$ varies, we estimate how the size of each horizon predicts those of its neighbors, i.e., how many nodes in generation $m\pm 1$ are connected to at least one of a typical family of $n$ nodes in generation $m$.  As explained  in Section \hyperref[sectiongeneralizations]{\ref{sectiongeneralizations}}, a reasonable estimate is
\begin{equation}\label{nexthorizon} h(n):= N-N\left(1-\frac{k}{N^2}\right)^n.\end{equation}
To detect a small world, we study asymptotic horizon size as $m\to\pm\infty$, ignoring slow average decay via fluctuations to zero, as discussed in Section \hyperref[subsecconnprob]{\ref{subsecconnprob}}.  This na\"{i}ve asymptotic size may be estimated by finding a {\it fixed point} of \eqref{nexthorizon}, i.e., a value of $n$ for which $h(n)=n$.  A trivial fixed point is $n=0$.  Since $\frac{d}{dn}\left(h(n)-n\right)\big|_{n=0}>0$ and $h(N)-N<0$ for $k\ge N$, there is another fixed point strictly between $0$ and $N$, i.e., at $n=\gamma N$ for some $\gamma$ between $0$ and $1$.  Solving for $k$ yields $k=N^2\left(1-\left(1-\gamma\right)^{\frac{1}{\gamma N}}\right)$.  Qualitatively, $\frac{1}{\gamma}$ estimates how small the small world is in units of the na\"{i}ve asymptotic horizon size, while $k$ estimates the average connectivity producing this size.  The ratio $\frac{k}{N}$ corresponding to a given $\gamma$ tends to $\frac{1}{\gamma}\log\left(\frac{1}{1-\gamma}\right)$ for large $N$. By \eqref{nexthorizon}, the $m$-fold iterate $H(m)$ of $h(1)$, which estimates $|\sigma_m(x)|$, satisfies the Abel functional equation
\begin{equation}\label{functionalequ}H(m+1)=N\left(1-q^{H(m)}\right),\end{equation}
where $q:=\left(1-\frac{k}{N^2}\right)$, and where $\lim_{m\to\infty} H(m)=\gamma N$ is the na\"{i}ve asymptotic horizon size.  Numerical evidence suggests that $\frac{H}{N}$ approximately maintains its shape and shifts to the right linearly with $\log N$ as $N$ increases.  A rough estimate of the $m$-dependent horizon size for all but very large $m$ is
\begin{equation}\label{swcurveshape}|\sigma_m(x)|\approx\gamma N \left(1-\exp\left(-\frac{k^{m-1}}{N^{m}}\right)\right),\end{equation}
given by replacing fixed $M$ with variable $m$ in the estimate \eqref{Mgenintro} of $P_N^M(K)$, substituting for $K$ in terms of $k$, and scaling by the na\"{i}ve asymptotic horizon size $\gamma N$.  The left-hand diagram in Figure \hyperref[finalsize]{\ref{finalsize}} compares an example of actual relative horizon growth (blue) to the estimate $\frac{H}{N}$ via \eqref{functionalequ} (green) and the rough estimate \eqref{swcurveshape} (black, scaled by $\frac{1}{N}$) for $\gamma=\frac{9}{10}$ and $N=1000$.  The red line indicates saturation; i.e., inclusion of every node in a given generation.  The estimates \eqref{functionalequ} and \eqref{swcurveshape} should be viewed with caution since they involve averaging together radically different types of behavior.  While some actual horizons grow faster than these estimates, others become trapped at size zero after a few generations.  For reasonably large $N$, horizons that survive the first few generations seem to behave similarly to \eqref{functionalequ} and \eqref{swcurveshape}, except for rare late fluctuations to zero.   We may therefore tentatively extrapolate to obtain a rough picture of horizon growth in networks too large to simulate without significant computing resources.  The right-hand diagram compares $\frac{H}{N}$ (colors) to \eqref{swcurveshape} (black, scaled by $\frac{1}{N}$) for  $\gamma=\frac{1}{10},\frac{1}{2}$ and $\frac{9}{10}$, and $N=10^{12}$.    Small world phase transitions are early and sharp for large $\gamma$. 

\vspace*{-.3cm}

More generally, small-world phase transitions may occur due to non-random generation-dependent changes in network structure. Late generations may exhibit  different average connectivity and randomness than  early generations.  For example, adding a dense layer to a convolutional neural network automatically produces a small world.  A subtler scenario occurs when network structure changes from random to geometric, or vice versa, over an interval of generations.  This induces a change from exponential to power-law scaling of horizons, or vice versa, as mentioned at the end of Section \hyperref[subsecconnprob]{\ref{subsecconnprob}}.  In a finite mostly acyclic network, exponential horizon scaling quickly leads to a small world unless the average connectivity is quite low.  As explained in Section \hyperref[sectionhorizon]{\ref{sectionhorizon}},  this effect may explain why the early universe appears to be a small world despite inadequate horizon scaling suggested by na\"{i}ve extrapolation from the present Lorentzian geometric epoch.

\section{Mostly Acyclic Directed Networks}
\label{sectiongeneralizations}
 
We now explain the reasoning behind the estimates in Section \hyperref[sectionintro]{\ref{sectionintro}}, and outline methods to extend these estimates to arbitrary mostly acyclic directed networks.  This requires generalizing the $M$-generation graphs $G_N^M(K)$ and $\ms{G}_N^M(p)$ to allow  generation-skipping edges, generations of different sizes, multiple edges between a given pair of nodes, and small cycles.  Some of these generalizations are amenable to tricks such as graph decomposition methods and superposition of edge-addition or removal processes.


%

\subsection{Basic Exponential Estimates}
\label{Mgen}

\begin{figure}[h]
\centering
\begin{subfigure}{.5\textwidth}
  \centering
  \includegraphics[width=1.1\linewidth]{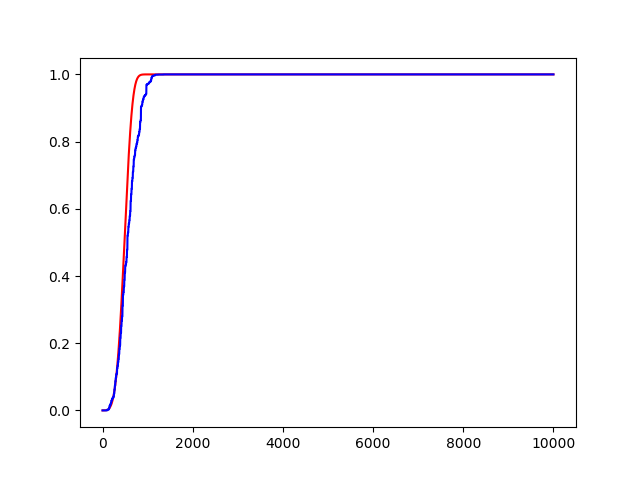}
\end{subfigure}%
\begin{subfigure}{.5\textwidth}
  \centering
\includegraphics[width=1.1\linewidth]{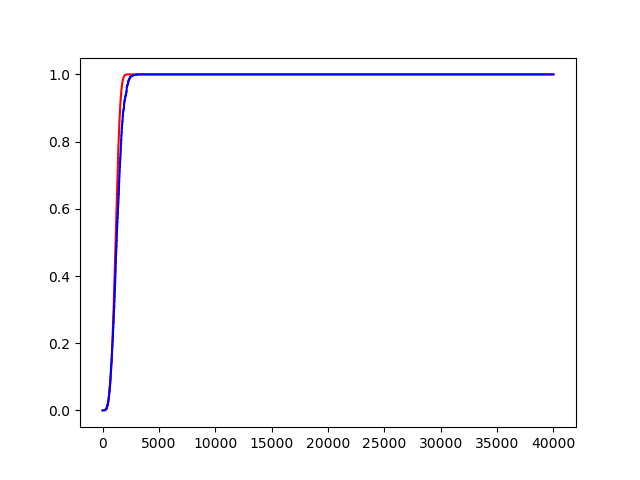}
\end{subfigure}
\caption{Estimates of $P_N^5(K)$ versus $F_N^5(K)$ for $N=50$ and $N=100$.}
\label{fig4and5}
\end{figure}

In Section \hyperref[subsection3pt]{\ref{subsection3pt}}, we offered large-$N$ estimates \eqref{Mgenintro} for the connection probabilities $P_N^M(K)$ and $\ms{P}_N^M(p)$ between initial and final-generation nodes in $G_N^M(K)$ and $\ms{G}_N^M(p)$.  We now sketch an argument supporting these estimates, focusing on $G_N^M(K)$, the harder of the two cases.  We assume that the edge distribution is relatively uniform, meaning that each neighboring pair of generations is connected by roughly $k=\frac{K}{M-1}$ edges.  The probability of connection between typical nodes $x_m$ and $x_{m+1}$ in generations $m$ and $m+1$ is then roughly $\frac{k}{N^2}$, and we assume that $k<<N^2$ in the regime of principal interest.  Loosely defined, this regime includes the neighborhood of the accessibility phase transition, but does not extend into territory where connection probabilities are indistinguishable from unity and horizons immediately expand to $O(N)$  in either direction from a typical node.  Under these assumptions, the probability that $x_{m+1}$ is connected to at least one of a family of $n$ such $x_m$ is approximately
\begin{equation}\label{yprobguess}1-\left(1-\frac{k}{N^2}\right)^n=\sum_{j=1}^n(-1)^{j+1}\binom{n}{j}\left(\frac{k}{N^2}\right)^{j}.\end{equation}
How applicable is this approximation?  For {\it exactly} $k$ edges between neighboring generations, inclusion-exclusion yields the probability 
\begin{equation}\label{yatleastonex}\frac{1}{\binom{N^2}{k}}\sum_{j=1}^n(-1)^{j+1} \binom{n}{j}\binom{N^2-j}{k-j}=\sum_{j=1}^n(-1)^{j+1}\binom{n}{j}\prod_{\ell=0}^{j-1}\frac{k-\ell}{N^2-\ell}.\end{equation}
The large-$N$ limit sharply favors uniform edge distributions. Approximating each factor $\frac{k-\ell}{N^2-\ell}$ in \eqref{yatleastonex} by $\frac{k}{N^2}$ yields \eqref{yprobguess}, which is plausible since $\ell<j\le n<<N$ in the regime of principal interest.  The number of such $x_{m+1}$ connected to at least one of $n$ such $x_m$ is then roughly $h(n):=N-N\left(1-\frac{k}{N^2}\right)^n$, which we used in \eqref{nexthorizon} to estimate na\"{i}ve asymptotic horizon sizes, ignoring slow average decay via fluctuations to zero.  We now apply the same idea to recover the estimate \eqref{Mgenintro} for the connection probability $P_N^M(K)$.  Setting $n=1$ in \eqref{nexthorizon} reproduces the expected degree-$1$ horizon size $|\sigma_1(x_0)|\approx h(1)=\frac{k}{N}$.  Feeding this back into \eqref{nexthorizon} yields $|\sigma_2(x_0)|\approx N-N\left(1-\frac{k}{N^2}\right)^{\frac{k}{N}}$, which has a positive limit when $k=\beta N^{\frac{3}{2}}$ for some $\beta>0$.  Dividing by $N$ to obtain a probability estimate reproduces the $3$-generation case of \eqref{Mgenintro}, to be proven in  \cite{DribusNHDII19}.  Assuming inductively that $|\sigma_{M-2}(x)|\approx N\Big(1-\exp\big(-\frac{k^{M-2}}{N^{M-1}}\big)\Big)$ and applying \eqref{yprobguess} again, 
\begin{equation}\label{iterativem}P_N^M(K)\approx1-\left(1-\frac{k}{N^2}\right)^{N\Big(1-\exp\big(-\frac{k^{M-2}}{N^{M-1}}\big)\Big)}\approx1-\left(1-\frac{k}{N^2}\right)^{\big(\frac{k}{N}\big)^{M-2}},\end{equation}
via first-order approximation of the exponent.  This expression has a positive limit when $k=\beta N^{\frac{M}{M-1}}$ for some $\beta>0$.  Substituting for $k$, \eqref{iterativem} tends to $1-e^{-\beta^{M-1}}$ for large $N$, reproducing the estimate \eqref{Mgenintro}, which works reasonably well numerically despite some overestimation.  It is worthwhile to pursue finer estimates, however, since even modest overestimation can be significant in applications such as the design of artificial neural networks.    Figure \hyperref[fig4and5]{\ref{fig4and5}} illustrates the $5$-generation case, comparing \eqref{Mgenintro} (red) to counting functions $F_N^5(K)$ (blue) for edge-addition processes with $N=50$ and $N=100$.  

\vspace*{-.3cm}

\subsection{Generation-Skipping Edges}
\label{Genskip}

We now generalize beyond the case of strictly-graded networks with uniform generation size, first by allowing edges that skip generations.  Such edges play an important role in certain artificial neural network architectures such as {\it ResNet} \cite{HeResnet16} and {\it DenseNet} \cite{HuangDenseNet2016}.  In physics, they represent {\it direct influence from the distant past,} which has been proposed as a ``moderately nonlocal" phenomenon in causal set theory \cite{Moore88,BLMSReply88,GlaserLocalityFails14}.  Here, we estimate connection probabilities $P(x_0\to x_{M-1})$ for initial and final-generation nodes $x_0$ and $x_{M-1}$ via {\it superposition of edge-addition processes.}  Expected horizon sizes may then be inferred in the obvious way.  In this context, it is actually easier to work with $G_N^M(K)$ than $\ms{G}_N^M(p)$.  Using the $3$-generation case as an example, random edge addition may be viewed as a superposition of two processes, one adding $2N^2$ edges between neighboring generations at relative rate $\frac{2}{3}$, and the other adding $N^2$ generation-skipping edges at relative rate $\frac{1}{3}$.  To estimate $P(x_0\to x_2)$, we may therefore replace $K$ by $\frac{2K}{3}$ in \eqref{Mgenintro} for $M=3$, and by $\frac{K}{3}$ in the exact bipartite connection probability  $\frac{K}{N^2}$.  Since $P(x_0\to x_2)$ is the complement of the probability of connection via {\it neither} process, this yields the estimate $P(x_0\to x_2)\approx\left(1-\frac{K}{3N^2}\right)\exp\left(-\frac{K^2}{9N^3}\right)$.  Similar reasoning yields estimates for $M$-generation graphs in which edges may skip up to $L$ generations.  If $L=M-2$, then edges may connect any pair of generations, and the estimate is 
\begin{equation}\label{alllayers}P(x_0\to x_{M-1})\approx1- \left(1-\frac{K}{\binom{M}{2}N^2}\right)\exp\left(-\sum_{m=0}^{M-3}\frac{\binom{M-2}{m}K^{M-m-1}}{\binom{M}{2}^{M-m-1}N^{M-m}}\right),\end{equation}
where the initial factor arises from the possible ``direct" edge $x_0x_{M-1}$, while the $m$th summand in the exponent arises from edges skipping $m<M-2$ generations.  If $L<M-2$, then the estimate is 
\begin{equation}\label{skipL}P(x_0\to x_{M-1})\approx1- \exp\left(-\sum_{m=0}^{L}\frac{\binom{M-2}{m}K^{M-m-1}}{\binom{M}{2}^{M-m-1}N^{M-m}}\right).\end{equation}


\vspace*{-.6cm}

\begin{figure}[h]
\centering
\begin{subfigure}{.5\textwidth}
  \centering
  \includegraphics[width=1.1\linewidth]{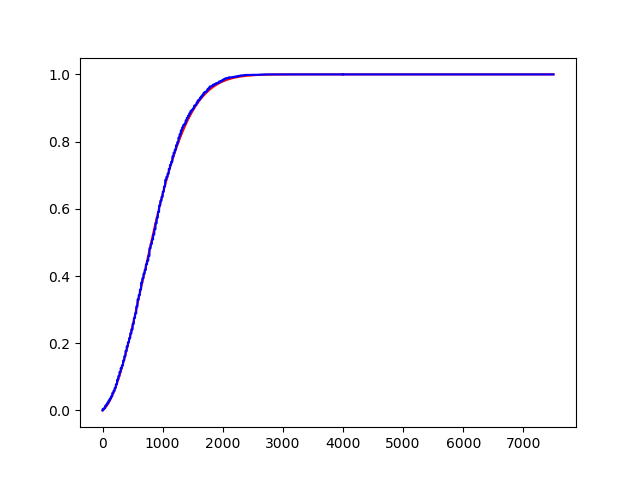}
\end{subfigure}%
\begin{subfigure}{.5\textwidth}
  \centering
\includegraphics[width=1.1\linewidth]{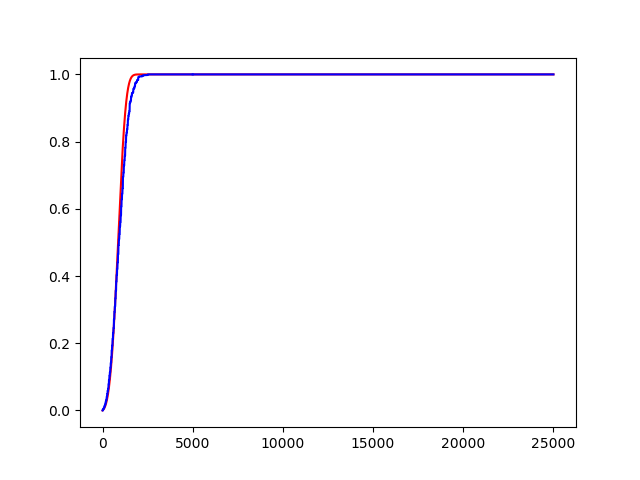}
\end{subfigure}
\caption{$3$-generation and $5$-generation edge-skipping processes for $N=50$.}
\label{figskip}
\end{figure}

In some applications, we may choose to add {\it weights} to \eqref{alllayers} or \eqref{skipL} to gradually damp out ``longer" edges, for example, to deprecate direct influence from the distant past. Numerical evidence suggests that these formulae inherit modest overestimation from \eqref{Mgenintro}, but still perform reasonably well for large $N$.  Figure \hyperref[figskip]{\ref{figskip}} compares \eqref{alllayers} (red) to actual edge-addition processes (blue) for $N=50$ and $M=3$ and $5$.

\subsection{Generations of Different Sizes}
\label{Gensize}

\begin{figure}[h]
\centering
\begin{subfigure}{.5\textwidth}
 \centering
  \includegraphics[width=1.1\linewidth]{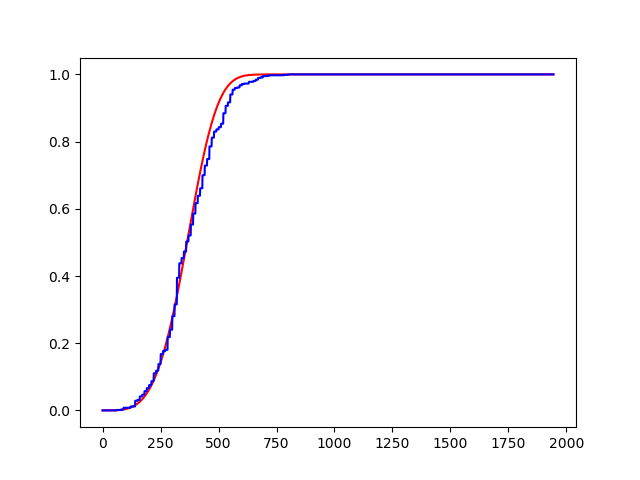}
 \end{subfigure}%
\begin{subfigure}{.5\textwidth}
 \centering
 \includegraphics[width=1.1\linewidth]{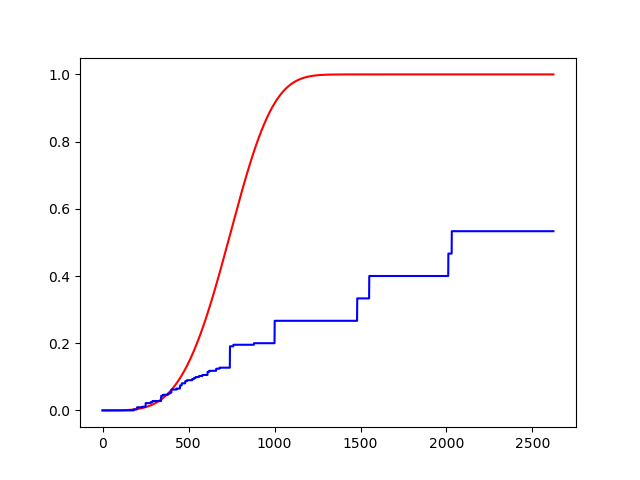}
\end{subfigure}
\caption{Generation sizes: $(70,49,19,51,11)$ and $(15,2,73,67,43)$.}
\label{figbestworst5gen}
\end{figure}

Next, we generalize to networks with independent generation sizes $N_0,...,N_{M-1}$.  In general, we denote by $k_{mn}$ the number of edges between generations $m$ and $n$ for any $m<n$, but we omit generation-skipping edges at present since they may be easily incorporated via superposition of edge-addition processes.   Deep neural networks often exhibit different generation sizes, governed by modeling considerations.  In toy examples involving digit recognition using the {\it MNIST dataset} \cite{Chollet2018,3B1B2017}, a few hundred initial-generation nodes are needed to receive the input data, while just ten final-generation nodes suffice for classification.  More serious image-analysis problems require hundreds of thousands of initial-generation nodes and hundreds of final generation nodes, but next-generation applications will increase these numbers significantly.  In network-based models of spacetime, increasing generation sizes can represent the familiar notion of {\it spacetime expansion,} though other network-theoretic phenomena can also mimic such expansion, as explained in Section \hyperref[sectionhorizon]{\ref{sectionhorizon}}.  Returning to the abstract setting, there are $N_0N_1$ permissible edges in the bipartite case, so the connection probability $P(x_0\to x_1)$ is $\frac{k_{01}}{N_0N_1}$, and the expected size $|\sigma_1(x_0)|$ of the degree-$1$ horizon of a generation-zero node $x_0$ is $\frac{k_{01}}{N_0}$. Adding a third generation with $N_2$ nodes yields an approximate expected horizon size
\begin{equation}\label{est123}|\sigma_2(x_0)|\approx N_2-N_2\left(1-\frac{k_{12}}{N_1N_2}\right)^{\frac{k_{01}}{N_0}}.\end{equation}
If we assume a roughly uniform edge distribution, meaning that the ratios $\frac{k_{mn}}{N_mN_n}$ are roughly equal to the fixed constant $p:=\frac{K}{\sum_{m=0}^{M-2} N_mN_{m+1}}$, then \eqref{est123} may be rewritten as $|\sigma_2(x_0)|\approx N_2-N_2\big(1-p\big)^{N_1p}$.  Since $p$ in this expression is the probability that a randomly-chosen permissible edge is included in the graph, we deliberately use the same notation as the inclusion probability $p$ in the model $\ms{G}_N^M(p)$. Dividing by $N_2$ and taking the limit as $N_1\to\infty$ under the assumption that $p$ is proportional to $N_1^{-\frac{1}{2}}$ yields the $3$-generation analogue of \eqref{Mgenintro}
\begin{equation}\label{connprob3gendiff}P(x_0\to x_2)\approx 1-\exp\left(-N_1p^2\right)\approx 1-\exp\left(-\frac{K^2}{N_1(N_0+N_2)^2}\right).\end{equation}
If we assume inductively that $|\sigma_{M-2}(x_0)|\approx N_{M-2}\left(1-\exp\left(p^{M-2}\prod_{j=1}^{M-3}N_j\right)\right)$, then 
\begin{equation}\label{est12M}|\sigma_{M-1}(x_0)|\approx N_{M-1}-N_{M-1}\big(1-p\big)^{N_{M-2}\Big(1-\exp\big(p^{M-2}\prod_{j=1}^{M-3}N_j\big)\Big)}.\end{equation}
Dropping higher-order terms in the exponent, dividing by $N_{M-1}$, and taking the limit as $\prod_{j=1}^{M-2}N_j\to\infty$ under the assumption that $p$ is proportional to $\prod_{j=1}^{M-2} N_j^{-\frac{1}{M-1}}$ yields the general analogue of \eqref{Mgenintro}
\[P(x_0\to x_{M-1})\approx1-\exp\left(-p^{M-2}\prod_{j=1}^{M-2}N_j\right)\]
\begin{equation}\label{limprobK}\approx1-\exp\left(-K^{M-1}\prod_{j=1}^{M-2}N_j\Big/\Big(\sum_{j=0}^{M-2}N_jN_{j+1}\Big)^{M-1}\right).\end{equation}
Evidence is mixed regarding the asymptotic accuracy of \eqref{limprobK}. Figure \hyperref[figbestworst5gen]{\ref{figbestworst5gen}} shows the best and worst estimates (red) for $P(x_0\to x_{M-1})$, compared to random edge-addition processes (blue), for ten graphs with $200$ nodes randomly partitioned into five generations.  The bad estimate is largely due to the $2$-node bottleneck for $m=1$.  Like  \eqref{Mgenintro}, \eqref{alllayers} and \eqref{skipL}, \eqref{limprobK} seems to overestimate the actual probability, but its accuracy may improve as all generations become large.  Also, estimates for networks with generation-skipping edges derived from \eqref{limprobK} via superposition of edge-addition processes may outperform \eqref{limprobK} itself, since generation-skipping edges can bypass small-generation bottlenecks.

\vspace*{-.3cm}

\subsection{General Acyclic Directed Networks}
\label{Genacyclic}

\begin{figure}[h]
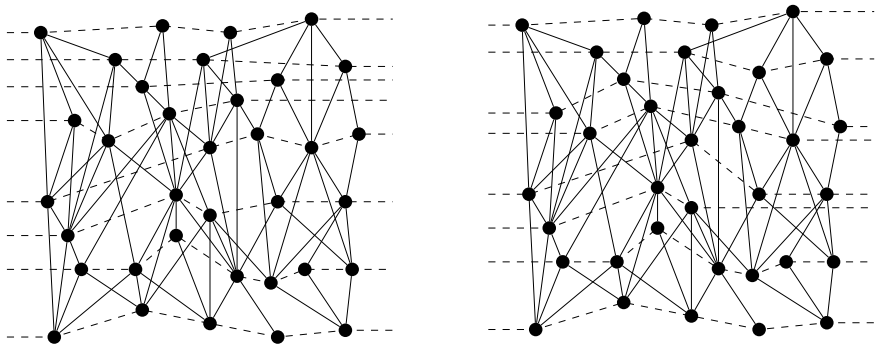

\begin{pgfpicture}{0cm}{0cm}{14cm}{4.6cm}
\begin{pgftranslate}{\pgfpoint{.4cm}{-.4cm}}
\begin{pgfmagnify}{.9}{.9}
\begin{pgfscope}
\pgfsetdash{{3pt}{3pt}}{0pt}
\pgfmoveto{\pgfxy(0,.5)}
\pgflineto{\pgfxy(.7,.5)}
\pgflineto{\pgfxy(2,.9)}
\pgflineto{\pgfxy(3,.7)}
\pgflineto{\pgfxy(4,.5)}
\pgflineto{\pgfxy(5,.6)}
\pgflineto{\pgfxy(5.7,.6)}
\pgfstroke
\pgfmoveto{\pgfxy(0,1.5)}
\pgflineto{\pgfxy(1.1,1.5)}
\pgflineto{\pgfxy(1.9,1.5)}
\pgflineto{\pgfxy(2.5,2)}
\pgflineto{\pgfxy(3.4,1.4)}
\pgflineto{\pgfxy(3.9,1.3)}
\pgflineto{\pgfxy(4.4,1.5)}
\pgflineto{\pgfxy(5.1,1.5)}
\pgflineto{\pgfxy(5.7,1.5)}
\pgfstroke
\pgfmoveto{\pgfxy(0,2)}
\pgflineto{\pgfxy(.9,2)}
\pgflineto{\pgfxy(2.5,2.6)}
\pgflineto{\pgfxy(3,2.3)}
\pgflineto{\pgfxy(4,2.5)}
\pgflineto{\pgfxy(5,2.5)}
\pgflineto{\pgfxy(5.7,2.5)}
\pgfstroke
\pgfmoveto{\pgfxy(0,2.5)}
\pgflineto{\pgfxy(.6,2.5)}
\pgflineto{\pgfxy(3,3.3)}
\pgflineto{\pgfxy(3.7,3.5)}
\pgflineto{\pgfxy(4.5,3.3)}
\pgflineto{\pgfxy(5.2,3.5)}
\pgflineto{\pgfxy(5.7,3.5)}
\pgfstroke
\pgfmoveto{\pgfxy(0,3.7)}
\pgflineto{\pgfxy(1,3.7)}
\pgflineto{\pgfxy(1.5,3.4)}
\pgflineto{\pgfxy(2.4,3.8)}
\pgflineto{\pgfxy(3.4,4)}
\pgflineto{\pgfxy(5.7,4)}
\pgfstroke
\pgfmoveto{\pgfxy(0,4.2)}
\pgflineto{\pgfxy(2,4.2)}
\pgflineto{\pgfxy(4,4.3)}
\pgflineto{\pgfxy(5.7,4.3)}
\pgfstroke
\pgfmoveto{\pgfxy(0,4.6)}
\pgflineto{\pgfxy(1.6,4.6)}
\pgflineto{\pgfxy(2.9,4.6)}
\pgflineto{\pgfxy(5,4.5)}
\pgflineto{\pgfxy(5.7,4.5)}
\pgfstroke
\pgfmoveto{\pgfxy(0,5)}
\pgflineto{\pgfxy(.5,5)}
\pgflineto{\pgfxy(2.3,5.1)}
\pgflineto{\pgfxy(3.3,5)}
\pgflineto{\pgfxy(4.5,5.2)}
\pgflineto{\pgfxy(5.7,5.2)}
\pgfstroke
\end{pgfscope}
%
\pgfnodecircle{Node1}[fill]{\pgfxy(.7,.5)}{.10cm}
\pgfnodecircle{Node2}[fill]{\pgfxy(.6,2.5)}{.10cm}
\pgfnodecircle{Node3}[fill]{\pgfxy(.5,5)}{.10cm}
\pgfnodecircle{Node4}[fill]{\pgfxy(1.1,1.5)}{.10cm}
\pgfnodecircle{Node5}[fill]{\pgfxy(.9,2)}{.10cm}
\pgfnodecircle{Node6}[fill]{\pgfxy(1,3.7)}{.10cm}
\pgfnodecircle{Node7}[fill]{\pgfxy(1.5,3.4)}{.10cm}
\pgfnodecircle{Node8}[fill]{\pgfxy(1.6,4.6)}{.10cm}
\pgfnodecircle{Node9}[fill]{\pgfxy(2,.9)}{.10cm}
\pgfnodecircle{Node10}[fill]{\pgfxy(1.9,1.5)}{.10cm}
\pgfnodecircle{Node11}[fill]{\pgfxy(2.5,2)}{.10cm}
\pgfnodecircle{Node12}[fill]{\pgfxy(2,4.2)}{.10cm}
\pgfnodecircle{Node13}[fill]{\pgfxy(2.5,2.6)}{.10cm}
\pgfnodecircle{Node14}[fill]{\pgfxy(2.4,3.8)}{.10cm}
\pgfnodecircle{Node15}[fill]{\pgfxy(2.3,5.1)}{.10cm}
\pgfnodecircle{Node16}[fill]{\pgfxy(3,.7)}{.10cm}
\pgfnodecircle{Node17}[fill]{\pgfxy(3,2.3)}{.10cm}
\pgfnodecircle{Node18}[fill]{\pgfxy(3,3.3)}{.10cm}
\pgfnodecircle{Node19}[fill]{\pgfxy(2.9,4.6)}{.10cm}
\pgfnodecircle{Node20}[fill]{\pgfxy(3.4,1.4)}{.10cm}
\pgfnodecircle{Node21}[fill]{\pgfxy(3.7,3.5)}{.10cm}
\pgfnodecircle{Node22}[fill]{\pgfxy(3.4,4)}{.10cm}
\pgfnodecircle{Node23}[fill]{\pgfxy(3.3,5)}{.10cm}
\pgfnodecircle{Node24}[fill]{\pgfxy(4,.5)}{.10cm}
\pgfnodecircle{Node25}[fill]{\pgfxy(3.9,1.3)}{.10cm}
\pgfnodecircle{Node26}[fill]{\pgfxy(4,2.5)}{.10cm}
\pgfnodecircle{Node27}[fill]{\pgfxy(4,4.3)}{.10cm}
\pgfnodecircle{Node28}[fill]{\pgfxy(4.4,1.5)}{.10cm}
\pgfnodecircle{Node29}[fill]{\pgfxy(4.5,3.3)}{.10cm}
\pgfnodecircle{Node30}[fill]{\pgfxy(4.5,5.2)}{.10cm}
\pgfnodecircle{Node31}[fill]{\pgfxy(5,.6)}{.10cm}
\pgfnodecircle{Node32}[fill]{\pgfxy(5.1,1.5)}{.10cm}
\pgfnodecircle{Node33}[fill]{\pgfxy(5,2.5)}{.10cm}
\pgfnodecircle{Node34}[fill]{\pgfxy(5.2,3.5)}{.10cm}
\pgfnodecircle{Node35}[fill]{\pgfxy(5,4.5)}{.10cm}
\pgfnodeconnline{Node1}{Node2}
\pgfnodeconnline{Node1}{Node4}
\pgfnodeconnline{Node1}{Node5}
\pgfnodeconnline{Node1}{Node10}
\pgfnodeconnline{Node2}{Node3}
\pgfnodeconnline{Node2}{Node6}
\pgfnodeconnline{Node2}{Node7}
\pgfnodeconnline{Node4}{Node5}
\pgfnodeconnline{Node4}{Node14}
\pgfnodeconnline{Node5}{Node2}
\pgfnodeconnline{Node5}{Node6}
\pgfnodeconnline{Node5}{Node7}
\pgfnodeconnline{Node5}{Node8}
\pgfnodeconnline{Node5}{Node14}
\pgfnodeconnline{Node6}{Node3}
\pgfnodeconnline{Node7}{Node3}
\pgfnodeconnline{Node7}{Node8}
\pgfnodeconnline{Node8}{Node3}
\pgfnodeconnline{Node9}{Node4}
\pgfnodeconnline{Node9}{Node10}
\pgfnodeconnline{Node9}{Node13}
\pgfnodeconnline{Node9}{Node17}
\pgfnodeconnline{Node10}{Node7}
\pgfnodeconnline{Node10}{Node13}
\pgfnodeconnline{Node11}{Node13}
\pgfnodeconnline{Node12}{Node8}
\pgfnodeconnline{Node12}{Node15}
\pgfnodeconnline{Node13}{Node7}
\pgfnodeconnline{Node13}{Node12}
\pgfnodeconnline{Node13}{Node14}
\pgfnodeconnline{Node13}{Node18}
\pgfnodeconnline{Node13}{Node19}
\pgfnodeconnline{Node14}{Node12}
\pgfnodeconnline{Node14}{Node15}
\pgfnodeconnline{Node16}{Node10}
\pgfnodeconnline{Node16}{Node11}
\pgfnodeconnline{Node16}{Node17}
\pgfnodeconnline{Node16}{Node20}
\pgfnodeconnline{Node17}{Node14}
\pgfnodeconnline{Node18}{Node14}
\pgfnodeconnline{Node18}{Node19}
\pgfnodeconnline{Node18}{Node22}
\pgfnodeconnline{Node18}{Node23}
\pgfnodeconnline{Node19}{Node30}
\pgfnodeconnline{Node20}{Node13}
\pgfnodeconnline{Node20}{Node17}
\pgfnodeconnline{Node20}{Node18}
\pgfnodeconnline{Node20}{Node22}
\pgfnodeconnline{Node20}{Node26}
\pgfnodeconnline{Node21}{Node22}
\pgfnodeconnline{Node21}{Node27}
\pgfnodeconnline{Node22}{Node19}
\pgfnodeconnline{Node22}{Node23}
\pgfnodeconnline{Node24}{Node20}
\pgfnodeconnline{Node25}{Node17}
\pgfnodeconnline{Node25}{Node21}
\pgfnodeconnline{Node25}{Node29}
\pgfnodeconnline{Node25}{Node33}
\pgfnodeconnline{Node26}{Node21}
\pgfnodeconnline{Node26}{Node29}
\pgfnodeconnline{Node27}{Node30}
\pgfnodeconnline{Node28}{Node33}
\pgfnodeconnline{Node29}{Node27}
\pgfnodeconnline{Node29}{Node30}
\pgfnodeconnline{Node29}{Node35}
\pgfnodeconnline{Node31}{Node25}
\pgfnodeconnline{Node31}{Node28}
\pgfnodeconnline{Node31}{Node32}
\pgfnodeconnline{Node32}{Node26}
\pgfnodeconnline{Node32}{Node29}
\pgfnodeconnline{Node32}{Node33}
\pgfnodeconnline{Node33}{Node29}
\pgfnodeconnline{Node33}{Node34}
\pgfnodeconnline{Node34}{Node35}
\pgfnodeconnline{Node35}{Node30}
\end{pgfmagnify}
\end{pgftranslate}
\begin{pgftranslate}{\pgfpoint{6.8cm}{-.3cm}}
\begin{pgfmagnify}{.9}{.9}
\begin{pgfscope}
\pgfsetdash{{3pt}{3pt}}{0pt}
\pgfmoveto{\pgfxy(0,.5)}
\pgflineto{\pgfxy(.7,.5)}
\pgflineto{\pgfxy(2,.9)}
\pgflineto{\pgfxy(3,.7)}
\pgflineto{\pgfxy(4,.5)}
\pgflineto{\pgfxy(5,.6)}
\pgflineto{\pgfxy(5.7,.6)}
\pgfstroke
\pgfmoveto{\pgfxy(0,1.5)}
\pgflineto{\pgfxy(1.1,1.5)}
\pgflineto{\pgfxy(1.9,1.5)}
\pgflineto{\pgfxy(2.5,2)}
\pgflineto{\pgfxy(3.4,1.4)}
\pgflineto{\pgfxy(3.9,1.3)}
\pgflineto{\pgfxy(4.4,1.5)}
\pgflineto{\pgfxy(5.1,1.5)}
\pgflineto{\pgfxy(5.7,1.5)}
\pgfstroke
\pgfmoveto{\pgfxy(0,2)}
\pgflineto{\pgfxy(.9,2)}
\pgflineto{\pgfxy(2.5,2.6)}
\pgflineto{\pgfxy(3,2.3)}
\pgflineto{\pgfxy(5.7,2.3)}
\pgfstroke
\pgfmoveto{\pgfxy(0,2.5)}
\pgflineto{\pgfxy(.6,2.5)}
\pgflineto{\pgfxy(3,3.3)}
\pgflineto{\pgfxy(4,2.5)}
\pgflineto{\pgfxy(5,2.5)}
\pgflineto{\pgfxy(5.7,2.5)}
\pgfstroke
\pgfmoveto{\pgfxy(0,3.4)}
\pgflineto{\pgfxy(1.5,3.4)}
\pgflineto{\pgfxy(2.4,3.8)}
\pgflineto{\pgfxy(3.7,3.5)}
\pgflineto{\pgfxy(4.5,3.3)}
\pgflineto{\pgfxy(5.7,3.3)}
\pgfstroke
\pgfmoveto{\pgfxy(0,3.7)}
\pgflineto{\pgfxy(1,3.7)}
\pgflineto{\pgfxy(2,4.2)}
\pgflineto{\pgfxy(3.4,4)}
\pgflineto{\pgfxy(5.2,3.5)}
\pgflineto{\pgfxy(5.7,3.5)}
\pgfstroke
\pgfmoveto{\pgfxy(0,4.6)}
\pgflineto{\pgfxy(1.6,4.6)}
\pgflineto{\pgfxy(2.9,4.6)}
\pgflineto{\pgfxy(4,4.3)}
\pgflineto{\pgfxy(5,4.5)}
\pgflineto{\pgfxy(5.7,4.5)}
\pgfstroke
\pgfmoveto{\pgfxy(0,5)}
\pgflineto{\pgfxy(.5,5)}
\pgflineto{\pgfxy(2.3,5.1)}
\pgflineto{\pgfxy(3.3,5)}
\pgflineto{\pgfxy(4.5,5.2)}
\pgflineto{\pgfxy(5.7,5.2)}
\pgfstroke
\end{pgfscope}
%
\pgfnodecircle{Node1}[fill]{\pgfxy(.7,.5)}{.10cm}
\pgfnodecircle{Node2}[fill]{\pgfxy(.6,2.5)}{.10cm}
\pgfnodecircle{Node3}[fill]{\pgfxy(.5,5)}{.10cm}
\pgfnodecircle{Node4}[fill]{\pgfxy(1.1,1.5)}{.10cm}
\pgfnodecircle{Node5}[fill]{\pgfxy(.9,2)}{.10cm}
\pgfnodecircle{Node6}[fill]{\pgfxy(1,3.7)}{.10cm}
\pgfnodecircle{Node7}[fill]{\pgfxy(1.5,3.4)}{.10cm}
\pgfnodecircle{Node8}[fill]{\pgfxy(1.6,4.6)}{.10cm}
\pgfnodecircle{Node9}[fill]{\pgfxy(2,.9)}{.10cm}
\pgfnodecircle{Node10}[fill]{\pgfxy(1.9,1.5)}{.10cm}
\pgfnodecircle{Node11}[fill]{\pgfxy(2.5,2)}{.10cm}
\pgfnodecircle{Node12}[fill]{\pgfxy(2,4.2)}{.10cm}
\pgfnodecircle{Node13}[fill]{\pgfxy(2.5,2.6)}{.10cm}
\pgfnodecircle{Node14}[fill]{\pgfxy(2.4,3.8)}{.10cm}
\pgfnodecircle{Node15}[fill]{\pgfxy(2.3,5.1)}{.10cm}
\pgfnodecircle{Node16}[fill]{\pgfxy(3,.7)}{.10cm}
\pgfnodecircle{Node17}[fill]{\pgfxy(3,2.3)}{.10cm}
\pgfnodecircle{Node18}[fill]{\pgfxy(3,3.3)}{.10cm}
\pgfnodecircle{Node19}[fill]{\pgfxy(2.9,4.6)}{.10cm}
\pgfnodecircle{Node20}[fill]{\pgfxy(3.4,1.4)}{.10cm}
\pgfnodecircle{Node21}[fill]{\pgfxy(3.7,3.5)}{.10cm}
\pgfnodecircle{Node22}[fill]{\pgfxy(3.4,4)}{.10cm}
\pgfnodecircle{Node23}[fill]{\pgfxy(3.3,5)}{.10cm}
\pgfnodecircle{Node24}[fill]{\pgfxy(4,.5)}{.10cm}
\pgfnodecircle{Node25}[fill]{\pgfxy(3.9,1.3)}{.10cm}
\pgfnodecircle{Node26}[fill]{\pgfxy(4,2.5)}{.10cm}
\pgfnodecircle{Node27}[fill]{\pgfxy(4,4.3)}{.10cm}
\pgfnodecircle{Node28}[fill]{\pgfxy(4.4,1.5)}{.10cm}
\pgfnodecircle{Node29}[fill]{\pgfxy(4.5,3.3)}{.10cm}
\pgfnodecircle{Node30}[fill]{\pgfxy(4.5,5.2)}{.10cm}
\pgfnodecircle{Node31}[fill]{\pgfxy(5,.6)}{.10cm}
\pgfnodecircle{Node32}[fill]{\pgfxy(5.1,1.5)}{.10cm}
\pgfnodecircle{Node33}[fill]{\pgfxy(5,2.5)}{.10cm}
\pgfnodecircle{Node34}[fill]{\pgfxy(5.2,3.5)}{.10cm}
\pgfnodecircle{Node35}[fill]{\pgfxy(5,4.5)}{.10cm}
\pgfnodeconnline{Node1}{Node2}
\pgfnodeconnline{Node1}{Node4}
\pgfnodeconnline{Node1}{Node5}
\pgfnodeconnline{Node1}{Node10}
\pgfnodeconnline{Node2}{Node3}
\pgfnodeconnline{Node2}{Node6}
\pgfnodeconnline{Node2}{Node7}
\pgfnodeconnline{Node4}{Node5}
\pgfnodeconnline{Node4}{Node14}
\pgfnodeconnline{Node5}{Node2}
\pgfnodeconnline{Node5}{Node6}
\pgfnodeconnline{Node5}{Node7}
\pgfnodeconnline{Node5}{Node8}
\pgfnodeconnline{Node5}{Node14}
\pgfnodeconnline{Node6}{Node3}
\pgfnodeconnline{Node7}{Node3}
\pgfnodeconnline{Node7}{Node8}
\pgfnodeconnline{Node8}{Node3}
\pgfnodeconnline{Node9}{Node4}
\pgfnodeconnline{Node9}{Node10}
\pgfnodeconnline{Node9}{Node13}
\pgfnodeconnline{Node9}{Node17}
\pgfnodeconnline{Node10}{Node7}
\pgfnodeconnline{Node10}{Node13}
\pgfnodeconnline{Node11}{Node13}
\pgfnodeconnline{Node12}{Node8}
\pgfnodeconnline{Node12}{Node15}
\pgfnodeconnline{Node13}{Node7}
\pgfnodeconnline{Node13}{Node12}
\pgfnodeconnline{Node13}{Node14}
\pgfnodeconnline{Node13}{Node18}
\pgfnodeconnline{Node13}{Node19}
\pgfnodeconnline{Node14}{Node12}
\pgfnodeconnline{Node14}{Node15}
\pgfnodeconnline{Node16}{Node10}
\pgfnodeconnline{Node16}{Node11}
\pgfnodeconnline{Node16}{Node17}
\pgfnodeconnline{Node16}{Node20}
\pgfnodeconnline{Node17}{Node14}
\pgfnodeconnline{Node18}{Node14}
\pgfnodeconnline{Node18}{Node19}
\pgfnodeconnline{Node18}{Node22}
\pgfnodeconnline{Node18}{Node23}
\pgfnodeconnline{Node19}{Node30}
\pgfnodeconnline{Node20}{Node13}
\pgfnodeconnline{Node20}{Node17}
\pgfnodeconnline{Node20}{Node18}
\pgfnodeconnline{Node20}{Node22}
\pgfnodeconnline{Node20}{Node26}
\pgfnodeconnline{Node21}{Node22}
\pgfnodeconnline{Node21}{Node27}
\pgfnodeconnline{Node22}{Node19}
\pgfnodeconnline{Node22}{Node23}
\pgfnodeconnline{Node24}{Node20}
\pgfnodeconnline{Node25}{Node17}
\pgfnodeconnline{Node25}{Node21}
\pgfnodeconnline{Node25}{Node29}
\pgfnodeconnline{Node25}{Node33}
\pgfnodeconnline{Node26}{Node21}
\pgfnodeconnline{Node26}{Node29}
\pgfnodeconnline{Node27}{Node30}
\pgfnodeconnline{Node28}{Node33}
\pgfnodeconnline{Node29}{Node27}
\pgfnodeconnline{Node29}{Node30}
\pgfnodeconnline{Node29}{Node35}
\pgfnodeconnline{Node31}{Node25}
\pgfnodeconnline{Node31}{Node28}
\pgfnodeconnline{Node31}{Node32}
\pgfnodeconnline{Node32}{Node26}
\pgfnodeconnline{Node32}{Node29}
\pgfnodeconnline{Node32}{Node33}
\pgfnodeconnline{Node33}{Node29}
\pgfnodeconnline{Node33}{Node34}
\pgfnodeconnline{Node34}{Node35}
\pgfnodeconnline{Node35}{Node30}
\end{pgfmagnify}
\end{pgftranslate}
\end{pgfpicture}
\caption{Different generational partitions of an acyclic directed network.}
\label{figdifferentgen}
\end{figure}

Since the node set of any acyclic directed network may be partitioned into generations, our methods are general enough to estimate connection probabilities and horizon sizes for any such network.  Such partitions are non-canonical  \cite{BleybelFoliation18}, constrained only by the requirement that passage along a directed edge must increase generation number. Physically, partitions are analogous to relativistic {\it frames of reference.}  Figure \hyperref[figdifferentgen]{\ref{figdifferentgen}} illustrates two different partitions of an acyclic directed network.  Edges are understood to point up the page.  Dashed lines indicate choices of partition. 

A node $x$ of a directed graph $G$ is called {\it minimal} if no edge begins at $x$, {\it maximal} if no edge terminates at $x$, and {\it extremal} if it is either minimal or maximal.  To generalize our previous results, we focus on connections between minimal and maximal nodes.  While $x$ may lose extremal status during an edge-addition process leading to $G$, we are typically interested only in the status of $x$ in $G$ itself, since such a process is usually a contextual tool rather than a realistic description of network genesis.  Assuming complete knowledge of the  structure of $G$, we describe how to choose generations and permissible edges for estimating connection probabilities and horizon sizes. In applications, we typically lack such complete knowledge, but this is also true for canonically graded networks. The procedure described here merely demonstrates how a general acyclic directed network {\it may be viewed} as a graded network.  With this in mind, we first use the $K$ edges of $G$ to partition its node set into generations, then forget the edges and perform random edge-addition processes, with permissible edges determined by our choice of generations.  Some of these processes produce $G$ when $K$ permissible edges are added, and our methods then yield partition-dependent estimates for connection probabilities and horizon sizes.  We typically choose as few generations as possible. Processes to superpose, and weights for these processes, are chosen based on the distribution of generation-skipping edges for each choice of partition.  Numerical evidence suggests that these choices often converge to similar results for large networks.  However, applications often impose structural constraints that allow for stronger and more specific results.  Hence, our goal is not so much to prove theorems about arbitrary networks as to provide tools for analyzing whatever class of networks our applications present us with.  Section \hyperref[sectionapplications]{\ref{sectionapplications}} highlights two such classes.

\subsection{Cycles; Acyclic Reduction}
\label{subsectioncyc}

\begin{figure}[h]
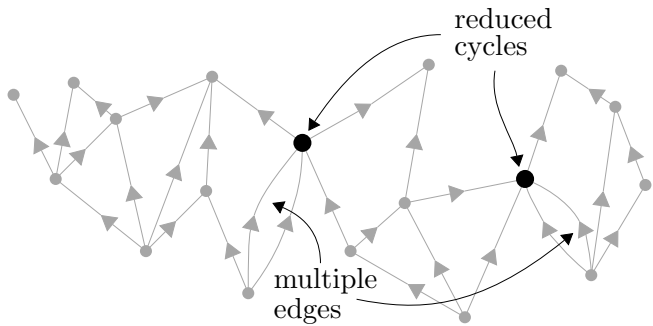

\begin{pgfpicture}{0cm}{0cm}{15cm}{4.3cm}
\begin{pgftranslate}{\pgfpoint{1.5cm}{0cm}}
\begin{pgftranslate}{\pgfpoint{1cm}{-.2cm}}
\begin{pgfmagnify}{.8}{.8}
\color{black!35}
\pgfnodecircle{Node01}[fill]{\pgfxy(-1,2.7)}{0.1cm}
\pgfnodecircle{Node02}[fill]{\pgfxy(-1.7,4.1)}{0.1cm}
\pgfnodecircle{Node03}[fill]{\pgfxy(-.7,4.3)}{0.1cm}
\pgfnodecircle{Node1}[fill]{\pgfxy(0,3.7)}{0.1cm}
\pgfnodecircle{Node2}[fill]{\pgfxy(.5,1.5)}{0.1cm}
\pgfnodecircle{Node3}[fill]{\pgfxy(1.5,2.5)}{0.1cm}
\pgfnodecircle{Node4}[fill]{\pgfxy(1.6,4.4)}{0.1cm}
\pgfnodecircle{Node5}[fill]{\pgfxy(2.2,.8)}{0.1cm}
\pgfnodecircle{Node9}[fill]{\pgfxy(3.9,1.5)}{0.1cm}
\pgfnodecircle{Node11}[fill]{\pgfxy(4.8,2.3)}{0.1cm}
\pgfnodecircle{Node12}[fill]{\pgfxy(5.2,4.6)}{0.1cm}
\pgfnodecircle{Node13}[fill]{\pgfxy(5.8,.4)}{0.1cm}
\pgfnodecircle{Node16}[fill]{\pgfxy(7.4,4.5)}{0.1cm}
\pgfnodecircle{Node18}[fill]{\pgfxy(7.9,1.1)}{0.1cm}
\pgfnodecircle{Node19}[fill]{\pgfxy(8.3,3.9)}{0.1cm}
\pgfnodecircle{Node20}[fill]{\pgfxy(8.8,2.6)}{0.1cm}
\pgfnodecircle{Node21}[fill]{\pgfxy(3.1,3.3)}{0.1cm}
\pgfnodecircle{Node22}[fill]{\pgfxy(6.8,2.7)}{0.1cm}
\pgfxyline(-.85,3.5)(-.7,4.3)
\pgfxyline(-1.35,3.4)(-1.7,4.1)
\pgfxyline(-.5,3.2)(0,3.7)
\pgfxyline(-.35,4)(-.7,4.3)
\pgfxyline(-.25,2.1)(-1,2.7)
\pgfxyline(.25,2.6)(0,3.7)
\pgfxyline(1,2)(1.5,2.5)
\pgfxyline(1.05,2.95)(1.6,4.4)
\pgfxyline(.8,4.05)(1.6,4.4)
\pgfxyline(1.55,3.45)(1.6,4.4)
\pgfxyline(1.85,1.65)(1.5,2.5)
\pgfmoveto{\pgfxy(2.3,2.1)}
\pgfcurveto{\pgfxy(2.4,2.4)}{\pgfxy(2.6,2.7)}{\pgfxy(3.1,3.3)}
\pgfstroke
\pgfxyline(2.35,3.85)(1.6,4.4)
\pgfmoveto{\pgfxy(2.9,2.1)}
\pgfcurveto{\pgfxy(3.0,2.4)}{\pgfxy(3.1,2.7)}{\pgfxy(3.1,3.3)}
\pgfstroke
\pgfxyline(3.5,2.4)(3.1,3.3)
\pgfxyline(4.25,3.95)(5.2,4.6)
\pgfxyline(4.35,1.9)(4.8,2.3)
\pgfxyline(5,3.45)(5.2,4.6)
\pgfxyline(5.8,2.5)(6.8,2.7)
\pgfxyline(4.85,.95)(3.9,1.5)
\pgfxyline(5.3,1.35)(4.8,2.3)
\pgfxyline(6.3,1.55)(6.8,2.7)
\pgfxyline(7.1,3.6)(7.4,4.5)
\pgfmoveto{\pgfxy(7.7,2)}
\pgfcurveto{\pgfxy(7.6,2.2)}{\pgfxy(7.3,2.5)}{\pgfxy(6.8,2.7)}
\pgfstroke
\pgfxyline(8.1,2.5)(8.3,3.9)
\pgfxyline(7.85,4.2)(7.4,4.5)
\pgfxyline(8.35,1.85)(8.8,2.6)
\pgfxyline(8.55,3.25)(8.3,3.9)
\pgfmoveto{\pgfxy(7.1,2)}
\pgfcurveto{\pgfxy(7,2.2)}{\pgfxy(6.8,2.5)}{\pgfxy(6.8,2.7)}
\pgfstroke
\begin{pgfscope}
\pgfsetarrowsend{Triangle[scale=2pt]}
\pgfxyline(-1,2.7)(-.85,3.5)
\pgfxyline(-1,2.7)(-1.35,3.4)
\pgfxyline(-1,2.7)(-.5,3.2)
\pgfxyline(0,3.7)(-.35,4)
\pgfxyline(.5,1.5)(-.25,2.1)
\pgfxyline(.5,1.5)(.25,2.6)
\pgfxyline(.5,1.5)(1,2)
\pgfxyline(.5,1.5)(1.05,2.95)
\pgfxyline(0,3.7)(.8,4.05)
\pgfxyline(1.5,2.5)(1.55,3.45)
\pgfxyline(2.2,.8)(1.85,1.65)
\pgfmoveto{\pgfxy(2.2,.8)}
\pgfcurveto{\pgfxy(2.2,1.3)}{\pgfxy(2.2,1.7)}{\pgfxy(2.3,2.1)}
\pgfstroke
\pgfxyline(3.1,3.3)(2.35,3.85)
\pgfmoveto{\pgfxy(2.2,.8)}
\pgfcurveto{\pgfxy(2.6,1.3)}{\pgfxy(2.75,1.7)}{\pgfxy(2.9,2.1)}
\pgfstroke
\pgfxyline(3.9,1.5)(3.5,2.4)
\pgfxyline(3.1,3.3)(4.25,3.95)
\pgfxyline(3.9,1.5)(4.35,1.9)
\pgfxyline(4.8,2.3)(5,3.45)
\pgfxyline(4.8,2.3)(5.8,2.5)
\pgfxyline(5.8,.4)(4.85,.95)
\pgfxyline(5.8,.4)(5.3,1.35)
\pgfxyline(5.8,.4)(6.3,1.55)
\pgfxyline(6.8,2.7)(7.1,3.6)
\pgfmoveto{\pgfxy(7.9,1.1)}
\pgfcurveto{\pgfxy(7.9,1.4)}{\pgfxy(7.8,1.8)}{\pgfxy(7.7,2)}
\pgfstroke
\pgfxyline(7.9,1.1)(8.1,2.5)
\pgfxyline(8.3,3.9)(7.85,4.2)
\pgfxyline(7.9,1.1)(8.35,1.85)
\pgfxyline(8.8,2.6)(8.55,3.25)
\pgfmoveto{\pgfxy(7.9,1.1)}
\pgfcurveto{\pgfxy(7.5,1.4)}{\pgfxy(7.2,1.8)}{\pgfxy(7.1,2)}
\pgfstroke
\end{pgfscope}
\color{black}
\pgfnodecircle{Node21}[fill]{\pgfxy(3.1,3.3)}{0.15cm}
\pgfnodecircle{Node22}[fill]{\pgfxy(6.8,2.7)}{0.15cm}
\pgfsetarrowsend{Triangle[scale=1.5pt]}
\pgfmoveto{\pgfxy(6.3,4.5)}
\pgfcurveto{\pgfxy(6.2,3.9)}{\pgfxy(6.6,3.4)}{\pgfxy(6.7,2.9)}
\pgfstroke
\pgfmoveto{\pgfxy(5.4,5.1)}
\pgfcurveto{\pgfxy(4.7,5.1)}{\pgfxy(4,4.7)}{\pgfxy(3.2,3.5)}
\pgfstroke

\pgfmoveto{\pgfxy(3.4,1.3)}
\pgfcurveto{\pgfxy(3.4,1.6)}{\pgfxy(3.3,2)}{\pgfxy(2.6,2.3)}
\pgfstroke
\pgfmoveto{\pgfxy(4,.7)}
\pgfcurveto{\pgfxy(5,.5)}{\pgfxy(6,.5)}{\pgfxy(7.6,1.8)}
\pgfstroke

\end{pgfmagnify}
\end{pgftranslate}
\pgfputat{\pgfxy(5.5,4.1)}{\pgfbox[left,center]{reduced}}
\pgfputat{\pgfxy(5.5,3.7)}{\pgfbox[left,center]{cycles}}

\pgfputat{\pgfxy(3.1,0.6)}{\pgfbox[left,center]{multiple}}
\pgfputat{\pgfxy(3.1,0.2)}{\pgfbox[left,center]{edges}}
\end{pgftranslate}

\end{pgfpicture}
\caption{Acyclic reduction of the network from Figure \hyperref[figmostlyacyclic]{\ref{figmostlyacyclic}}.}
\label{figCAD}
\end{figure}

Any directed graph $G$ may be reduced to an acyclic directed {\it multigraph} $G'$, its {\it acyclic reduction,} by merging each strongly-connected component into a single node, as illustrated in Figure \hyperref[figCAD]{\ref{figCAD}}.  For discrete spacetime models, this is an example of {\it causal atomic decomposition} \cite{DribusDCT}.   More generally, it falls into a broad category of graph decomposition methods called {\it community detection} \cite{NewmanNetworks18}.  $G'$ may be further reduced to a simple graph $G''$, the {\it simple acyclic reduction} of $G$, by merging multiple edges.  If the strongly-connected components of $G$ are small and few in number, then the connectivity properties of $G''$ closely approximate those of $G$.  Comparing such properties offers one way to quantify the degree to which $G$ is ``mostly acyclic."

We briefly elaborate on multigraphs such as $G'$.  In a multigraph, pairs of nodes may be connected by multiple edges in the same direction, as illustrated in Figure \hyperref[figCAD]{\ref{figCAD}}.  Such edges can be important for a variety of reasons.  First, they can encode real-world structural nuances in applications; for example, connected pairs of neurons in the mammalian brain often communicate along three to eight different synapses \cite{FauthSynapses15}.  Second, they are ``tolerated" in influential network models such as the {\it configuration model} \cite{NewmanNetworks18}, to avoid the bookkeeping nuisance of excluding them artificially.  Third, they are equivalent to positive integer {\it edge weights,} which may be combined and scaled to approximate real or complex weights in applications such as deep neural networks or discrete quantum systems.   

Multiple edges rarely obstruct analysis of network connectivity, since their redundancy can be handled by replacing the number of edges $K$ with the {\it expected number of occupied edge sites} $E(K)$.  However, passage from $G$ to $G''$ can significantly affect connection probabilities and horizon sizes if $G$ has significant strongly-connected components.  Merging such components to form $G'$ reduces the overall number of nodes, but introduces new nodes of unusually high degree, which may also be extremal.  Merging multiple edges to form $G''$ modestly reduces the degrees of the new nodes.  Overall, $G''$ typically has fewer nodes than $G$, more extremal nodes, and higher average node degree, which is no longer random.   Extremal nodes of $G$ survive in $G''$, so changes to connection probabilities involve only new extremal nodes.  These typically have unusually high degree, and are therefore likely to be connected both to each other and to extremal nodes of $G$.  Analysis based on $G''$ therefore tends to overestimate connection probabilities for $G$.  By contrast, horizon sizes may be underestimated because $G''$ has fewer nodes.  {\it Relative} horizon sizes may be either underestimated or overestimated.  

Other methods for reducing $G$ to an acyclic directed graph are possible.  One may simply {\it delete} strongly-connected components, along with all incoming and outgoing edges.  This avoids the creation of nodes of unusually high degree, but may drastically reduce connectivity.  For example, applying this method to the network from Figure \hyperref[figmostlyacyclic]{\ref{figmostlyacyclic}} fragments it into a three-component network with much different properties. Alternatively, one may replace each strongly-connected component with a randomly-constructed acyclic ``patch" consisting of the same number of nodes and edges.  This preserves important parameters of $G$, but reduces connectivity by eliminating ``hub-like" effects of strongly-connected components.  In particular, it affects connections between extremal nodes of $G$, which are unaffected by passage to $G''$.  The practical utility of $G''$ is determined by how well its connectivity properties approximate those of $G$ in applications. Many interesting networks are {\it very} close to being acyclic in this sense.  For example, in a citation network, pairs of scientific papers rarely cite each other due to temporal constraints, despite some spectacular opportunities \cite{Dicke65,Penzias65}, though pending results may be cited in a series of papers \cite{DribusNHDI19,DribusNHDII19}.  Similar  constraints prohibit cycles in most networks of physical events.  Spacetime cycles may occur near black holes or at the fundamental scale, but a rigid arrow of time seems to prevail in ordinary regimes.  Artificial neural networks can depend crucially on local recurrence \cite{ChungSequence14}, but often exhibit a {\it globally feed-forward structure} \cite{Tsoi94}, or may be transformed into such a structure via {\it unrolling.}  Overall, use of $G''$ seems useful for a wide variety of applications.

\section{Applications}
\label{sectionapplications}

The prominence of network connectivity properties in computer science, biology, social science, and physics leads to many potential applications for the results outlined in this paper.  Foremost among these are those involving random network structure, which is non-geometric, non-local, and subject to abrupt changes in connection probabilities and horizon sizes.   We focus on two such applications: {\it deep learning} in artificial neural networks, and the {\it horizon problem} in the cosmology of the early universe.  Since machine learning and cosmology appeal to different audiences, we include enough expository material to render the necessary topical facets broadly accessible.

\subsection{Deep Learning}
\label{sectionNN}

In an {\it artificial neural network} (ANN) \cite{Schmidhuber15,Aggarwal18,Chollet2018}, nodes represent {\it artificial neurons,} minimal processing units analogous to biological neurons.  They convert families of real inputs to real outputs via {\it activation functions.} Nonlinear activation functions enable approximation of arbitrary functions of the input variables \cite{HornikApproximators89}.  Directed edges represent {\it artificial synapses,} connections between artificial neurons.  We focus on acyclic networks,  called {\it feedforward networks,} though {\it recurrent neural networks} (RNNs) offer advantages such as {\it Turing completeness} \cite{SiegelmannTuring95} and efficient {\it sequence modeling}  \cite{ChungSequence14}. Generations in an ANN are called {\it layers.} Non-input layers are often specified along with their incoming edges, allowing modular description of network architectures in terms of layer types such as {\it dense,} {\it convolutional,} and {\it pooling} layers.  While biological brains remain far more complex \cite{LicataMind15},  ANNs mimic enough of their properties to play a central role in modern artificial intelligence.

\begin{figure}[h]
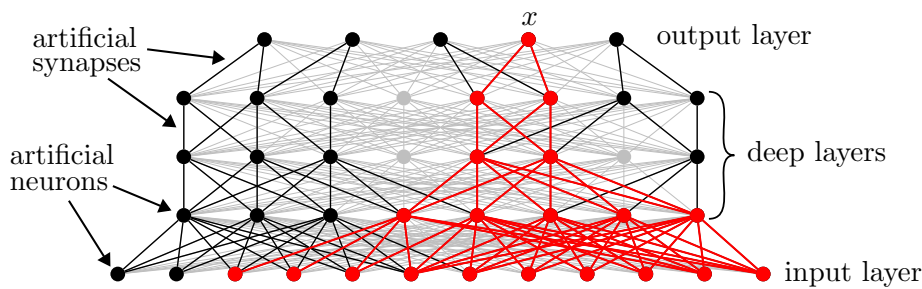


\caption{Deep neural network; training-induced sparseness (black and red); small-world behavior (red).}
\label{figdeep}
\end{figure}

{\bf Deep neural networks.} Figure \hyperref[figdeep]{\ref{figdeep}} illustrates a {\it deep neural network} (DNN), characterized by the presence of {\it deep} or {\it hidden} layers between its inputs and outputs.  Deep layers reduce the number of parameters necessary for accurate function approximation \cite{MontufarDeep14}.  They also enable accessibility phase transitions and small-world phase transitions, with consequences described below.  The illustrated DNN has $M=5$ layers, labeled $m=0$ to $4$, though typically only deep layers are counted.  A left-to-right convention for information flow is common, but we maintain our spacetime-inspired ``up the page" convention.  Each edge in the network is assigned a real {\it weight,} which determines strength of influence between its initial and terminal neurons.  


{\bf Network architectures: dense, convolutional.} The network in Figure \hyperref[figdeep]{\ref{figdeep}} is {\it dense,} meaning that each non-input neuron is connected to every neuron in the previous layer. The network is therefore non-geometric and non-local.  It has the structure of a strictly graded graph of the type studied in Section \hyperref[Gensize]{\ref{Gensize}}, with independent generation sizes $N_0,...,N_{M-1}$, and with every permissible edge included.  These edges are shown in gray; colors are for future reference.  Contrasting with this dense architecture are {\it convolutional neural networks} (CNNs) \cite{FukushimaNeocog82,LeCun89,KrizhevskyCNN12}, inspired by biological vision systems \cite{HubelCat59}, which are highly {\it local} and {\it sparse}, with each neuron connected to only a small {\it kernel} of neurons in the previous layer.  More generally, the term ``CNN" is used to describe any DNN that makes significant use of convolutional layers.  CNNs are the most successful DNNs in applications, and are central to our discussion here.  However, many other architectures are possible, including RNNs and even {\it networks of networks} \cite{WinbergSuper02}.  Absolute connectivity properties of DNNs have traditionally remained fixed following design, though subsequent alterations such as {\it pruning} have recently become common \cite{DaiDeformable17,NVIDIAPruning17,Checkik98,Checkik99,Alford18}.  However, {\it functional} connectivity is determined not only by network topology, but also by edge weights, which change as the network {\it learns.} This leads to interesting horizon dynamics. 


{\bf Applications; feature extraction.} DNNs excel at qualitative pattern recognition involving complicated data sets. Famous examples include image analysis \cite{KrizhevskyCNN12}, speech recognition \cite{GravesSpeech13}, vehicle navigation \cite{HuvalDriving15}, and game strategy \cite{GoogleAlphaZero17,GoogleAlphaZero18}.  Such problems require the network to extract, synthesize, and interpret definitive {\it features} of its input data.  Different network layers often perform different tasks in this context.  For example, early convolutional layers in a large CNN might {\it convert information to geometry} by extracting simple local features from neighboring pixel values in an image, while late dense layers might {\it convert geometry to information} by synthesizing complex features into overall shapes and classifying them.  A useful toy image analysis problem, already mentioned briefly in Section \hyperref[Gensize]{\ref{Gensize}}, is to classify handwritten digits from the {\it MNIST dataset,} encoded as grayscale images \cite{Chollet2018,3B1B2017}.  Features in this context consist of lines, curves, and loops.  



{\bf Network training: parameter space, gradient descent.}  Input data for a DNN is converted into numerical {\it tensors} for processing.   Most such data is irrelevant for tasks such as image classification, but discriminating between relevant and irrelevant data has been one of the principal long-term challenges of artificial intelligence.  DNNs address this challenge via a trial-and-error {\it training} process to minimize a {\it loss function} $L$, which measures the error of actual outputs for training samples with known ``correct" outputs.  The arguments of $L$ are the {\it parameters} characterizing the network, which consist of the edge weights and possibly a few other quantities.  After each training {\it batch,} the parameters are updated and the process is repeated.   Analytical minimization of $L$ is typically infeasible due to the large number of variables involved, which is why numerical trial and error is used.  {\it Gradient descent} and its variations are the principal methods.  In gradient descent, the graph of $L$ is regarded as a hypersurface in {\it parameter space,} whose downward slope $-\nabla L$ is followed step-by-step to find a local minimum.  The step size used for each parameter update is called the {\it learning rate.}  For simplicity, we call the family of local minima the {\it solution space,} though solutions are only approximate.  However, local minima can yield surprisingly accurate results for a well-designed network.  For example, a modest digit-recognition DNN can achieve $99\%$ classification accuracy.   Training is successful if the DNN can {\it generalize,} i.e., accurately process previously-unseen inputs.


{\bf Dimension reduction; pruning; phase transitions.}    Weights in a trained DNN can be highly correlated, which implies that the solution space is concentrated near a relatively low-dimensional submanifold of parameter space.  This {\it dimension reduction} phenomenon is common in applications.  The simplest type of correlation occurs when many weights approach zero.  Training methods explicitly favoring small weights, such as {\it Tikhonov regularization} \cite{Aggarwal18}, have produced favorable results.  In principle, arbitrary correlations may be re-expressed in this manner via coordinate transformations.  Remaining edges with significant weights then constitute a sparse residual network, like the one illustrated in black and red in Figure \hyperref[figdeep]{\ref{figdeep}}.  From this viewpoint, training induces a functional edge-removal process. {\it Pruning,} inspired by biological {\it synaptic pruning,} may also be applied to eliminate superfluous edges entirely \cite{NVIDIAPruning17,Checkik98,Checkik99,Alford18}.  Both processes can induce accessibility phase transitions similar to those induced by random processes, with potentially important consequences.  Since they involve edge {\it removal,} such transitions are time-reversed, manifesting as abrupt {\it decreases} in functional connectivity.   Topological phase transitions are less relevant in this context, since functional fragmentation of a DNN is possible only for specialized datasets that can often be separated {\it a priori}.  By contrast, a broad variety of training scenarios can induce accessibility phase transitions.


{\bf Locality; sparseness; physics connections.} While empirical dimension reduction offers ample justification for studying such transitions, low-dimensional solution spaces are also favored on theoretical grounds. Recent work linking deep learning to {\it Hamiltonian mechanics} \cite{Tegmark2017} systematically demonstrates how basic structural properties such as {\it symmetries} and {\it locality} lead to functionally sparse networks.  Applications often incorporate such properties for physical reasons.  For example, photographic images inherit locality from the physical space of their subjects.  Training a network for image classification therefore encourages sparseness, which can induce an accessibility phase transition if sufficiently pronounced. Training processes for which such a transition predicts optimization of the loss function $L$ may be loosely characterized as satisfying a {\it descent-transition criterion.}   Training-induced transitions may manifest differently than those induced by random processes.  Functional sparseness may appear in some layers while others remain relatively dense, and transitions may occur in subnetworks rather than globally.  Also, as discussed below, {\it networks may be designed to anticipate structural properties that might otherwise have arisen naturally via a phase transition.}  For example, CNNs leverage built-in locality, rather than ``rediscovering" it during training.  Layers that might otherwise have become sparse are sometimes chosen small enough to remain relatively dense, or are pruned down to a small size during training.  A compelling reason to limit layer size is to avoid {\it overfitting,} in which a network ascribes undue significance to artifacts of its training data. 


{\bf Applied horizon dynamics: exceeding biological limitations.}  Training is often the ``bottleneck" for deep learning applications, since it is time-consuming and computationally expensive.  For example, Google DeepMind's overnight training of its multipurpose DNN program {\it AlphaZero} to play superhuman chess required an immense hardware array involving 5000 tensor processing units \cite{GoogleAlphaZero17}.  An open-source analogue called {\it Leela Chess Zero} has required more than a year to achieve comparable results with more modest resources \cite{LC0}.   Our principal motivation for applying phase transition theory to deep learning is to reduce training cost by improving network design and training diagnostics. We focus first on design.  A key objective is to reduce the initial parameter space dimension without sacrificing performance, but this can be difficult, tantamount to partially solving the deep learning problem itself.  Established approaches include {\it feature engineering} to jumpstart the sorting process prior to input, and specific {\it sparse designs} such as CNNs to pare down a larger na\"{i}ve parameter space.  A limitation of both approaches is reduced generality and expressive power.  Recently, significant progress has been made via connectivity-enhancing notions such as generation-skipping edges \cite{HeResnet16,HuangDenseNet2016} and {\it expander graphs} \cite{VadhanPseudo12,Prabhu18,Robinett18}.  We propose a general approach that leverages the horizon dynamics of {\it hybrid local/random networks.}  This approach complements the CNN-exemplified method of ``mining" biology for design clues by {\it deliberately exploiting structural features that biological systems cannot employ for physical reasons.}  For example, the human brain cannot utilize the nonlocal connectivity properties and rapid horizon growth of random networks, due to its embedding in three-dimensional space.  


  
{\bf Optimizing sparseness.} We begin our discussion of DNN design by posing a general question: {\it how sparse can a DNN with specified layer sizes be without significantly reducing its training potential for interesting problems?}  A biological motivation for sparseness is the recent observation that only a small proportion of neurons in the auditory cortex fire in response to any given stimulus \cite{Hromadka08,Ovsepian19}. Perhaps surprisingly, the idea of constructing general sparse network topologies {\it a priori,} rather than achieving sparseness by pruning denser networks, seems to have been implemented only very recently, with such architectures as {\it X-Nets} \cite{Prabhu18} and {\it RadiX-Nets} \cite{Robinett18}.    An obvious condition limiting sparseness is that each output neuron must be sensitive to all, or at least most, of the input data, and it is desirable in practice to have relatively high connectivity even at shorter ranges.  Connection probabilities and horizon sizes are therefore of central interest, especially since highly-connected sparse networks like those mentioned above have already produced accurate results with good efficiency \cite{Alford18,Prabhu18,Robinett18}.   A lower bound for acceptable sparseness in random networks is given by the accessibility phase transition, since sparser random networks exhibit low connectivity.  However, this leaves room for networks significantly sparser than biological neural networks or typical non-random ANN architectures.  In particular, using randomness as a deliberate design feature, in the same spirit as the pseudorandom X-Net designs, offers potentially decisive advantages.  As we demonstrate below, hybrid local/random networks can exceed even random networks in connectivity, while preserving the local feature-extraction potential necessary for applications such as image classification.

\begin{figure}[h]
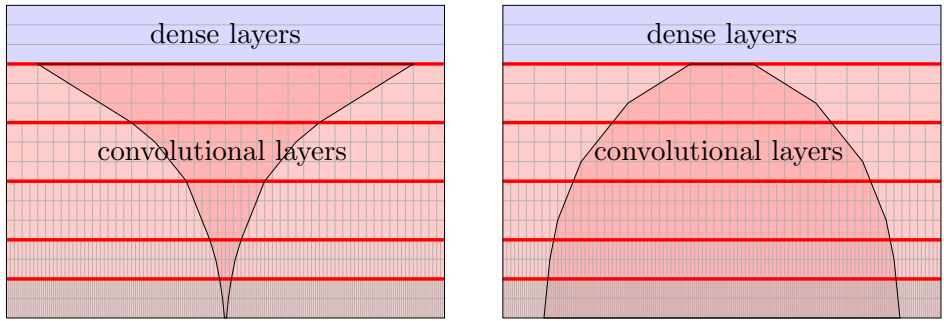

\begin{pgfpicture}{0cm}{0cm}{15cm}{4.3cm}
\begin{pgftranslate}{\pgfpoint{3cm}{0cm}}
\begin{pgfmagnify}{1.3}{1.3}

\color{blue!15}
\pgfmoveto{\pgfxy(-2.24,2.6)}
\pgflineto{\pgfxy(2.24,2.6)}
\pgflineto{\pgfxy(2.24,3.2)}
\pgflineto{\pgfxy(-2.24,3.2)}
\pgflineto{\pgfxy(-2.24,2.6)}
\pgffill
\color{red!20}
\pgfmoveto{\pgfxy(-2.24,2.6)}
\pgflineto{\pgfxy(2.24,2.6)}
\pgflineto{\pgfxy(2.24,0)}
\pgflineto{\pgfxy(-2.24,0)}
\pgflineto{\pgfxy(-2.24,2.6)}
\pgffill

\color{red!30}
\pgfmoveto{\pgfxy(.01,0)}
\pgflineto{\pgfxy(.03,.2)}
\pgflineto{\pgfxy(.06,.4)}
\pgflineto{\pgfxy(.10,.6)}
\pgflineto{\pgfxy(.16,.8)}
\pgflineto{\pgfxy(.24,1)}
\pgflineto{\pgfxy(.32,1.2)}
\pgflineto{\pgfxy(.40,1.4)}
\pgflineto{\pgfxy(.56,1.6)}
\pgflineto{\pgfxy(.72,1.8)}
\pgflineto{\pgfxy(.96,2)}
\pgflineto{\pgfxy(1.28,2.2)}
\pgflineto{\pgfxy(1.60,2.4)}
\pgflineto{\pgfxy(1.92,2.6)}
\pgflineto{\pgfxy(-1.92,2.6)}
\pgflineto{\pgfxy(-1.60,2.4)}
\pgflineto{\pgfxy(-1.28,2.2)}
\pgflineto{\pgfxy(-.96,2)}
\pgflineto{\pgfxy(-.72,1.8)}
\pgflineto{\pgfxy(-.56,1.6)}
\pgflineto{\pgfxy(-.40,1.4)}
\pgflineto{\pgfxy(-.32,1.2)}
\pgflineto{\pgfxy(-.24,1)}
\pgflineto{\pgfxy(-.16,.8)}
\pgflineto{\pgfxy(-.10,.6)}
\pgflineto{\pgfxy(-.06,.4)}
\pgflineto{\pgfxy(-.03,.2)}
\pgflineto{\pgfxy(-.01,0)}
\pgflineto{\pgfxy(.01,0)}
\pgffill
\color{black!30}
\pgfsetlinewidth{.2pt}
\pgfxyline(-2.24,.2)(2.24,.2)
\pgfxyline(-2.24,.4)(2.24,.4)
\pgfxyline(-2.24,.6)(2.24,.6)
\pgfxyline(-2.24,.8)(2.24,.8)
\pgfxyline(-2.24,1.0)(2.24,1.0)
\pgfxyline(-2.24,1.2)(2.24,1.2)
\pgfxyline(-2.24,1.4)(2.24,1.4)
\pgfxyline(-2.24,1.6)(2.24,1.6)
\pgfxyline(-2.24,1.8)(2.24,1.8)
\pgfxyline(-2.24,2.0)(2.24,2.0)
\pgfxyline(-2.24,2.2)(2.24,2.2)
\pgfxyline(-2.24,2.4)(2.24,2.4)
\pgfxyline(-2.24,2.8)(2.24,2.8)
\pgfxyline(-2.24,3.0)(2.24,3.0)
\pgfxyline(-1.92,0)(-1.92,2.6)
\pgfxyline(-1.60,0)(-1.60,2.6)
\pgfxyline(-1.28,0)(-1.28,2.6)
\pgfxyline(-.96,0)(-.96,2.6)
\pgfxyline(-.64,0)(-.64,2.6)
\pgfxyline(-.32,0)(-.32,2.6)
\pgfxyline(0,0)(0,2.6)
\pgfxyline(.32,0)(.32,2.6)
\pgfxyline(.64,0)(.64,2.6)
\pgfxyline(.96,0)(.96,2.6)
\pgfxyline(1.28,0)(1.28,2.6)
\pgfxyline(1.60,0)(1.60,2.6)
\pgfxyline(1.92,0)(1.92,2.6)
\pgfxyline(-2.08,0)(-2.08,2)
\pgfxyline(-1.76,0)(-1.76,2)
\pgfxyline(-1.44,0)(-1.44,2)
\pgfxyline(-1.12,0)(-1.12,2)
\pgfxyline(-.80,0)(-.80,2)
\pgfxyline(-.48,0)(-.48,2)
\pgfxyline(-.16,0)(-.16,2)
\pgfxyline(-2.16,0)(-2.16,1.4)
\pgfxyline(-2.00,0)(-2.00,1.4)
\pgfxyline(-1.84,0)(-1.84,1.4)
\pgfxyline(-1.68,0)(-1.68,1.4)
\pgfxyline(-1.52,0)(-1.52,1.4)
\pgfxyline(-1.36,0)(-1.36,1.4)
\pgfxyline(-1.20,0)(-1.20,1.4)
\pgfxyline(-1.04,0)(-1.04,1.4)
\pgfxyline(-.88,0)(-.88,1.4)
\pgfxyline(-.72,0)(-.72,1.4)
\pgfxyline(-.56,0)(-.56,1.4)
\pgfxyline(-.40,0)(-.40,1.4)
\pgfxyline(-.24,0)(-.24,1.4)
\pgfxyline(-.08,0)(-.08,1.4)
\pgfxyline(-2.20,0)(-2.20,.8)
\pgfxyline(-2.12,0)(-2.12,.8)
\pgfxyline(-2.04,0)(-2.04,.8)
\pgfxyline(-1.96,0)(-1.96,.8)
\pgfxyline(-1.88,0)(-1.88,.8)
\pgfxyline(-1.80,0)(-1.80,.8)
\pgfxyline(-1.72,0)(-1.72,.8)
\pgfxyline(-1.64,0)(-1.64,.8)
\pgfxyline(-1.56,0)(-1.56,.8)
\pgfxyline(-1.48,0)(-1.48,.8)
\pgfxyline(-1.40,0)(-1.40,.8)
\pgfxyline(-1.32,0)(-1.32,.8)
\pgfxyline(-1.24,0)(-1.24,.8)
\pgfxyline(-1.16,0)(-1.16,.8)
\pgfxyline(-1.08,0)(-1.08,.8)
\pgfxyline(-1.00,0)(-1.00,.8)
\pgfxyline(-.92,0)(-.92,.8)
\pgfxyline(-.84,0)(-.84,.8)
\pgfxyline(-.76,0)(-.76,.8)
\pgfxyline(-.68,0)(-.68,.8)
\pgfxyline(-.60,0)(-.60,.8)
\pgfxyline(-.52,0)(-.52,.8)
\pgfxyline(-.44,0)(-.44,.8)
\pgfxyline(-.36,0)(-.36,.8)
\pgfxyline(-.28,0)(-.28,.8)
\pgfxyline(-.20,0)(-.20,.8)
\pgfxyline(-.12,0)(-.12,.8)
\pgfxyline(-.04,0)(-.04,.8)
\pgfxyline(-2.22,0)(-2.22,.4)
\pgfxyline(-2.18,0)(-2.18,.4)
\pgfxyline(-2.14,0)(-2.14,.4)
\pgfxyline(-2.10,0)(-2.10,.4)
\pgfxyline(-2.06,0)(-2.06,.4)
\pgfxyline(-2.02,0)(-2.02,.4)
\pgfxyline(-1.98,0)(-1.98,.4)
\pgfxyline(-1.94,0)(-1.94,.4)
\pgfxyline(-1.90,0)(-1.90,.4)
\pgfxyline(-1.86,0)(-1.86,.4)
\pgfxyline(-1.82,0)(-1.82,.4)
\pgfxyline(-1.78,0)(-1.78,.4)
\pgfxyline(-1.74,0)(-1.74,.4)
\pgfxyline(-1.70,0)(-1.70,.4)
\pgfxyline(-1.66,0)(-1.66,.4)
\pgfxyline(-1.62,0)(-1.62,.4)
\pgfxyline(-1.58,0)(-1.58,.4)
\pgfxyline(-1.54,0)(-1.54,.4)
\pgfxyline(-1.50,0)(-1.50,.4)
\pgfxyline(-1.46,0)(-1.46,.4)
\pgfxyline(-1.42,0)(-1.42,.4)
\pgfxyline(-1.38,0)(-1.38,.4)
\pgfxyline(-1.34,0)(-1.34,.4)
\pgfxyline(-1.30,0)(-1.30,.4)
\pgfxyline(-1.26,0)(-1.26,.4)
\pgfxyline(-1.22,0)(-1.22,.4)
\pgfxyline(-1.18,0)(-1.18,.4)
\pgfxyline(-1.14,0)(-1.14,.4)
\pgfxyline(-1.10,0)(-1.10,.4)
\pgfxyline(-1.06,0)(-1.06,.4)
\pgfxyline(-1.02,0)(-1.02,.4)
\pgfxyline(-.98,0)(-.98,.4)
\pgfxyline(-.94,0)(-.94,.4)
\pgfxyline(-.90,0)(-.90,.4)
\pgfxyline(-.86,0)(-.86,.4)
\pgfxyline(-.82,0)(-.82,.4)
\pgfxyline(-.78,0)(-.78,.4)
\pgfxyline(-.74,0)(-.74,.4)
\pgfxyline(-.70,0)(-.70,.4)
\pgfxyline(-.66,0)(-.66,.4)
\pgfxyline(-.62,0)(-.62,.4)
\pgfxyline(-.58,0)(-.58,.4)
\pgfxyline(-.54,0)(-.54,.4)
\pgfxyline(-.50,0)(-.50,.4)
\pgfxyline(-.46,0)(-.46,.4)
\pgfxyline(-.42,0)(-.42,.4)
\pgfxyline(-.38,0)(-.38,.4)
\pgfxyline(-.34,0)(-.34,.4)
\pgfxyline(-.30,0)(-.30,.4)
\pgfxyline(-.26,0)(-.26,.4)
\pgfxyline(-.22,0)(-.22,.4)
\pgfxyline(-.18,0)(-.18,.4)
\pgfxyline(-.14,0)(-.14,.4)
\pgfxyline(-.10,0)(-.10,.4)
\pgfxyline(-.06,0)(-.06,.4)
\pgfxyline(-.02,0)(-.02,.4)
\pgfxyline(2.08,0)(2.08,2)
\pgfxyline(1.76,0)(1.76,2)
\pgfxyline(1.44,0)(1.44,2)
\pgfxyline(1.12,0)(1.12,2)
\pgfxyline(.80,0)(.80,2)
\pgfxyline(.48,0)(.48,2)
\pgfxyline(.16,0)(.16,2)
\pgfxyline(2.16,0)(2.16,1.4)
\pgfxyline(2.00,0)(2.00,1.4)
\pgfxyline(1.84,0)(1.84,1.4)
\pgfxyline(1.68,0)(1.68,1.4)
\pgfxyline(1.52,0)(1.52,1.4)
\pgfxyline(1.36,0)(1.36,1.4)
\pgfxyline(1.20,0)(1.20,1.4)
\pgfxyline(1.04,0)(1.04,1.4)
\pgfxyline(.88,0)(.88,1.4)
\pgfxyline(.72,0)(.72,1.4)
\pgfxyline(.56,0)(.56,1.4)
\pgfxyline(.40,0)(.40,1.4)
\pgfxyline(.24,0)(.24,1.4)
\pgfxyline(.08,0)(.08,1.4)
\pgfxyline(2.20,0)(2.20,.8)
\pgfxyline(2.12,0)(2.12,.8)
\pgfxyline(2.04,0)(2.04,.8)
\pgfxyline(1.96,0)(1.96,.8)
\pgfxyline(1.88,0)(1.88,.8)
\pgfxyline(1.80,0)(1.80,.8)
\pgfxyline(1.72,0)(1.72,.8)
\pgfxyline(1.64,0)(1.64,.8)
\pgfxyline(1.56,0)(1.56,.8)
\pgfxyline(1.48,0)(1.48,.8)
\pgfxyline(1.40,0)(1.40,.8)
\pgfxyline(1.32,0)(1.32,.8)
\pgfxyline(1.24,0)(1.24,.8)
\pgfxyline(1.16,0)(1.16,.8)
\pgfxyline(1.08,0)(1.08,.8)
\pgfxyline(1.00,0)(1.00,.8)
\pgfxyline(.92,0)(.92,.8)
\pgfxyline(.84,0)(.84,.8)
\pgfxyline(.76,0)(.76,.8)
\pgfxyline(.68,0)(.68,.8)
\pgfxyline(.60,0)(.60,.8)
\pgfxyline(.52,0)(.52,.8)
\pgfxyline(.44,0)(.44,.8)
\pgfxyline(.36,0)(.36,.8)
\pgfxyline(.28,0)(.28,.8)
\pgfxyline(.20,0)(.20,.8)
\pgfxyline(.12,0)(.12,.8)
\pgfxyline(.04,0)(.04,.8)
\pgfxyline(2.22,0)(2.22,.4)
\pgfxyline(2.18,0)(2.18,.4)
\pgfxyline(2.14,0)(2.14,.4)
\pgfxyline(2.10,0)(2.10,.4)
\pgfxyline(2.06,0)(2.06,.4)
\pgfxyline(2.02,0)(2.02,.4)
\pgfxyline(1.98,0)(1.98,.4)
\pgfxyline(1.94,0)(1.94,.4)
\pgfxyline(1.90,0)(1.90,.4)
\pgfxyline(1.86,0)(1.86,.4)
\pgfxyline(1.82,0)(1.82,.4)
\pgfxyline(1.78,0)(1.78,.4)
\pgfxyline(1.74,0)(1.74,.4)
\pgfxyline(1.70,0)(1.70,.4)
\pgfxyline(1.66,0)(1.66,.4)
\pgfxyline(1.62,0)(1.62,.4)
\pgfxyline(1.58,0)(1.58,.4)
\pgfxyline(1.54,0)(1.54,.4)
\pgfxyline(1.50,0)(1.50,.4)
\pgfxyline(1.46,0)(1.46,.4)
\pgfxyline(1.42,0)(1.42,.4)
\pgfxyline(1.38,0)(1.38,.4)
\pgfxyline(1.34,0)(1.34,.4)
\pgfxyline(1.30,0)(1.30,.4)
\pgfxyline(1.26,0)(1.26,.4)
\pgfxyline(1.22,0)(1.22,.4)
\pgfxyline(1.18,0)(1.18,.4)
\pgfxyline(1.14,0)(1.14,.4)
\pgfxyline(1.10,0)(1.10,.4)
\pgfxyline(1.06,0)(1.06,.4)
\pgfxyline(1.02,0)(1.02,.4)
\pgfxyline(.98,0)(.98,.4)
\pgfxyline(.94,0)(.94,.4)
\pgfxyline(.90,0)(.90,.4)
\pgfxyline(.86,0)(.86,.4)
\pgfxyline(.82,0)(.82,.4)
\pgfxyline(.78,0)(.78,.4)
\pgfxyline(.74,0)(.74,.4)
\pgfxyline(.70,0)(.70,.4)
\pgfxyline(.66,0)(.66,.4)
\pgfxyline(.62,0)(.62,.4)
\pgfxyline(.58,0)(.58,.4)
\pgfxyline(.54,0)(.54,.4)
\pgfxyline(.50,0)(.50,.4)
\pgfxyline(.46,0)(.46,.4)
\pgfxyline(.42,0)(.42,.4)
\pgfxyline(.38,0)(.38,.4)
\pgfxyline(.34,0)(.34,.4)
\pgfxyline(.30,0)(.30,.4)
\pgfxyline(.26,0)(.26,.4)
\pgfxyline(.22,0)(.22,.4)
\pgfxyline(.18,0)(.18,.4)
\pgfxyline(.14,0)(.14,.4)
\pgfxyline(.10,0)(.10,.4)
\pgfxyline(.06,0)(.06,.4)
\pgfxyline(.02,0)(.02,.4)
\color{red}
\pgfsetlinewidth{1pt}
\pgfxyline(-2.24,.4)(2.24,.4)
\pgfxyline(-2.24,.8)(2.24,.8)
\pgfxyline(-2.24,1.4)(2.24,1.4)
\pgfxyline(-2.24,2)(2.24,2)
\pgfxyline(-2.24,2.6)(2.24,2.6)
\color{black}
\pgfsetlinewidth{.2pt}
\pgfmoveto{\pgfxy(-2.24,0)}
\pgflineto{\pgfxy(2.24,0)}
\pgflineto{\pgfxy(2.24,3.2)}
\pgflineto{\pgfxy(-2.24,3.2)}
\pgflineto{\pgfxy(-2.24,0)}
\pgfstroke
\pgfmoveto{\pgfxy(.01,0)}
\pgflineto{\pgfxy(.03,.2)}
\pgflineto{\pgfxy(.06,.4)}
\pgflineto{\pgfxy(.10,.6)}
\pgflineto{\pgfxy(.16,.8)}
\pgflineto{\pgfxy(.24,1)}
\pgflineto{\pgfxy(.32,1.2)}
\pgflineto{\pgfxy(.40,1.4)}
\pgflineto{\pgfxy(.56,1.6)}
\pgflineto{\pgfxy(.72,1.8)}
\pgflineto{\pgfxy(.96,2)}
\pgflineto{\pgfxy(1.28,2.2)}
\pgflineto{\pgfxy(1.60,2.4)}
\pgflineto{\pgfxy(1.92,2.6)}
\pgflineto{\pgfxy(-1.92,2.6)}
\pgflineto{\pgfxy(-1.60,2.4)}
\pgflineto{\pgfxy(-1.28,2.2)}
\pgflineto{\pgfxy(-.96,2)}
\pgflineto{\pgfxy(-.72,1.8)}
\pgflineto{\pgfxy(-.56,1.6)}
\pgflineto{\pgfxy(-.40,1.4)}
\pgflineto{\pgfxy(-.32,1.2)}
\pgflineto{\pgfxy(-.24,1)}
\pgflineto{\pgfxy(-.16,.8)}
\pgflineto{\pgfxy(-.10,.6)}
\pgflineto{\pgfxy(-.06,.4)}
\pgflineto{\pgfxy(-.03,.2)}
\pgflineto{\pgfxy(-.01,0)}
\pgflineto{\pgfxy(.01,0)}
\pgfstroke
\end{pgfmagnify}
\end{pgftranslate}
\begin{pgftranslate}{\pgfpoint{9.6cm}{0cm}}
\begin{pgfmagnify}{1.3}{1.3}
\color{blue!15}
\pgfmoveto{\pgfxy(-2.24,2.6)}
\pgflineto{\pgfxy(2.24,2.6)}
\pgflineto{\pgfxy(2.24,3.2)}
\pgflineto{\pgfxy(-2.24,3.2)}
\pgflineto{\pgfxy(-2.24,2.6)}
\pgffill
\color{red!20}
\pgfmoveto{\pgfxy(-2.24,2.6)}
\pgflineto{\pgfxy(2.24,2.6)}
\pgflineto{\pgfxy(2.24,0)}
\pgflineto{\pgfxy(-2.24,0)}
\pgflineto{\pgfxy(-2.24,2.6)}
\pgffill
\color{red!30}
\pgfmoveto{\pgfxy(.32,2.6)}
\pgflineto{\pgfxy(.64,2.4)}
\pgflineto{\pgfxy(.96,2.2)}
\pgflineto{\pgfxy(1.12,2.0)}
\pgflineto{\pgfxy(1.28,1.8)}
\pgflineto{\pgfxy(1.44,1.6)}
\pgflineto{\pgfxy(1.52,1.4)}
\pgflineto{\pgfxy(1.60,1.2)}
\pgflineto{\pgfxy(1.68,1.0)}
\pgflineto{\pgfxy(1.72,.8)}
\pgflineto{\pgfxy(1.76,.6)}
\pgflineto{\pgfxy(1.78,.4)}
\pgflineto{\pgfxy(1.80,.2)}
\pgflineto{\pgfxy(1.82,0)}
\pgflineto{\pgfxy(-1.82,0)}
\pgflineto{\pgfxy(-1.80,.2)}
\pgflineto{\pgfxy(-1.78,.4)}
\pgflineto{\pgfxy(-1.76,.6)}
\pgflineto{\pgfxy(-1.72,.8)}
\pgflineto{\pgfxy(-1.68,1.0)}
\pgflineto{\pgfxy(-1.60,1.2)}
\pgflineto{\pgfxy(-1.52,1.4)}
\pgflineto{\pgfxy(-1.44,1.6)}
\pgflineto{\pgfxy(-1.28,1.8)}
\pgflineto{\pgfxy(-1.12,2.0)}
\pgflineto{\pgfxy(-.96,2.2)}
\pgflineto{\pgfxy(-.64,2.4)}
\pgflineto{\pgfxy(-.32,2.6)}
\pgflineto{\pgfxy(.32,2.6)}
\pgffill
\color{black!30}
\pgfsetlinewidth{.2pt}
\pgfxyline(-2.24,.2)(2.24,.2)
\pgfxyline(-2.24,.4)(2.24,.4)
\pgfxyline(-2.24,.6)(2.24,.6)
\pgfxyline(-2.24,.8)(2.24,.8)
\pgfxyline(-2.24,1.0)(2.24,1.0)
\pgfxyline(-2.24,1.2)(2.24,1.2)
\pgfxyline(-2.24,1.4)(2.24,1.4)
\pgfxyline(-2.24,1.6)(2.24,1.6)
\pgfxyline(-2.24,1.8)(2.24,1.8)
\pgfxyline(-2.24,2.0)(2.24,2.0)
\pgfxyline(-2.24,2.2)(2.24,2.2)
\pgfxyline(-2.24,2.4)(2.24,2.4)
\pgfxyline(-2.24,2.8)(2.24,2.8)
\pgfxyline(-2.24,3.0)(2.24,3.0)
\pgfxyline(-1.92,0)(-1.92,2.6)
\pgfxyline(-1.60,0)(-1.60,2.6)
\pgfxyline(-1.28,0)(-1.28,2.6)
\pgfxyline(-.96,0)(-.96,2.6)
\pgfxyline(-.64,0)(-.64,2.6)
\pgfxyline(-.32,0)(-.32,2.6)
\pgfxyline(0,0)(0,2.6)
\pgfxyline(.32,0)(.32,2.6)
\pgfxyline(.64,0)(.64,2.6)
\pgfxyline(.96,0)(.96,2.6)
\pgfxyline(1.28,0)(1.28,2.6)
\pgfxyline(1.60,0)(1.60,2.6)
\pgfxyline(1.92,0)(1.92,2.6)
\pgfxyline(-2.08,0)(-2.08,2)
\pgfxyline(-1.76,0)(-1.76,2)
\pgfxyline(-1.44,0)(-1.44,2)
\pgfxyline(-1.12,0)(-1.12,2)
\pgfxyline(-.80,0)(-.80,2)
\pgfxyline(-.48,0)(-.48,2)
\pgfxyline(-.16,0)(-.16,2)
\pgfxyline(-2.16,0)(-2.16,1.4)
\pgfxyline(-2.00,0)(-2.00,1.4)
\pgfxyline(-1.84,0)(-1.84,1.4)
\pgfxyline(-1.68,0)(-1.68,1.4)
\pgfxyline(-1.52,0)(-1.52,1.4)
\pgfxyline(-1.36,0)(-1.36,1.4)
\pgfxyline(-1.20,0)(-1.20,1.4)
\pgfxyline(-1.04,0)(-1.04,1.4)
\pgfxyline(-.88,0)(-.88,1.4)
\pgfxyline(-.72,0)(-.72,1.4)
\pgfxyline(-.56,0)(-.56,1.4)
\pgfxyline(-.40,0)(-.40,1.4)
\pgfxyline(-.24,0)(-.24,1.4)
\pgfxyline(-.08,0)(-.08,1.4)
\pgfxyline(-2.20,0)(-2.20,.8)
\pgfxyline(-2.12,0)(-2.12,.8)
\pgfxyline(-2.04,0)(-2.04,.8)
\pgfxyline(-1.96,0)(-1.96,.8)
\pgfxyline(-1.88,0)(-1.88,.8)
\pgfxyline(-1.80,0)(-1.80,.8)
\pgfxyline(-1.72,0)(-1.72,.8)
\pgfxyline(-1.64,0)(-1.64,.8)
\pgfxyline(-1.56,0)(-1.56,.8)
\pgfxyline(-1.48,0)(-1.48,.8)
\pgfxyline(-1.40,0)(-1.40,.8)
\pgfxyline(-1.32,0)(-1.32,.8)
\pgfxyline(-1.24,0)(-1.24,.8)
\pgfxyline(-1.16,0)(-1.16,.8)
\pgfxyline(-1.08,0)(-1.08,.8)
\pgfxyline(-1.00,0)(-1.00,.8)
\pgfxyline(-.92,0)(-.92,.8)
\pgfxyline(-.84,0)(-.84,.8)
\pgfxyline(-.76,0)(-.76,.8)
\pgfxyline(-.68,0)(-.68,.8)
\pgfxyline(-.60,0)(-.60,.8)
\pgfxyline(-.52,0)(-.52,.8)
\pgfxyline(-.44,0)(-.44,.8)
\pgfxyline(-.36,0)(-.36,.8)
\pgfxyline(-.28,0)(-.28,.8)
\pgfxyline(-.20,0)(-.20,.8)
\pgfxyline(-.12,0)(-.12,.8)
\pgfxyline(-.04,0)(-.04,.8)
\pgfxyline(-2.22,0)(-2.22,.4)
\pgfxyline(-2.18,0)(-2.18,.4)
\pgfxyline(-2.14,0)(-2.14,.4)
\pgfxyline(-2.10,0)(-2.10,.4)
\pgfxyline(-2.06,0)(-2.06,.4)
\pgfxyline(-2.02,0)(-2.02,.4)
\pgfxyline(-1.98,0)(-1.98,.4)
\pgfxyline(-1.94,0)(-1.94,.4)
\pgfxyline(-1.90,0)(-1.90,.4)
\pgfxyline(-1.86,0)(-1.86,.4)
\pgfxyline(-1.82,0)(-1.82,.4)
\pgfxyline(-1.78,0)(-1.78,.4)
\pgfxyline(-1.74,0)(-1.74,.4)
\pgfxyline(-1.70,0)(-1.70,.4)
\pgfxyline(-1.66,0)(-1.66,.4)
\pgfxyline(-1.62,0)(-1.62,.4)
\pgfxyline(-1.58,0)(-1.58,.4)
\pgfxyline(-1.54,0)(-1.54,.4)
\pgfxyline(-1.50,0)(-1.50,.4)
\pgfxyline(-1.46,0)(-1.46,.4)
\pgfxyline(-1.42,0)(-1.42,.4)
\pgfxyline(-1.38,0)(-1.38,.4)
\pgfxyline(-1.34,0)(-1.34,.4)
\pgfxyline(-1.30,0)(-1.30,.4)
\pgfxyline(-1.26,0)(-1.26,.4)
\pgfxyline(-1.22,0)(-1.22,.4)
\pgfxyline(-1.18,0)(-1.18,.4)
\pgfxyline(-1.14,0)(-1.14,.4)
\pgfxyline(-1.10,0)(-1.10,.4)
\pgfxyline(-1.06,0)(-1.06,.4)
\pgfxyline(-1.02,0)(-1.02,.4)
\pgfxyline(-.98,0)(-.98,.4)
\pgfxyline(-.94,0)(-.94,.4)
\pgfxyline(-.90,0)(-.90,.4)
\pgfxyline(-.86,0)(-.86,.4)
\pgfxyline(-.82,0)(-.82,.4)
\pgfxyline(-.78,0)(-.78,.4)
\pgfxyline(-.74,0)(-.74,.4)
\pgfxyline(-.70,0)(-.70,.4)
\pgfxyline(-.66,0)(-.66,.4)
\pgfxyline(-.62,0)(-.62,.4)
\pgfxyline(-.58,0)(-.58,.4)
\pgfxyline(-.54,0)(-.54,.4)
\pgfxyline(-.50,0)(-.50,.4)
\pgfxyline(-.46,0)(-.46,.4)
\pgfxyline(-.42,0)(-.42,.4)
\pgfxyline(-.38,0)(-.38,.4)
\pgfxyline(-.34,0)(-.34,.4)
\pgfxyline(-.30,0)(-.30,.4)
\pgfxyline(-.26,0)(-.26,.4)
\pgfxyline(-.22,0)(-.22,.4)
\pgfxyline(-.18,0)(-.18,.4)
\pgfxyline(-.14,0)(-.14,.4)
\pgfxyline(-.10,0)(-.10,.4)
\pgfxyline(-.06,0)(-.06,.4)
\pgfxyline(-.02,0)(-.02,.4)
\pgfxyline(2.08,0)(2.08,2)
\pgfxyline(1.76,0)(1.76,2)
\pgfxyline(1.44,0)(1.44,2)
\pgfxyline(1.12,0)(1.12,2)
\pgfxyline(.80,0)(.80,2)
\pgfxyline(.48,0)(.48,2)
\pgfxyline(.16,0)(.16,2)
\pgfxyline(2.16,0)(2.16,1.4)
\pgfxyline(2.00,0)(2.00,1.4)
\pgfxyline(1.84,0)(1.84,1.4)
\pgfxyline(1.68,0)(1.68,1.4)
\pgfxyline(1.52,0)(1.52,1.4)
\pgfxyline(1.36,0)(1.36,1.4)
\pgfxyline(1.20,0)(1.20,1.4)
\pgfxyline(1.04,0)(1.04,1.4)
\pgfxyline(.88,0)(.88,1.4)
\pgfxyline(.72,0)(.72,1.4)
\pgfxyline(.56,0)(.56,1.4)
\pgfxyline(.40,0)(.40,1.4)
\pgfxyline(.24,0)(.24,1.4)
\pgfxyline(.08,0)(.08,1.4)
\pgfxyline(2.20,0)(2.20,.8)
\pgfxyline(2.12,0)(2.12,.8)
\pgfxyline(2.04,0)(2.04,.8)
\pgfxyline(1.96,0)(1.96,.8)
\pgfxyline(1.88,0)(1.88,.8)
\pgfxyline(1.80,0)(1.80,.8)
\pgfxyline(1.72,0)(1.72,.8)
\pgfxyline(1.64,0)(1.64,.8)
\pgfxyline(1.56,0)(1.56,.8)
\pgfxyline(1.48,0)(1.48,.8)
\pgfxyline(1.40,0)(1.40,.8)
\pgfxyline(1.32,0)(1.32,.8)
\pgfxyline(1.24,0)(1.24,.8)
\pgfxyline(1.16,0)(1.16,.8)
\pgfxyline(1.08,0)(1.08,.8)
\pgfxyline(1.00,0)(1.00,.8)
\pgfxyline(.92,0)(.92,.8)
\pgfxyline(.84,0)(.84,.8)
\pgfxyline(.76,0)(.76,.8)
\pgfxyline(.68,0)(.68,.8)
\pgfxyline(.60,0)(.60,.8)
\pgfxyline(.52,0)(.52,.8)
\pgfxyline(.44,0)(.44,.8)
\pgfxyline(.36,0)(.36,.8)
\pgfxyline(.28,0)(.28,.8)
\pgfxyline(.20,0)(.20,.8)
\pgfxyline(.12,0)(.12,.8)
\pgfxyline(.04,0)(.04,.8)
\pgfxyline(2.22,0)(2.22,.4)
\pgfxyline(2.18,0)(2.18,.4)
\pgfxyline(2.14,0)(2.14,.4)
\pgfxyline(2.10,0)(2.10,.4)
\pgfxyline(2.06,0)(2.06,.4)
\pgfxyline(2.02,0)(2.02,.4)
\pgfxyline(1.98,0)(1.98,.4)
\pgfxyline(1.94,0)(1.94,.4)
\pgfxyline(1.90,0)(1.90,.4)
\pgfxyline(1.86,0)(1.86,.4)
\pgfxyline(1.82,0)(1.82,.4)
\pgfxyline(1.78,0)(1.78,.4)
\pgfxyline(1.74,0)(1.74,.4)
\pgfxyline(1.70,0)(1.70,.4)
\pgfxyline(1.66,0)(1.66,.4)
\pgfxyline(1.62,0)(1.62,.4)
\pgfxyline(1.58,0)(1.58,.4)
\pgfxyline(1.54,0)(1.54,.4)
\pgfxyline(1.50,0)(1.50,.4)
\pgfxyline(1.46,0)(1.46,.4)
\pgfxyline(1.42,0)(1.42,.4)
\pgfxyline(1.38,0)(1.38,.4)
\pgfxyline(1.34,0)(1.34,.4)
\pgfxyline(1.30,0)(1.30,.4)
\pgfxyline(1.26,0)(1.26,.4)
\pgfxyline(1.22,0)(1.22,.4)
\pgfxyline(1.18,0)(1.18,.4)
\pgfxyline(1.14,0)(1.14,.4)
\pgfxyline(1.10,0)(1.10,.4)
\pgfxyline(1.06,0)(1.06,.4)
\pgfxyline(1.02,0)(1.02,.4)
\pgfxyline(.98,0)(.98,.4)
\pgfxyline(.94,0)(.94,.4)
\pgfxyline(.90,0)(.90,.4)
\pgfxyline(.86,0)(.86,.4)
\pgfxyline(.82,0)(.82,.4)
\pgfxyline(.78,0)(.78,.4)
\pgfxyline(.74,0)(.74,.4)
\pgfxyline(.70,0)(.70,.4)
\pgfxyline(.66,0)(.66,.4)
\pgfxyline(.62,0)(.62,.4)
\pgfxyline(.58,0)(.58,.4)
\pgfxyline(.54,0)(.54,.4)
\pgfxyline(.50,0)(.50,.4)
\pgfxyline(.46,0)(.46,.4)
\pgfxyline(.42,0)(.42,.4)
\pgfxyline(.38,0)(.38,.4)
\pgfxyline(.34,0)(.34,.4)
\pgfxyline(.30,0)(.30,.4)
\pgfxyline(.26,0)(.26,.4)
\pgfxyline(.22,0)(.22,.4)
\pgfxyline(.18,0)(.18,.4)
\pgfxyline(.14,0)(.14,.4)
\pgfxyline(.10,0)(.10,.4)
\pgfxyline(.06,0)(.06,.4)
\pgfxyline(.02,0)(.02,.4)
\color{red}
\pgfsetlinewidth{1pt}
\pgfxyline(-2.24,.4)(2.24,.4)
\pgfxyline(-2.24,.8)(2.24,.8)
\pgfxyline(-2.24,1.4)(2.24,1.4)
\pgfxyline(-2.24,2)(2.24,2)
\pgfxyline(-2.24,2.6)(2.24,2.6)
\color{black}
\pgfsetlinewidth{.2pt}
\pgfmoveto{\pgfxy(-2.24,0)}
\pgflineto{\pgfxy(2.24,0)}
\pgflineto{\pgfxy(2.24,3.2)}
\pgflineto{\pgfxy(-2.24,3.2)}
\pgflineto{\pgfxy(-2.24,0)}
\pgfstroke
\pgfmoveto{\pgfxy(.32,2.6)}
\pgflineto{\pgfxy(.64,2.4)}
\pgflineto{\pgfxy(.96,2.2)}
\pgflineto{\pgfxy(1.12,2.0)}
\pgflineto{\pgfxy(1.28,1.8)}
\pgflineto{\pgfxy(1.44,1.6)}
\pgflineto{\pgfxy(1.52,1.4)}
\pgflineto{\pgfxy(1.60,1.2)}
\pgflineto{\pgfxy(1.68,1.0)}
\pgflineto{\pgfxy(1.72,.8)}
\pgflineto{\pgfxy(1.76,.6)}
\pgflineto{\pgfxy(1.78,.4)}
\pgflineto{\pgfxy(1.80,.2)}
\pgflineto{\pgfxy(1.82,0)}
\pgflineto{\pgfxy(-1.82,0)}
\pgflineto{\pgfxy(-1.80,.2)}
\pgflineto{\pgfxy(-1.78,.4)}
\pgflineto{\pgfxy(-1.76,.6)}
\pgflineto{\pgfxy(-1.72,.8)}
\pgflineto{\pgfxy(-1.68,1.0)}
\pgflineto{\pgfxy(-1.60,1.2)}
\pgflineto{\pgfxy(-1.52,1.4)}
\pgflineto{\pgfxy(-1.44,1.6)}
\pgflineto{\pgfxy(-1.28,1.8)}
\pgflineto{\pgfxy(-1.12,2.0)}
\pgflineto{\pgfxy(-.96,2.2)}
\pgflineto{\pgfxy(-.64,2.4)}
\pgflineto{\pgfxy(-.32,2.6)}
\pgflineto{\pgfxy(.32,2.6)}
\pgfstroke
\end{pgfmagnify}
\end{pgftranslate}

\pgfputat{\pgfxy(2.0,3.75)}{\pgfbox[left,center]{dense layers}}
\pgfputat{\pgfxy(1.3,2.2)}{\pgfbox[left,center]{convolutional layers}}
\pgfputat{\pgfxy(8.6,3.75)}{\pgfbox[left,center]{dense layers}}
\pgfputat{\pgfxy(7.9,2.2)}{\pgfbox[left,center]{convolutional layers}}

\end{pgfpicture}
\caption{Horizon growth in a large CNN.}
\label{figVGG}
\end{figure}

{\bf Background: dense layers in a CNN.}  We next highlight the cost of using dense layers near the output in a large CNN, a common design method. Figure \hyperref[figVGG]{\ref{figVGG}} illustrates past and future horizon growth in such a network, loosely based on the influential {\it VGG network} \cite{SimonyanVGG14}.  Progress in DNN construction is so rapid that the original VGG designs have been surpassed, but these designs still provide useful comparisons since they have been so deeply analyzed.  Each convolutional layer in our network is arranged to form a three-dimensional ``spatial section" of the network, where two dimensions involve actual spatial geometry of input images, and the third represents different {\it filters} used for feature extraction.  The entire network is therefore four-dimensional, with a single dimension of {\it process time} pointing up the page.  Like Minkowski spacetime diagrams, these diagrams  suppress two ``spatial" dimensions.  Each cell in the lower (red) regions represents a neuron in a $1$-dimensional cross-section of a convolutional layer.  Cells grow larger near the top of these regions to represent {\it pooling,} whereby activation functions are combined over small patches of neurons.  Red lines indicate pooling layers.  Blue regions indicate stacks of dense layers, used to synthesize, interpret, and classify combinations of features extracted by the convolutional layers.  We focus for the moment on horizon growth in the convolutional layers, looking forward in process time from the input layer and backward from the final pooling layer.  The top cross section in the right-hand diagram consists of two cells rather than one, since the final pooling layer combines these cells prior to the first dense layer.  While there is reasonably high connectivity between the input layer and the final pooling layer, shorter-range connectivity between pairs of convolutional layers is rather poor.  This limits early exchange of information and leaves the entire synthesis process to the dense layers.  

We now pose a second question: {\it can the synthesis process in a large CNN be partially superposed with the extraction process, allowing reduction or elimination of one or more dense layers?}  Recent results suggest that the answer is affirmative \cite{Alford18,Prabhu18,Robinett18}. The reason for this question is that dense layers are costly in terms of parameter space dimensions.  To be specific, the final pooling layer in configuration D of the VGG network has $7\times7\times2^9\approx 2.5\times10^4$ neurons, due to the use of $2^9$ filters in the final convolutional layer \cite{Aggarwal18,Karpathy16}.  The first dense layer has $2^{12}$ neurons, so the number of connections between the two is roughly $10^8$, contributing more than $74\%$ of all network parameters. The remaining dense layers contribute a further $15\%$.  This strongly incentivizes replacing dense layers with sparser structure.  We mention four out of many possible methods to accomplish this.  First, one may simply remove a large proportion of edges in the dense layers, leaving behind sparse random layers, similar to pseudorandom {\it X-Linear layers} \cite{Prabhu18}.  The success of various pruning and {\it dropout} methods hints at this approach.  Second, one may employ non-random sparse architectures, as in {\it RadiX-Nets} \cite{Robinett18}.  Third, one may employ {\it fuzzy} or {\it smeared kernels} to obtain a hybrid local/random structure, in which connections to a given neuron are locally concentrated, but not sharply cut off outside a hyperrectangular window.  Fourth, as a special case of the third method, one may augment a pure CNN by superposing sparse random structure.  We illustrate a toy example of this method below.   In general, such methods may also incorporate generation-skipping edges, as in {\it ResNet} \cite{HeResnet16} and {\it DenseNet} \cite{HuangDenseNet2016}. 

\begin{figure}[h]
\centering
\begin{subfigure}{.5\textwidth}
 \centering
  \includegraphics[width=1.1\linewidth]{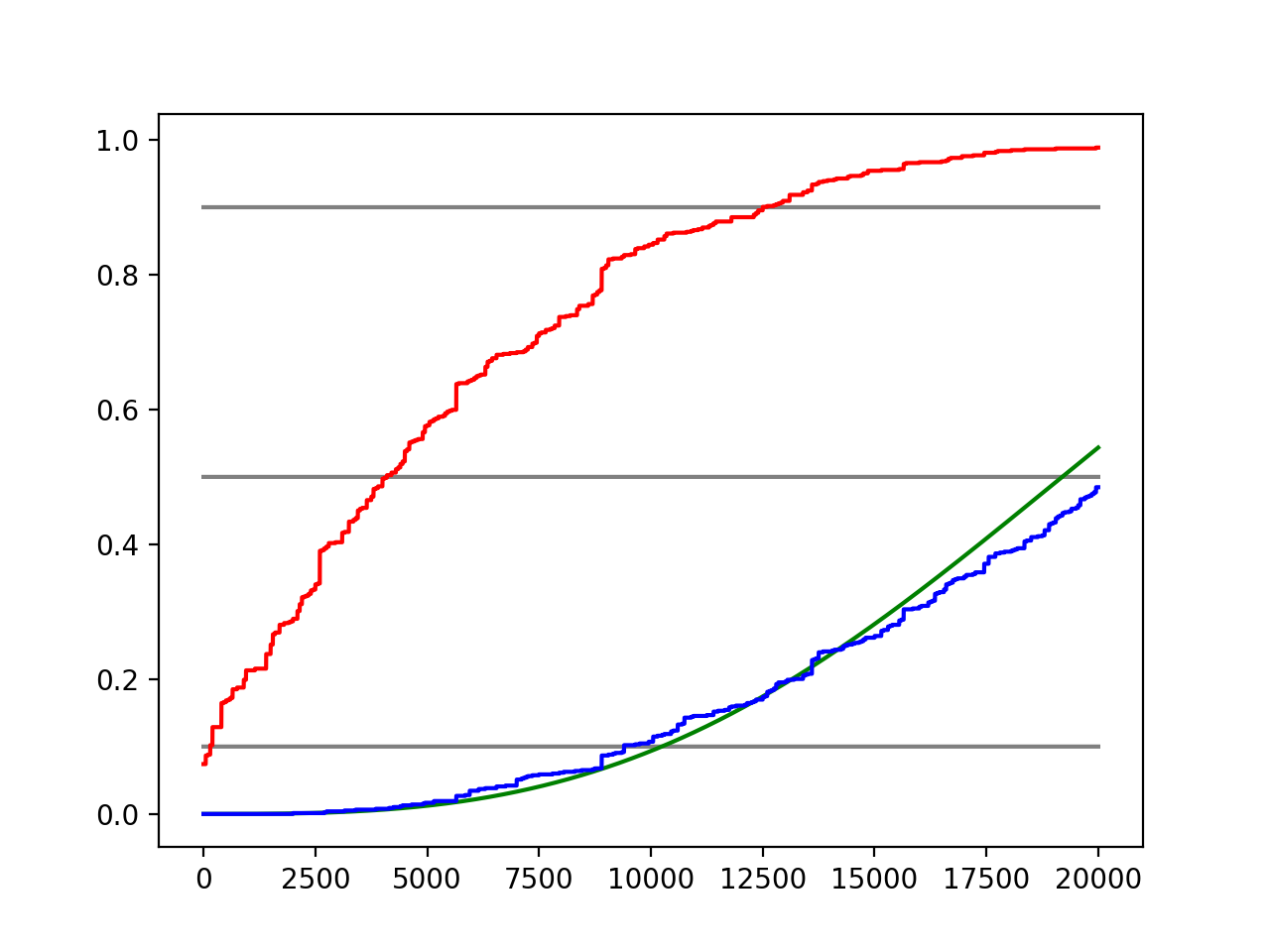}
 \end{subfigure}%
\begin{subfigure}{.5\textwidth}
 \centering
 \includegraphics[width=1.1\linewidth]{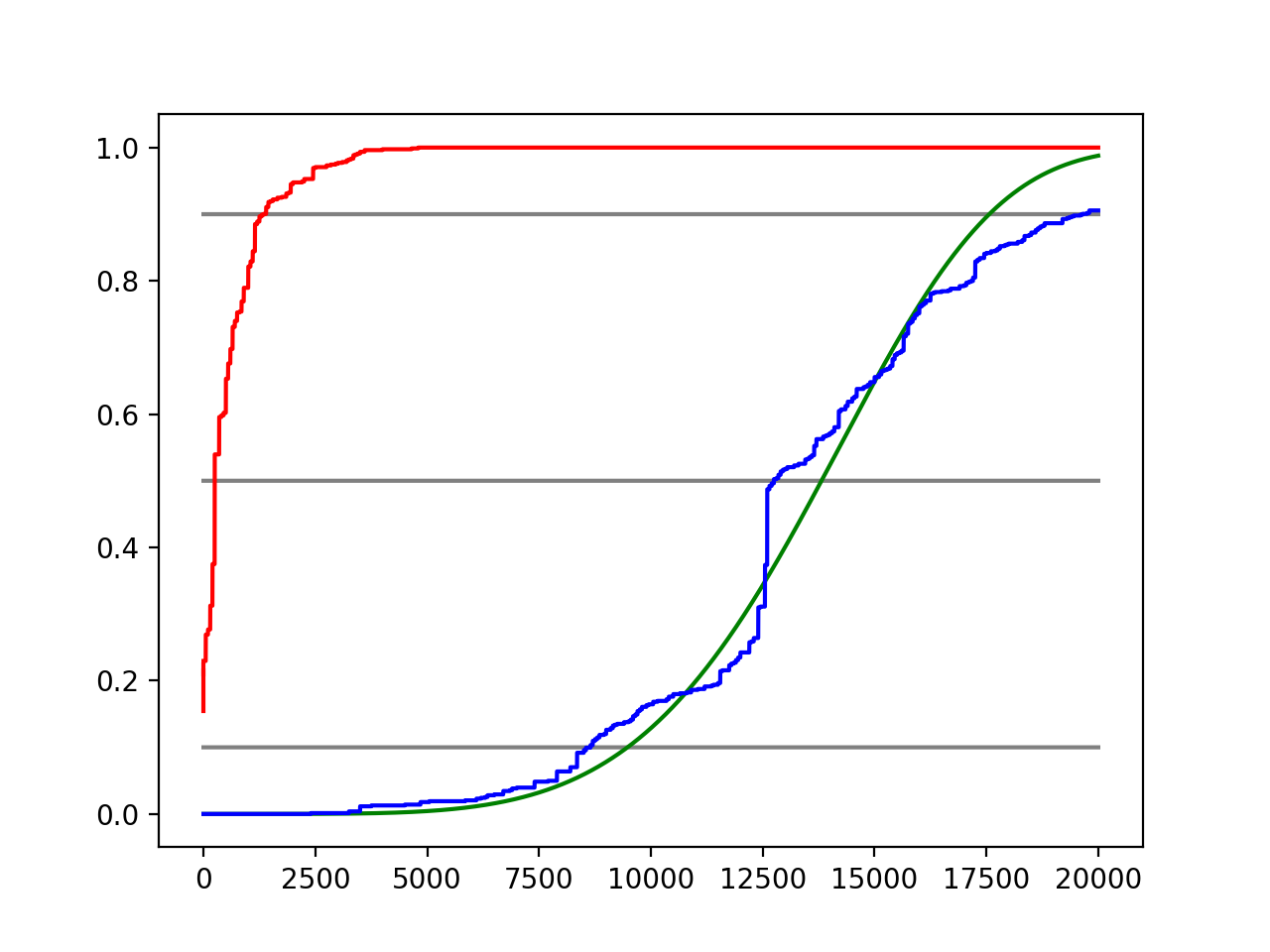}
\end{subfigure}
\caption{Augmenting CNNs with sparse random structure.}
\label{figaugmentedCNNs}
\end{figure}

{\bf Example: augmented CNNs.}  Figure \hyperref[figaugmentedCNNs]{\ref{figaugmentedCNNs}} illustrates the effect of augmenting toy CNNs by superposing sparse random structure.  In the left-hand diagram, the CNN has three layers (four generations, including input), while in the right-hand diagram it has five layers (six generations).  Each generation has $28\times28=784$ neurons, the input size for MNIST digit recognition.  The {\it kernel size} of each CNN is $3\times 3$, and the {\it stride length} is $1$.  The total numbers of permissible edges are about $1.8\times10^6$ and $3.1\times10^6$, but the CNNs have only about  $2\times 10^4$ and $3.4\times 10^4$ edges, with $6\%$ and $15\%$ connectivity between initial and final generations.  Green curves show the connection probability estimates \eqref{Mgenintro} for random edge-addition processes with the same generation sizes, beginning with zero edges.  Blue curves snapshot actual connectivity for two such processes, measuring the proportion of connections between a central initial-generation neuron and the final layer.  Edge and corner neurons behave similarly. Horizontal lines indicate $10\%$, $50\%$, and $90\%$ connectivity.  For example, in the left-hand diagram, $K\approx2\times 10^4$ random edges produce about $50\%$ connectivity.  Accessibility phase transitions occur near $K\approx 2\times10^4$ and $K\approx 1.5\times10^4$.   Red curves show the effect of superposing the same edges on the CNNs, forming hybrid local/random structures.  

\vspace*{-.3cm}

\begin{figure}[h]
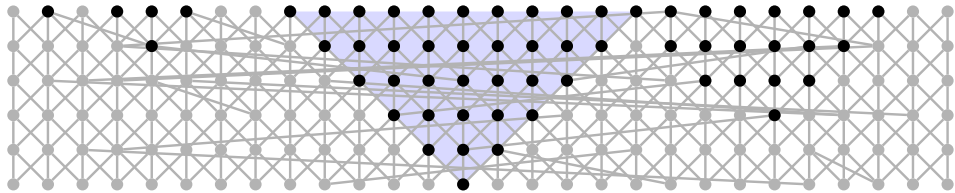


\caption{Horizon growth in a hybrid augmented CNN.}
\label{figaugmentedconv}
\end{figure}

Figure \hyperref[figaugmentedconv]{\ref{figaugmentedconv}} shows a $2$-dimensional cross section of the right-hand hybrid network from Figure \hyperref[figaugmentedCNNs]{\ref{figaugmentedCNNs}} after a few hundred random edges are added.  The shaded triangle illustrates horizon growth in the original CNN, while the black nodes illustrate the faster horizon growth in the hybrid network.  Adding about $2500$ random edges increases the connectivity from $15\%$ to $95\%$, even though the independent connectivity of these random edges is negligible.  A single dense layer, meanwhile, costs roughly $6\times10^5$ edges.

\begin{figure}[h]
\centering
\begin{subfigure}{.5\textwidth}
 \centering
  \includegraphics[width=1.1\linewidth]{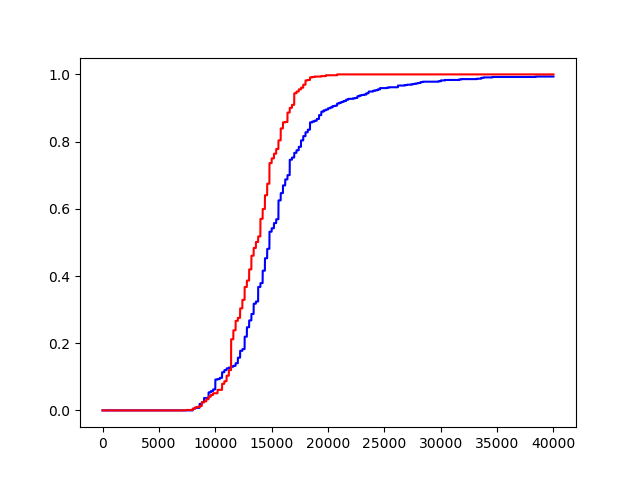}
 \end{subfigure}%
\begin{subfigure}{.5\textwidth}
 \centering
 \includegraphics[width=1.1\linewidth]{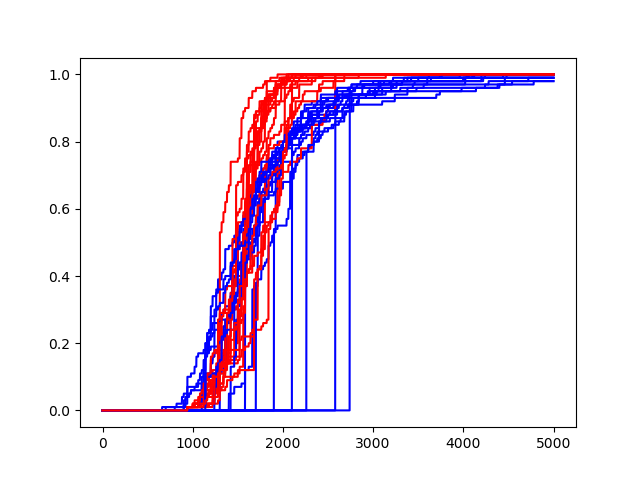}
\end{subfigure}
\caption{Local/random connectivity beats purely random connectivity.}
\label{figcake}
\end{figure}

{\bf Hybrid synergy: more connected} {\it and} {\bf more local.} Simulations reveal a fortuitous synergy for large networks: $K$ local/random edges can produce better connectivity than $K$ random edges, even though purely random structure offers much better connectivity than purely local structure.  Hybrid design therefore offers a ``cake and eat it too" possibility for combining the structural advantages of convolutional and dense layers.  The left-hand diagram in Figure \hyperref[figcake]{\ref{figcake}} compares a random network (blue) to a hybrid local/random network (red) with the same number of edges.  Each curve shows the proportion of connections between a central first-generation neuron and the final generation.  The random network is constructed by adding $40000$ random edges to $10$ layers ($11$ generations, including input) of $784$ nodes each.  The hybrid network is constructed by first adding $7840$ edges in a minimal convolutional structure with kernel size $1$, then adding the final $31160$ edges from the random network.  The hybrid network requires only half as many edges to reach $100\%$ connectivity, while retaining local properties absent in the random network. 


\begin{figure}[h]
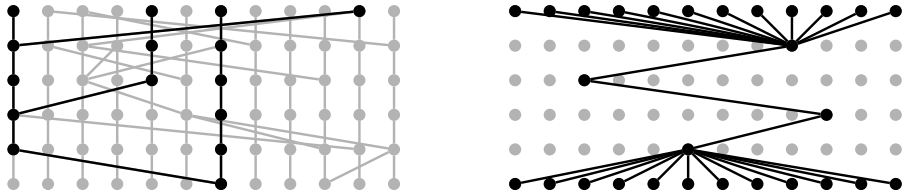

\begin{pgfpicture}{0cm}{0cm}{15cm}{2.8cm}


\begin{pgfmagnify}{2.3}{2.3}
\begin{pgftranslate}{\pgfpoint{.2cm}{.1cm}}
\color{black!30}
\pgfnodecircle{Node0101}[fill]{\pgfxy(0,0)}{0.035cm}
\pgfnodecircle{Node0102}[fill]{\pgfxy(.2,0)}{0.035cm}
\pgfnodecircle{Node0103}[fill]{\pgfxy(.4,0)}{0.035cm}
\pgfnodecircle{Node0104}[fill]{\pgfxy(.6,0)}{0.035cm}
\pgfnodecircle{Node0105}[fill]{\pgfxy(.8,0)}{0.035cm}
\pgfnodecircle{Node0106}[fill]{\pgfxy(1,0)}{0.035cm}
\pgfnodecircle{Node0107}[fill]{\pgfxy(1.2,0)}{0.035cm}
\pgfnodecircle{Node0108}[fill]{\pgfxy(1.4,0)}{0.035cm}
\pgfnodecircle{Node0109}[fill]{\pgfxy(1.6,0)}{0.035cm}
\pgfnodecircle{Node0110}[fill]{\pgfxy(1.8,0)}{0.035cm}
\pgfnodecircle{Node0111}[fill]{\pgfxy(2,0)}{0.035cm}
\pgfnodecircle{Node0112}[fill]{\pgfxy(2.2,0)}{0.035cm}
\pgfnodecircle{Node0201}[fill]{\pgfxy(0,.2)}{0.035cm}
\pgfnodecircle{Node0202}[fill]{\pgfxy(.2,.2)}{0.035cm}
\pgfnodecircle{Node0203}[fill]{\pgfxy(.4,.2)}{0.035cm}
\pgfnodecircle{Node0204}[fill]{\pgfxy(.6,.2)}{0.035cm}
\pgfnodecircle{Node0205}[fill]{\pgfxy(.8,.2)}{0.035cm}
\pgfnodecircle{Node0206}[fill]{\pgfxy(1,.2)}{0.035cm}
\pgfnodecircle{Node0207}[fill]{\pgfxy(1.2,.2)}{0.035cm}
\pgfnodecircle{Node0208}[fill]{\pgfxy(1.4,.2)}{0.035cm}
\pgfnodecircle{Node0209}[fill]{\pgfxy(1.6,.2)}{0.035cm}
\pgfnodecircle{Node0210}[fill]{\pgfxy(1.8,.2)}{0.035cm}
\pgfnodecircle{Node0211}[fill]{\pgfxy(2,.2)}{0.035cm}
\pgfnodecircle{Node0212}[fill]{\pgfxy(2.2,.2)}{0.035cm}
\pgfnodecircle{Node0301}[fill]{\pgfxy(0,.4)}{0.035cm}
\pgfnodecircle{Node0302}[fill]{\pgfxy(.2,.4)}{0.035cm}
\pgfnodecircle{Node0303}[fill]{\pgfxy(.4,.4)}{0.035cm}
\pgfnodecircle{Node0304}[fill]{\pgfxy(.6,.4)}{0.035cm}
\pgfnodecircle{Node0305}[fill]{\pgfxy(.8,.4)}{0.035cm}
\pgfnodecircle{Node0306}[fill]{\pgfxy(1,.4)}{0.035cm}
\pgfnodecircle{Node0307}[fill]{\pgfxy(1.2,.4)}{0.035cm}
\pgfnodecircle{Node0308}[fill]{\pgfxy(1.4,.4)}{0.035cm}
\pgfnodecircle{Node0309}[fill]{\pgfxy(1.6,.4)}{0.035cm}
\pgfnodecircle{Node0310}[fill]{\pgfxy(1.8,.4)}{0.035cm}
\pgfnodecircle{Node0311}[fill]{\pgfxy(2,.4)}{0.035cm}
\pgfnodecircle{Node0312}[fill]{\pgfxy(2.2,.4)}{0.035cm}
\pgfnodecircle{Node0401}[fill]{\pgfxy(0,.6)}{0.035cm}
\pgfnodecircle{Node0402}[fill]{\pgfxy(.2,.6)}{0.035cm}
\pgfnodecircle{Node0403}[fill]{\pgfxy(.4,.6)}{0.035cm}
\pgfnodecircle{Node0404}[fill]{\pgfxy(.6,.6)}{0.035cm}
\pgfnodecircle{Node0405}[fill]{\pgfxy(.8,.6)}{0.035cm}
\pgfnodecircle{Node0406}[fill]{\pgfxy(1,.6)}{0.035cm}
\pgfnodecircle{Node0407}[fill]{\pgfxy(1.2,.6)}{0.035cm}
\pgfnodecircle{Node0408}[fill]{\pgfxy(1.4,.6)}{0.035cm}
\pgfnodecircle{Node0409}[fill]{\pgfxy(1.6,.6)}{0.035cm}
\pgfnodecircle{Node0410}[fill]{\pgfxy(1.8,.6)}{0.035cm}
\pgfnodecircle{Node0411}[fill]{\pgfxy(2,.6)}{0.035cm}
\pgfnodecircle{Node0412}[fill]{\pgfxy(2.2,.6)}{0.035cm}
\pgfnodecircle{Node0501}[fill]{\pgfxy(0,.8)}{0.035cm}
\pgfnodecircle{Node0502}[fill]{\pgfxy(.2,.8)}{0.035cm}
\pgfnodecircle{Node0503}[fill]{\pgfxy(.4,.8)}{0.035cm}
\pgfnodecircle{Node0504}[fill]{\pgfxy(.6,.8)}{0.035cm}
\pgfnodecircle{Node0505}[fill]{\pgfxy(.8,.8)}{0.035cm}
\pgfnodecircle{Node0506}[fill]{\pgfxy(1,.8)}{0.035cm}
\pgfnodecircle{Node0507}[fill]{\pgfxy(1.2,.8)}{0.035cm}
\pgfnodecircle{Node0508}[fill]{\pgfxy(1.4,.8)}{0.035cm}
\pgfnodecircle{Node0509}[fill]{\pgfxy(1.6,.8)}{0.035cm}
\pgfnodecircle{Node0510}[fill]{\pgfxy(1.8,.8)}{0.035cm}
\pgfnodecircle{Node0511}[fill]{\pgfxy(2,.8)}{0.035cm}
\pgfnodecircle{Node0512}[fill]{\pgfxy(2.2,.8)}{0.035cm}
\pgfnodecircle{Node0601}[fill]{\pgfxy(0,1)}{0.035cm}
\pgfnodecircle{Node0602}[fill]{\pgfxy(.2,1)}{0.035cm}
\pgfnodecircle{Node0603}[fill]{\pgfxy(.4,1)}{0.035cm}
\pgfnodecircle{Node0604}[fill]{\pgfxy(.6,1)}{0.035cm}
\pgfnodecircle{Node0605}[fill]{\pgfxy(.8,1)}{0.035cm}
\pgfnodecircle{Node0606}[fill]{\pgfxy(1,1)}{0.035cm}
\pgfnodecircle{Node0607}[fill]{\pgfxy(1.2,1)}{0.035cm}
\pgfnodecircle{Node0608}[fill]{\pgfxy(1.4,1)}{0.035cm}
\pgfnodecircle{Node0609}[fill]{\pgfxy(1.6,1)}{0.035cm}
\pgfnodecircle{Node0610}[fill]{\pgfxy(1.8,1)}{0.035cm}
\pgfnodecircle{Node0611}[fill]{\pgfxy(2,1)}{0.035cm}
\pgfnodecircle{Node0612}[fill]{\pgfxy(2.2,1)}{0.035cm}
\pgfnodeconnline{Node0101}{Node0201}
\pgfnodeconnline{Node0102}{Node0202}
\pgfnodeconnline{Node0103}{Node0203}
\pgfnodeconnline{Node0104}{Node0204}
\pgfnodeconnline{Node0105}{Node0205}
\pgfnodeconnline{Node0106}{Node0206}
\pgfnodeconnline{Node0107}{Node0207}
\pgfnodeconnline{Node0108}{Node0208}
\pgfnodeconnline{Node0109}{Node0209}
\pgfnodeconnline{Node0110}{Node0210}
\pgfnodeconnline{Node0111}{Node0211}
\pgfnodeconnline{Node0112}{Node0212}
\pgfnodeconnline{Node0201}{Node0301}
\pgfnodeconnline{Node0202}{Node0302}
\pgfnodeconnline{Node0203}{Node0303}
\pgfnodeconnline{Node0204}{Node0304}
\pgfnodeconnline{Node0205}{Node0305}
\pgfnodeconnline{Node0206}{Node0306}
\pgfnodeconnline{Node0207}{Node0307}
\pgfnodeconnline{Node0208}{Node0308}
\pgfnodeconnline{Node0209}{Node0309}
\pgfnodeconnline{Node0210}{Node0310}
\pgfnodeconnline{Node0211}{Node0311}
\pgfnodeconnline{Node0212}{Node0312}
\pgfnodeconnline{Node0301}{Node0401}
\pgfnodeconnline{Node0302}{Node0402}
\pgfnodeconnline{Node0303}{Node0403}
\pgfnodeconnline{Node0304}{Node0404}
\pgfnodeconnline{Node0305}{Node0405}
\pgfnodeconnline{Node0306}{Node0406}
\pgfnodeconnline{Node0307}{Node0407}
\pgfnodeconnline{Node0308}{Node0408}
\pgfnodeconnline{Node0309}{Node0409}
\pgfnodeconnline{Node0310}{Node0410}
\pgfnodeconnline{Node0311}{Node0411}
\pgfnodeconnline{Node0312}{Node0412}
\pgfnodeconnline{Node0401}{Node0501}
\pgfnodeconnline{Node0402}{Node0502}
\pgfnodeconnline{Node0403}{Node0503}
\pgfnodeconnline{Node0404}{Node0504}
\pgfnodeconnline{Node0405}{Node0505}
\pgfnodeconnline{Node0406}{Node0506}
\pgfnodeconnline{Node0407}{Node0507}
\pgfnodeconnline{Node0408}{Node0508}
\pgfnodeconnline{Node0409}{Node0509}
\pgfnodeconnline{Node0410}{Node0510}
\pgfnodeconnline{Node0411}{Node0511}
\pgfnodeconnline{Node0412}{Node0512}
\pgfnodeconnline{Node0501}{Node0601}
\pgfnodeconnline{Node0502}{Node0602}
\pgfnodeconnline{Node0503}{Node0603}
\pgfnodeconnline{Node0504}{Node0604}
\pgfnodeconnline{Node0505}{Node0605}
\pgfnodeconnline{Node0506}{Node0606}
\pgfnodeconnline{Node0507}{Node0607}
\pgfnodeconnline{Node0508}{Node0608}
\pgfnodeconnline{Node0509}{Node0609}
\pgfnodeconnline{Node0510}{Node0610}
\pgfnodeconnline{Node0511}{Node0611}
\pgfnodeconnline{Node0512}{Node0612}
\pgfnodeconnline{Node0306}{Node0403}
\pgfnodeconnline{Node0501}{Node0611}
\pgfnodeconnline{Node0110}{Node0212}
\pgfnodeconnline{Node0503}{Node0611}
\pgfnodeconnline{Node0212}{Node0306}
\pgfnodeconnline{Node0410}{Node0503}
\pgfnodeconnline{Node0403}{Node0504}
\pgfnodeconnline{Node0512}{Node0602}
\pgfnodeconnline{Node0508}{Node0603}
\pgfnodeconnline{Node0107}{Node0201}
\pgfnodeconnline{Node0406}{Node0502}
\pgfnodeconnline{Node0211}{Node0301}
\pgfnodeconnline{Node0301}{Node0405}
\pgfnodeconnline{Node0210}{Node0306}
\pgfnodeconnline{Node0403}{Node0507}
\color{black}
\pgfnodecircle{Node0107}[fill]{\pgfxy(1.2,0)}{0.035cm}
\pgfnodecircle{Node0201}[fill]{\pgfxy(0,.2)}{0.035cm}
\pgfnodecircle{Node0207}[fill]{\pgfxy(1.2,.2)}{0.035cm}
\pgfnodecircle{Node0301}[fill]{\pgfxy(0,.4)}{0.035cm}
\pgfnodecircle{Node0307}[fill]{\pgfxy(1.2,.4)}{0.035cm}
\pgfnodecircle{Node0401}[fill]{\pgfxy(0,.6)}{0.035cm}
\pgfnodecircle{Node0405}[fill]{\pgfxy(.8,.6)}{0.035cm}
\pgfnodecircle{Node0407}[fill]{\pgfxy(1.2,.6)}{0.035cm}
\pgfnodecircle{Node0501}[fill]{\pgfxy(0,.8)}{0.035cm}
\pgfnodecircle{Node0505}[fill]{\pgfxy(.8,.8)}{0.035cm}
\pgfnodecircle{Node0507}[fill]{\pgfxy(1.2,.8)}{0.035cm}
\pgfnodecircle{Node0601}[fill]{\pgfxy(0,1)}{0.035cm}
\pgfnodecircle{Node0605}[fill]{\pgfxy(.8,1)}{0.035cm}
\pgfnodecircle{Node0607}[fill]{\pgfxy(1.2,1)}{0.035cm}
\pgfnodecircle{Node0611}[fill]{\pgfxy(2.0,1)}{0.035cm}
\pgfnodeconnline{Node0107}{Node0201}
\pgfnodeconnline{Node0107}{Node0207}
\pgfnodeconnline{Node0201}{Node0301}
\pgfnodeconnline{Node0207}{Node0307}
\pgfnodeconnline{Node0301}{Node0401}
\pgfnodeconnline{Node0301}{Node0405}
\pgfnodeconnline{Node0307}{Node0407}
\pgfnodeconnline{Node0401}{Node0501}
\pgfnodeconnline{Node0405}{Node0505}
\pgfnodeconnline{Node0407}{Node0507}
\pgfnodeconnline{Node0501}{Node0601}
\pgfnodeconnline{Node0501}{Node0611}
\pgfnodeconnline{Node0505}{Node0605}
\pgfnodeconnline{Node0507}{Node0607}
\end{pgftranslate}


\begin{pgftranslate}{\pgfpoint{-.1cm}{.1cm}}
\color{black!30}
\pgfnodecircle{Node0117}[fill]{\pgfxy(3.2,0)}{0.035cm}
\pgfnodecircle{Node0118}[fill]{\pgfxy(3.4,0)}{0.035cm}
\pgfnodecircle{Node0119}[fill]{\pgfxy(3.6,0)}{0.035cm}
\pgfnodecircle{Node0120}[fill]{\pgfxy(3.8,0)}{0.035cm}
\pgfnodecircle{Node0121}[fill]{\pgfxy(4,0)}{0.035cm}
\pgfnodecircle{Node0122}[fill]{\pgfxy(4.2,0)}{0.035cm}
\pgfnodecircle{Node0123}[fill]{\pgfxy(4.4,0)}{0.035cm}
\pgfnodecircle{Node0124}[fill]{\pgfxy(4.6,0)}{0.035cm}
\pgfnodecircle{Node0125}[fill]{\pgfxy(4.8,0)}{0.035cm}
\pgfnodecircle{Node0126}[fill]{\pgfxy(5,0)}{0.035cm}
\pgfnodecircle{Node0127}[fill]{\pgfxy(5.2,0)}{0.035cm}
\pgfnodecircle{Node0128}[fill]{\pgfxy(5.4,0)}{0.035cm}
\pgfnodecircle{Node0217}[fill]{\pgfxy(3.2,.2)}{0.035cm}
\pgfnodecircle{Node0218}[fill]{\pgfxy(3.4,.2)}{0.035cm}
\pgfnodecircle{Node0219}[fill]{\pgfxy(3.6,.2)}{0.035cm}
\pgfnodecircle{Node0220}[fill]{\pgfxy(3.8,.2)}{0.035cm}
\pgfnodecircle{Node0221}[fill]{\pgfxy(4,.2)}{0.035cm}
\pgfnodecircle{Node0222}[fill]{\pgfxy(4.2,.2)}{0.035cm}
\pgfnodecircle{Node0223}[fill]{\pgfxy(4.4,.2)}{0.035cm}
\pgfnodecircle{Node0224}[fill]{\pgfxy(4.6,.2)}{0.035cm}
\pgfnodecircle{Node0225}[fill]{\pgfxy(4.8,.2)}{0.035cm}
\pgfnodecircle{Node0226}[fill]{\pgfxy(5,.2)}{0.035cm}
\pgfnodecircle{Node0227}[fill]{\pgfxy(5.2,.2)}{0.035cm}
\pgfnodecircle{Node0228}[fill]{\pgfxy(5.4,.2)}{0.035cm}
\pgfnodecircle{Node0317}[fill]{\pgfxy(3.2,.4)}{0.035cm}
\pgfnodecircle{Node0318}[fill]{\pgfxy(3.4,.4)}{0.035cm}
\pgfnodecircle{Node0319}[fill]{\pgfxy(3.6,.4)}{0.035cm}
\pgfnodecircle{Node0320}[fill]{\pgfxy(3.8,.4)}{0.035cm}
\pgfnodecircle{Node0321}[fill]{\pgfxy(4,.4)}{0.035cm}
\pgfnodecircle{Node0322}[fill]{\pgfxy(4.2,.4)}{0.035cm}
\pgfnodecircle{Node0323}[fill]{\pgfxy(4.4,.4)}{0.035cm}
\pgfnodecircle{Node0324}[fill]{\pgfxy(4.6,.4)}{0.035cm}
\pgfnodecircle{Node0325}[fill]{\pgfxy(4.8,.4)}{0.035cm}
\pgfnodecircle{Node0326}[fill]{\pgfxy(5,.4)}{0.035cm}
\pgfnodecircle{Node0327}[fill]{\pgfxy(5.2,.4)}{0.035cm}
\pgfnodecircle{Node0328}[fill]{\pgfxy(5.4,.4)}{0.035cm}
\pgfnodecircle{Node0417}[fill]{\pgfxy(3.2,.6)}{0.035cm}
\pgfnodecircle{Node0418}[fill]{\pgfxy(3.4,.6)}{0.035cm}
\pgfnodecircle{Node0419}[fill]{\pgfxy(3.6,.6)}{0.035cm}
\pgfnodecircle{Node0420}[fill]{\pgfxy(3.8,.6)}{0.035cm}
\pgfnodecircle{Node0421}[fill]{\pgfxy(4,.6)}{0.035cm}
\pgfnodecircle{Node0422}[fill]{\pgfxy(4.2,.6)}{0.035cm}
\pgfnodecircle{Node0423}[fill]{\pgfxy(4.4,.6)}{0.035cm}
\pgfnodecircle{Node0424}[fill]{\pgfxy(4.6,.6)}{0.035cm}
\pgfnodecircle{Node0425}[fill]{\pgfxy(4.8,.6)}{0.035cm}
\pgfnodecircle{Node0426}[fill]{\pgfxy(5,.6)}{0.035cm}
\pgfnodecircle{Node0427}[fill]{\pgfxy(5.2,.6)}{0.035cm}
\pgfnodecircle{Node0428}[fill]{\pgfxy(5.4,.6)}{0.035cm}
\pgfnodecircle{Node0517}[fill]{\pgfxy(3.2,.8)}{0.035cm}
\pgfnodecircle{Node0518}[fill]{\pgfxy(3.4,.8)}{0.035cm}
\pgfnodecircle{Node0519}[fill]{\pgfxy(3.6,.8)}{0.035cm}
\pgfnodecircle{Node0520}[fill]{\pgfxy(3.8,.8)}{0.035cm}
\pgfnodecircle{Node0521}[fill]{\pgfxy(4,.8)}{0.035cm}
\pgfnodecircle{Node0522}[fill]{\pgfxy(4.2,.8)}{0.035cm}
\pgfnodecircle{Node0523}[fill]{\pgfxy(4.4,.8)}{0.035cm}
\pgfnodecircle{Node0524}[fill]{\pgfxy(4.6,.8)}{0.035cm}
\pgfnodecircle{Node0525}[fill]{\pgfxy(4.8,.8)}{0.035cm}
\pgfnodecircle{Node0526}[fill]{\pgfxy(5,.8)}{0.035cm}
\pgfnodecircle{Node0527}[fill]{\pgfxy(5.2,.8)}{0.035cm}
\pgfnodecircle{Node0528}[fill]{\pgfxy(5.4,.8)}{0.035cm}
\pgfnodecircle{Node0617}[fill]{\pgfxy(3.2,1)}{0.035cm}
\pgfnodecircle{Node0618}[fill]{\pgfxy(3.4,1)}{0.035cm}
\pgfnodecircle{Node0619}[fill]{\pgfxy(3.6,1)}{0.035cm}
\pgfnodecircle{Node0620}[fill]{\pgfxy(3.8,1)}{0.035cm}
\pgfnodecircle{Node0621}[fill]{\pgfxy(4,1)}{0.035cm}
\pgfnodecircle{Node0622}[fill]{\pgfxy(4.2,1)}{0.035cm}
\pgfnodecircle{Node0623}[fill]{\pgfxy(4.4,1)}{0.035cm}
\pgfnodecircle{Node0624}[fill]{\pgfxy(4.6,1)}{0.035cm}
\pgfnodecircle{Node0625}[fill]{\pgfxy(4.8,1)}{0.035cm}
\pgfnodecircle{Node0626}[fill]{\pgfxy(5,1)}{0.035cm}
\pgfnodecircle{Node0627}[fill]{\pgfxy(5.2,1)}{0.035cm}
\pgfnodecircle{Node0628}[fill]{\pgfxy(5.4,1)}{0.035cm}
\color{black}
\pgfnodecircle{Node0117}[fill]{\pgfxy(3.2,0)}{0.035cm}
\pgfnodecircle{Node0118}[fill]{\pgfxy(3.4,0)}{0.035cm}
\pgfnodecircle{Node0119}[fill]{\pgfxy(3.6,0)}{0.035cm}
\pgfnodecircle{Node0120}[fill]{\pgfxy(3.8,0)}{0.035cm}
\pgfnodecircle{Node0121}[fill]{\pgfxy(4,0)}{0.035cm}
\pgfnodecircle{Node0122}[fill]{\pgfxy(4.2,0)}{0.035cm}
\pgfnodecircle{Node0123}[fill]{\pgfxy(4.4,0)}{0.035cm}
\pgfnodecircle{Node0124}[fill]{\pgfxy(4.6,0)}{0.035cm}
\pgfnodecircle{Node0125}[fill]{\pgfxy(4.8,0)}{0.035cm}
\pgfnodecircle{Node0126}[fill]{\pgfxy(5,0)}{0.035cm}
\pgfnodecircle{Node0127}[fill]{\pgfxy(5.2,0)}{0.035cm}
\pgfnodecircle{Node0128}[fill]{\pgfxy(5.4,0)}{0.035cm}
\pgfnodecircle{Node0222}[fill]{\pgfxy(4.2,.2)}{0.035cm}
\pgfnodecircle{Node0326}[fill]{\pgfxy(5,.4)}{0.035cm}
\pgfnodecircle{Node0419}[fill]{\pgfxy(3.6,.6)}{0.035cm}
\pgfnodecircle{Node0525}[fill]{\pgfxy(4.8,.8)}{0.035cm}
\pgfnodecircle{Node0617}[fill]{\pgfxy(3.2,1)}{0.035cm}
\pgfnodecircle{Node0618}[fill]{\pgfxy(3.4,1)}{0.035cm}
\pgfnodecircle{Node0619}[fill]{\pgfxy(3.6,1)}{0.035cm}
\pgfnodecircle{Node0620}[fill]{\pgfxy(3.8,1)}{0.035cm}
\pgfnodecircle{Node0621}[fill]{\pgfxy(4,1)}{0.035cm}
\pgfnodecircle{Node0622}[fill]{\pgfxy(4.2,1)}{0.035cm}
\pgfnodecircle{Node0623}[fill]{\pgfxy(4.4,1)}{0.035cm}
\pgfnodecircle{Node0624}[fill]{\pgfxy(4.6,1)}{0.035cm}
\pgfnodecircle{Node0625}[fill]{\pgfxy(4.8,1)}{0.035cm}
\pgfnodecircle{Node0626}[fill]{\pgfxy(5,1)}{0.035cm}
\pgfnodecircle{Node0627}[fill]{\pgfxy(5.2,1)}{0.035cm}
\pgfnodecircle{Node0628}[fill]{\pgfxy(5.4,1)}{0.035cm}
\pgfnodeconnline{Node0117}{Node0222}
\pgfnodeconnline{Node0118}{Node0222}
\pgfnodeconnline{Node0119}{Node0222}
\pgfnodeconnline{Node0120}{Node0222}
\pgfnodeconnline{Node0121}{Node0222}
\pgfnodeconnline{Node0122}{Node0222}
\pgfnodeconnline{Node0123}{Node0222}
\pgfnodeconnline{Node0124}{Node0222}
\pgfnodeconnline{Node0125}{Node0222}
\pgfnodeconnline{Node0126}{Node0222}
\pgfnodeconnline{Node0127}{Node0222}
\pgfnodeconnline{Node0128}{Node0222}
\pgfnodeconnline{Node0222}{Node0326}
\pgfnodeconnline{Node0326}{Node0419}
\pgfnodeconnline{Node0419}{Node0525}
\pgfnodeconnline{Node0525}{Node0617}
\pgfnodeconnline{Node0525}{Node0618}
\pgfnodeconnline{Node0525}{Node0619}
\pgfnodeconnline{Node0525}{Node0620}
\pgfnodeconnline{Node0525}{Node0621}
\pgfnodeconnline{Node0525}{Node0622}
\pgfnodeconnline{Node0525}{Node0623}
\pgfnodeconnline{Node0525}{Node0624}
\pgfnodeconnline{Node0525}{Node0625}
\pgfnodeconnline{Node0525}{Node0626}
\pgfnodeconnline{Node0525}{Node0627}
\pgfnodeconnline{Node0525}{Node0628}
\end{pgftranslate}

\end{pgfmagnify}
\pgfsetarrowsend{Triangle[scale=1.4pt]}
\end{pgfpicture}
\caption{Stable horizon growth in a hybrid network; ``short circuit" network.}
\label{fig2NplusM}
\end{figure}

The left-hand diagram in Figure \hyperref[fig2NplusM]{\ref{fig2NplusM}} illustrates the type of structure driving the rapid hybrid growth of connectivity in Figure \hyperref[figcake]{\ref{figcake}}.  Kernel size $1$ means that each neuron is connected to the corresponding neurons in neighboring generations.  Vertical flow of information is therefore assured, while superposed random edges open new conduits for this flow.  The comprehensive superiority of the hybrid edge-addition process prompts an interesting open optimization question: {\it what edge-addition process optimizes connectivity growth for given layer sizes?}  The right-hand diagram in Figure  \hyperref[fig2NplusM]{\ref{fig2NplusM}} shows an extreme case in which $O(N+M)$ edges produce $100\%$ connectivity for $M$ generations of $N$ neurons each, demonstrating that the $O(N^{\frac{M}{M-1}})$ edges required for similar connectivity in \eqref{Mgenintro} can be significantly improved.  However, this is an impractical ``short circuit" design.   We therefore restrict attention to {\it homogeneous} processes like those in Figure \hyperref[figcake]{\ref{figcake}}, which treat all parts of the network equally, excepting possible edge effects.  Such processes allow for networks with broad general training potential.  A similar emphasis prevails in the fundamental physics context discussed in Section \hyperref[sectionhorizon]{\ref{sectionhorizon}}, due to the empirical uniformities of spacetime at measurable scales.  We remark that maximizing input-output connectivity need not maximize typical horizon growth, as illustrated by the ``short-circuit" network in Figure  \hyperref[fig2NplusM]{\ref{fig2NplusM}}.  This network achieves high connectivity at low cost by spending only a few edges on intermediate layers, while typical nodes have empty horizons. Such pathologies are  less of an issue for homogeneous edge-addition processes. 

The right-hand diagram in Figure \hyperref[figcake]{\ref{figcake}}  compares the results of $20$ edge-addition processes for random (blue) and hybrid (red) networks.   Random networks are constructed by adding $5000$ random edges to $10$ layers ($11$ generations) of $100$ nodes each, while hybrid networks combine a minimal convolutional structure of $1000$ edges with the final $4000$ edges from the corresponding random networks. Two features stand out in this diagram.  First, the hybrid networks typically require only about half as many edges to reach $100\%$ connectivity as their random counterparts. Second, connectivity for individual neurons increases in a smoother and stabler manner in the hybrid networks, since the vertical information flow ensured by the minimal convolutional structure ``preserves the gains" in horizon size due to nonlocal connections in the random structure.  As a result, while $4$ out of $20$ random networks have zero connectivity between the chosen neuron and the final generation after $2000$ edges are added, all of the hybrid networks with at least $2000$ edges have at least $70\%$ connectivity.




{\bf Future applications: large populations.}  The toy CNNs in Figures \hyperref[figaugmentedCNNs]{\ref{figaugmentedCNNs}} through \hyperref[fig2NplusM]{\ref{fig2NplusM}} have uniform layer size, unlike typical large CNNs such as VGG, whose layer sizes shrink by hundreds of times between input and output layers. For such networks, reducing late dense layers can decrease the number of edges by perhaps one order of magnitude. However, future applications such as recognition of individuals among a large population may require at least as many outputs as inputs.  In this case, reducing late dense layers might decrease the number of edges by thousands of times or more.  While ethical use of such technology cannot be overemphasized, improved efficiency can at least democratize its use.  Due to weight sharing in CNNs, superposed random edges may contribute relatively heavily to parameter space dimensions, but this effect is unlikely to offset huge overall decreases in edge counts.  Further data is needed to determine how such sparse hybrid networks will perform compared to their denser conventional counterparts, but results involving similar architectures are encouraging \cite{Alford18,Prabhu18,Robinett18}.


{\bf Phase transitions and training diagnostics.} We now sketch how phase transition theory might contribute to training diagnostics, addressing issues such as management of learning rates, detection of peak performance, and avoidance of ``bad" minima of $L$.  Design improvements might reduce the scope of such analysis by anticipating transition-induced network properties, but the resulting methods might still apply to problems for which optimal design is unknown.  For brevity, we focus on training processes that satisfy a {\it descent-transition criterion,} already mentioned above, in which an accessibility phase transition heralds the approach of a local minimum of $L$.  The benefit of this scenario is that it allows {\it direct} application of knowledge about network structure, unmediated by $L$.   For example, candidate stopping points for training might be detectable in terms of average connectivity properties, reducing the role of costly performance-based testing procedures.   Learning rates might be adjusted similarly.  Methods such as {\it batch normalization} \cite{IoffeBatch15} and {\it layer normalization} \cite{BaLayer16} offer precedent for such direct analysis and manipulation of network properties. 


\newpage

{\bf Small-world behavior; spacetime analogy.}  We conclude this section by discussing small-world behavior in DNNs.  This leads to a profound physical analogy between DNNs and spacetime, foreshadowed in the context of general networks in Section \hyperref[subsection3pt]{\ref{subsection3pt}}.  Here, ``time" is the process time pointing from inputs to outputs, while each layer of the DNN represents a {\it spatial section of spacetime.}  For some networks, such as the convolutional layers in Figure \hyperref[figVGG]{\ref{figVGG}}, this analogy is obvious, since such layers encode actual spatial information.  Other networks, such as the dense network in Figure \hyperref[figdeep]{\ref{figdeep}}, are dissimilar to familiar spacetime, since they are non-geometric and non-local.  Small-world behavior depends on the interplay between local and nonlocal structure, with local structure favoring largeness and nonlocal structure favoring smallness.  Transition to a small world may occur in either direction of process time: in the future direction in Figure \hyperref[figVGG]{\ref{figVGG}}, and in the past direction in Figure \hyperref[figdeep]{\ref{figdeep}}.  As mentioned above, the distinction between these opposite structural trends may be conceptualized as a contrast between {\it converting geometry to information} and {\it converting information to geometry.}


We spotlight small-world behavior in a dense network like the one in Figure \hyperref[figdeep]{\ref{figdeep}}, both before and after training.  This behavior is not universal, but foreshadows the results of the next section in a useful way.   Assuming random initialization of weights, small-world phase transitions in both directions of process time are immediate prior to training, since neighboring layers exhibit many random functional connections.  This changes as the network is trained and weights become correlated, leading to functionally sparser structure.  Red nodes and edges in Figure \hyperref[figdeep]{\ref{figdeep}} illustrate horizon growth looking backward in process time from an output neuron $x$ in the trained network.  Concretely, this means tracing data arriving at $x$ back to its origins in the input layer.  Near the output layer, data is sufficiently sorted that $x$ receives significant input from only a few of its immediate predecessors.  If the network is trained for classification, this means that identification of a given cross section of input data as an instance of ``$x$" depends on a limited number of high-level features.  The network therefore exhibits locality in its final layers, and the world remains large.  Past horizons of $x$ occupy only a small proportion of these late layers, exhibiting relatively slow power-law growth.  Near the beginning of process time, however, this behavior breaks down.  Data on its way to $x$ is not yet significantly sorted, instead remaining distributed throughout the early layers. In structural terms, many different low-level features contribute to a typical high-level feature.  Na\"{i}ve extrapolation of horizon growth near the output layer then leads to the erroneous conclusion that only a small proportion of input data influences $x$.  This deceptive appearance based on local horizon growth bears an unmistakable similarity to a well-known physical problem examined in the next section.

\subsection{Cosmological Horizon Problem}
\label{sectionhorizon}

We now reexamine a famous cosmological problem, whose implications may support the novel hypothesis that spacetime itself emerges from network structure at the fundamental scale.  Emergence means that the familiar smooth four-dimensional Lorentzian geometry of general relativity is really just a low-resolution approximation, analogous to the illusion of smooth imagery on a pixelated display. Though we do not yet know exactly how this particular geometry came to dominate at large scales and late times in our universe, such emergence naturally favors a change from exponential to power-law scaling of horizons, since random structure {\it a priori} dominates the primordial sum over evolutionary pathways that generalizes Feynman's path integral in quantum gravity.  Standard astronomical data suggests that such a scaling change did in fact occur in the early universe, a conclusion broadly acknowledged on empirical grounds despite absence of a compelling mechanism to account for it in conventional relativistic cosmology and quantum field theory \cite{GuthPaper81,LindePaper82,Ellis12,Maartens11}.  This conclusion follows from evidence of interaction between regions of spacetime that appear causally disconnected on the basis of na\"{i}ve backward extrapolation of local Lorentzian horizon growth, first pointed out by Rindler \cite{RindlerHorizon56} in the middle 1950s.  Foremost among this evidence is the large-scale homogeneity of the observable universe, which may be inferred, under modest assumptions, from measurements of the {\it cosmic microwave background} (CMB) \cite{Dicke65,Penzias65,Maartens11}. The consensus explanation for this homogeneity is {\it equilibration} via a mixing process, since otherwise the properties of the early universe would appear to be coincidentally {\it fine-tuned.}  However, mixing requires causal contact, and therefore much faster horizon growth than what is observed today.  This clash between extrapolated and empirically-inferred degrees of interaction is called the {\it horizon problem.} 


\begin{figure}[h]
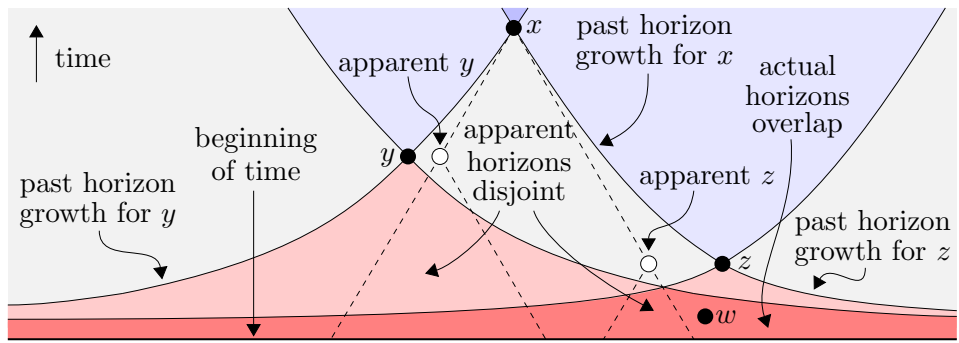

\begin{pgfpicture}{0cm}{0cm}{15cm}{4.6cm}
\begin{pgftranslate}{\pgfpoint{0cm}{.3cm}}
\begin{pgfmagnify}{.76}{.76}
\begin{pgfscope}
\color{black!5}
\pgfmoveto{\pgfxy(0,0)}
\pgflineto{\pgfxy(16.6,0)}
\pgflineto{\pgfxy(16.6,5.6)}
\pgflineto{\pgfxy(0,5.6)}
\pgflineto{\pgfxy(0,0)}
\pgffill
\color{blue!10}
\pgfmoveto{\pgfxy(7,3)}
\pgfcurveto{\pgfxy(6,4)}{\pgfxy(5.5,4.7)}{\pgfxy(4.9,5.6)}
\pgflineto{\pgfxy(9.1,5.6)}
\pgfcurveto{\pgfxy(8.5,4.7)}{\pgfxy(8,4)}{\pgfxy(7,3)}
\pgffill
\pgfmoveto{\pgfxy(12.5,1.12)}
\pgfcurveto{\pgfxy(11.5,1.75)}{\pgfxy(10.5,2.55)}{\pgfxy(8.64,5.6)}
\pgflineto{\pgfxy(16.4,5.6)}
\pgfcurveto{\pgfxy(14.5,2.5)}{\pgfxy(13.5,1.75)}{\pgfxy(12.5,1.12)}
\pgffill
\color{blue!25}
\pgfmoveto{\pgfxy(9.1,5.6)}
\pgflineto{\pgfxy(8.88,5.25)}
\pgflineto{\pgfxy(8.64,5.6)}
\pgffill
\color{red!20}
\pgfmoveto{\pgfxy(0,.4)}
\pgfcurveto{\pgfxy(.5,.4)}{\pgfxy(1,.45)}{\pgfxy(2,.6)}
\pgfcurveto{\pgfxy(5,1.2)}{\pgfxy(6,2)}{\pgfxy(7,3)}
\pgfcurveto{\pgfxy(8,2)}{\pgfxy(9,1.2)}{\pgfxy(12,.6)}
\pgfcurveto{\pgfxy(14,.3)}{\pgfxy(16,.2)}{\pgfxy(16.6,.2)}
\pgflineto{\pgfxy(16.6,0)}
\pgflineto{\pgfxy(0,0)}
\pgflineto{\pgfxy(0,.4)}
\pgffill
\color{red!20}
\pgfmoveto{\pgfxy(0,.1)}
\pgfcurveto{\pgfxy(10,.1)}{\pgfxy(11.5,.55)}{\pgfxy(12.5,1.12)}
\pgfcurveto{\pgfxy(13.5,.55)}{\pgfxy(15,.4)}{\pgfxy(16.6,.3)}
\pgflineto{\pgfxy(16.6,0)}
\pgflineto{\pgfxy(0,0)}
\pgflineto{\pgfxy(0,.1)}
\pgffill
\color{red!50}
\pgfmoveto{\pgfxy(0,.15)}
\pgfcurveto{\pgfxy(10,.15)}{\pgfxy(11,.6)}{\pgfxy(11.5,.7)}
\pgfcurveto{\pgfxy(12,.57)}{\pgfxy(15,.2)}{\pgfxy(16.6,.2)}
\pgflineto{\pgfxy(16.6,-.2)}
\pgflineto{\pgfxy(0,-.2)}
\pgflineto{\pgfxy(0,.1)}
\pgffill
\end{pgfscope}
\begin{pgfscope}
\pgfsetlinewidth{1pt}
\pgfxyline(0,-.2)(16.6,-.2)
\end{pgfscope}
\begin{pgftranslate}{\pgfpoint{2cm}{0cm}}
\pgfmoveto{\pgfxy(-2,.4)}
\pgfcurveto{\pgfxy(-1.5,.4)}{\pgfxy(-1,.45)}{\pgfxy(0,.6)}
\pgfstroke
\pgfmoveto{\pgfxy(0,.6)}
\pgfcurveto{\pgfxy(3,1.2)}{\pgfxy(4,2)}{\pgfxy(5,3)}
\pgfstroke
\pgfmoveto{\pgfxy(5,3)}
\pgfcurveto{\pgfxy(6,4)}{\pgfxy(6.5,4.7)}{\pgfxy(7.1,5.6)}
\pgfstroke
\pgfmoveto{\pgfxy(14.6,.2)}
\pgfcurveto{\pgfxy(14,.2)}{\pgfxy(12,.3)}{\pgfxy(10,.6)}
\pgfstroke
\pgfmoveto{\pgfxy(10,.6)}
\pgfcurveto{\pgfxy(7,1.2)}{\pgfxy(6,2)}{\pgfxy(5,3)}
\pgfstroke
\pgfmoveto{\pgfxy(5,3)}
\pgfcurveto{\pgfxy(4,4)}{\pgfxy(3.5,4.7)}{\pgfxy(2.9,5.6)}
\pgfstroke
\end{pgftranslate}
\begin{pgfscope}
\pgfsetdash{{3pt}{3pt}}{0pt}
\pgfxyline(8.85,5.25)(7.56,3)
\pgfxyline(7.56,3)(5.65,-.2)
\pgfxyline(7.56,3)(9.41,-.2)
\pgfxyline(8.85,5.25)(11.21,1.12)
\pgfxyline(11.21,1.12)(12,-.2)
\pgfxyline(11.21,1.12)(10.42,-.2)
\end{pgfscope}
\pgfmoveto{\pgfxy(0,.15)}
\pgfcurveto{\pgfxy(10,.15)}{\pgfxy(11.5,.55)}{\pgfxy(12.5,1.12)}
\pgfstroke
\pgfmoveto{\pgfxy(12.5,1.12)}
\pgfcurveto{\pgfxy(13.5,1.75)}{\pgfxy(14.5,2.5)}{\pgfxy(16.4,5.6)}
\pgfstroke
\pgfmoveto{\pgfxy(16.6,.3)}
\pgfcurveto{\pgfxy(15,.4)}{\pgfxy(13.5,.55)}{\pgfxy(12.5,1.12)}
\pgfstroke
\pgfmoveto{\pgfxy(12.5,1.12)}
\pgfcurveto{\pgfxy(11.5,1.75)}{\pgfxy(10.5,2.55)}{\pgfxy(8.64,5.6)}
\pgfstroke
\pgfnodecircle{Node0}[fill]{\pgfxy(12.2,.2)}{0.14cm}
\pgfnodecircle{Node1}[fill]{\pgfxy(7,3)}{0.14cm}
\pgfnodecircle{Node2}[fill]{\pgfxy(8.85,5.25)}{0.14cm}
\pgfnodecircle{Node3}[fill]{\pgfxy(12.5,1.12)}{0.14cm}
\color{white}
\pgfnodecircle{Node10}[fill]{\pgfxy(7.56,3)}{0.14cm}
\pgfnodecircle{Node20}[fill]{\pgfxy(11.21,1.12)}{0.14cm}
\color{black}
\pgfnodecircle{Node10}[stroke]{\pgfxy(7.56,3)}{0.14cm}
\pgfnodecircle{Node20}[stroke]{\pgfxy(11.21,1.12)}{0.14cm}
\pgfsetarrowsend{Triangle[scale=1.7pt]}
\pgfmoveto{\pgfxy(7,4.4)}
\pgfcurveto{\pgfxy(7.1,3.6)}{\pgfxy(7.56,3.9)}{\pgfxy(7.56,3.2)}
\pgfstroke
\pgfmoveto{\pgfxy(12,2.4)}
\pgfcurveto{\pgfxy(11.9,1.6)}{\pgfxy(11.21,1.9)}{\pgfxy(11.21,1.35)}
\pgfstroke
\pgfmoveto{\pgfxy(1.6,1.7)}
\pgfcurveto{\pgfxy(1.6,1.1)}{\pgfxy(2.6,1.5)}{\pgfxy(2.7,.85)}
\pgfstroke
\pgfmoveto{\pgfxy(15,1.05)}
\pgfcurveto{\pgfxy(15,.5)}{\pgfxy(14.4,1.2)}{\pgfxy(14.1,.67)}
\pgfstroke
\pgfmoveto{\pgfxy(11.2,4.45)}
\pgfcurveto{\pgfxy(11.2,4.1)}{\pgfxy(11.2,3.6)}{\pgfxy(10.35,3.2)}
\pgfstroke
\pgfxyline(.5,4.3)(.5,5.3)
\pgfxyline(4.3,2.4)(4.3,-.15)
\pgfmoveto{\pgfxy(8.6,2.05)}
\pgfcurveto{\pgfxy(8.5,1.7)}{\pgfxy(7.8,1.1)}{\pgfxy(7.3,.95)}
\pgfstroke
\pgfmoveto{\pgfxy(9.2,2.05)}
\pgfcurveto{\pgfxy(9.3,1.7)}{\pgfxy(10,1)}{\pgfxy(11.3,.3)}
\pgfstroke
\pgfmoveto{\pgfxy(13.7,3.3)}
\pgfcurveto{\pgfxy(13.8,2)}{\pgfxy(13,.9)}{\pgfxy(13.3,0)}
\pgfstroke
\end{pgfmagnify}
\pgfputat{\pgfxy(9.55,.15)}{\pgfbox[center,center]{$w$}}
\pgfputat{\pgfxy(5.3,3.5)}{\pgfbox[center,center]{apparent $y$}}
\pgfputat{\pgfxy(5.05,2.26)}{\pgfbox[center,center]{$y$}}
\pgfputat{\pgfxy(9.3,2)}{\pgfbox[center,center]{apparent $z$}}
\pgfputat{\pgfxy(9.8,.87)}{\pgfbox[center,center]{$z$}}
\pgfputat{\pgfxy(7,4)}{\pgfbox[center,center]{$x$}}
\pgfputat{\pgfxy(10.5,3.5)}{\pgfbox[center,center]{actual}}
\pgfputat{\pgfxy(10.5,3.1)}{\pgfbox[center,center]{horizons}}
\pgfputat{\pgfxy(10.5,2.7)}{\pgfbox[center,center]{overlap}}
\pgfputat{\pgfxy(1.2,1.9)}{\pgfbox[center,center]{past horizon}}
\pgfputat{\pgfxy(1.2,1.5)}{\pgfbox[center,center]{growth for $y$}}
\pgfputat{\pgfxy(11.5,1.4)}{\pgfbox[center,center]{past horizon}}
\pgfputat{\pgfxy(11.5,1.0)}{\pgfbox[center,center]{growth for $z$}}
\pgfputat{\pgfxy(8.6,4)}{\pgfbox[center,center]{past horizon}}
\pgfputat{\pgfxy(8.6,3.6)}{\pgfbox[center,center]{growth for $x$}}
\pgfputat{\pgfxy(6.8,2.6)}{\pgfbox[center,center]{apparent}}
\pgfputat{\pgfxy(6.8,2.22)}{\pgfbox[center,center]{horizons}}
\pgfputat{\pgfxy(6.8,1.8)}{\pgfbox[center,center]{disjoint}}
\pgfputat{\pgfxy(1,3.6)}{\pgfbox[center,center]{time}}
%
\pgfputat{\pgfxy(3.3,2.5)}{\pgfbox[center,center]{beginning}}
\pgfputat{\pgfxy(3.3,2.1)}{\pgfbox[center,center]{of time}}

\end{pgftranslate}

\end{pgfpicture}
\caption{Horizon problem: apparent horizons of $y$ and $z$ suggest no common ancestors, but actual horizons overlap due to different early structure.}
\label{fighorizonGR}
\end{figure}

{\bf General horizon problem.} Figure \hyperref[fighorizonGR]{\ref{fighorizonGR}} shows a schematic description of the horizon problem, independent of specific structural assumptions except for a beginning of time and an absence of large-scale cycles. Possible ``cosmological prehistory" such as {\it eternal inflation} is not included in this picture. Events $y$ and $z$ are {\it ancestors} of event $x$, lying on its {\it past horizon.}  In relativity, the past horizon of $x$ means the boundary of the set of events than can influence $x$, called its {\it causal past.}  By contrast, our horizons $\sigma_m(x)$ from Section \hyperref[subsecconnprob]{\ref{subsecconnprob}} are defined to be cross sections of the past or future of $x$ rather than boundaries, because the latter are harder to generalize to non-geometric settings.  Here, either notion suffices, due to a judicious choice of $y$ and $z$.   The {\it apparent} locations of $y$ and $z$ (white nodes) from the perspective of an observer at $x$ may differ from their actual locations due to factors such as expansion or contraction of spacetime, gravitational lensing, variation of the speed of light, and spatial anisotropy.   Since $y$ and $z$ share common ancestors such as $w$, it is not surprising if they exhibit common environmental features indicative of equilibration.  However, such features {\it would} surprise a na\"{i}ve observer at $x$, unaware of the drastic stretching of horizons near the beginning of time.  Such an observer would assume power-law horizon scaling derived from the local spacetime dimension and speed of light, yielding disjoint {\it apparent horizons} (dashed lines), and would therefore recognize a horizon problem. 

\begin{figure}[h]
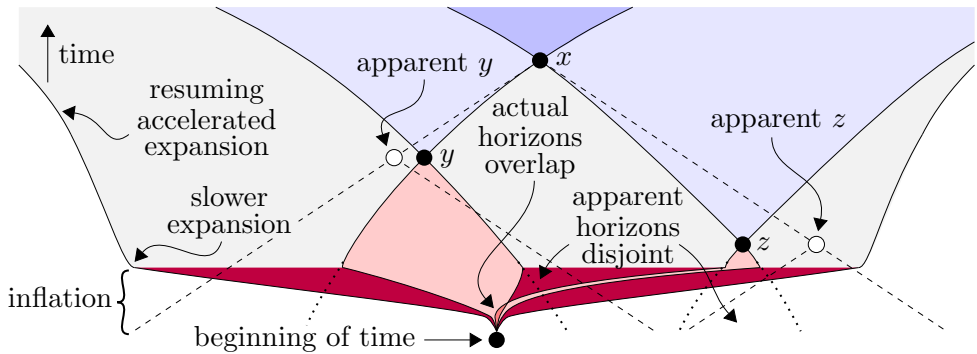

\begin{pgfpicture}{0cm}{0cm}{15cm}{4.6cm}


\begin{pgftranslate}{\pgfpoint{0cm}{.2cm}}
\begin{pgfmagnify}{.77}{.77}
\begin{pgfscope}
\color{black!5}
\pgfmoveto{\pgfxy(8.25,0)}
\pgfcurveto{\pgfxy(8.3,.1)}{\pgfxy(8.35,.2)}{\pgfxy(8.45,.25)}
\pgfcurveto{\pgfxy(8.85,.4)}{\pgfxy(12.5,.8)}{\pgfxy(14.5,1.1)}
\pgfcurveto{\pgfxy(14.7,1.1)}{\pgfxy(15,2.2)}{\pgfxy(15.6,3.4)}
\pgfcurveto{\pgfxy(15.8,3.8)}{\pgfxy(16.1,4.2)}{\pgfxy(16.5,4.6)}
\pgflineto{\pgfxy(16.5,5.6)}
\pgflineto{\pgfxy(0,5.6)}
\pgflineto{\pgfxy(0,4.6)}
\pgfcurveto{\pgfxy(.4,4.2)}{\pgfxy(.7,3.8)}{\pgfxy(.9,3.4)}
\pgfcurveto{\pgfxy(1.5,2.2)}{\pgfxy(1.8,1.1)}{\pgfxy(2,1.1)}
\pgfcurveto{\pgfxy(4,.8)}{\pgfxy(7.65,.4)}{\pgfxy(8.05,.25)}
\pgfcurveto{\pgfxy(8.15,.2)}{\pgfxy(8.2,.1)}{\pgfxy(8.25,0)}
\pgffill
\color{blue!10}
\pgfmoveto{\pgfxy(7,3)}
\pgfcurveto{\pgfxy(5.8,4.2)}{\pgfxy(5,4.7)}{\pgfxy(4.5,5)}
\pgfcurveto{\pgfxy(4.2,5.2)}{\pgfxy(3.8,5.4)}{\pgfxy(3.4,5.6)}
\pgflineto{\pgfxy(10.6,5.6)}
\pgfcurveto{\pgfxy(10.2,5.4)}{\pgfxy(9.8,5.2)}{\pgfxy(9.5,5)}
\pgfcurveto{\pgfxy(9,4.7)}{\pgfxy(8.2,4.2)}{\pgfxy(7,3)}
\pgffill
\pgfmoveto{\pgfxy(12.5,1.5)}
\pgfcurveto{\pgfxy(9.8,4.2)}{\pgfxy(9,4.7)}{\pgfxy(8.5,5)}
\pgfcurveto{\pgfxy(8.2,5.2)}{\pgfxy(7.8,5.4)}{\pgfxy(7.4,5.6)}
\pgflineto{\pgfxy(16.5,5.6)}
\pgflineto{\pgfxy(16.5,5)}
\pgfcurveto{\pgfxy(16,4.7)}{\pgfxy(15.2,4.2)}{\pgfxy(12.5,1.5)}
\pgffill
\color{blue!25}
\pgfmoveto{\pgfxy(7.4,5.6)}
\pgfcurveto{\pgfxy(8.2,5.2)}{\pgfxy(7.8,5.4)}{\pgfxy(9,4.7)}
\pgfcurveto{\pgfxy(9.4,4.9)}{\pgfxy(9.8,5.2)}{\pgfxy(10.6,5.6)}
\pgflineto{\pgfxy(7.4,5.6)}
\pgffill
\color{purple}
\pgfmoveto{\pgfxy(8.25,0)}
\pgfcurveto{\pgfxy(8.3,.1)}{\pgfxy(8.35,.2)}{\pgfxy(8.45,.25)}
\pgfcurveto{\pgfxy(8.85,.4)}{\pgfxy(12.5,.8)}{\pgfxy(14.5,1.1)}
\pgflineto{\pgfxy(2,1.1)}
\pgfcurveto{\pgfxy(4,.8)}{\pgfxy(7.65,.4)}{\pgfxy(8.05,.25)}
\pgfcurveto{\pgfxy(8.15,.2)}{\pgfxy(8.2,.1)}{\pgfxy(8.25,0)}
\pgffill
\color{red!20}
\pgfmoveto{\pgfxy(8.25,0)}
\pgfcurveto{\pgfxy(8.23,.1)}{\pgfxy(8.2,.2)}{\pgfxy(8.15,.25)}
\pgfcurveto{\pgfxy(7.9,.4)}{\pgfxy(6.8,.8)}{\pgfxy(5.6,1.1)}
\pgfcurveto{\pgfxy(5.5,1.1)}{\pgfxy(6,2)}{\pgfxy(7,3)}
\pgfcurveto{\pgfxy(8,2)}{\pgfxy(8.7,1.1)}{\pgfxy(8.7,1.1)}
\pgfcurveto{\pgfxy(8.7,.8)}{\pgfxy(8.35,.4)}{\pgfxy(8.3,.25)}
\pgfcurveto{\pgfxy(8.28,.2)}{\pgfxy(8.26,.1)}{\pgfxy(8.25,0)}
\pgffill
\pgfmoveto{\pgfxy(8.25,0)}
\pgfcurveto{\pgfxy(8.24,.1)}{\pgfxy(8.23,.2)}{\pgfxy(8.22,.25)}
\pgfcurveto{\pgfxy(8.25,.4)}{\pgfxy(8.3,.8)}{\pgfxy(12.2,1.1)}
\pgfcurveto{\pgfxy(12.2,1.2)}{\pgfxy(12.4,1.4)}{\pgfxy(12.5,1.5)}
\pgfcurveto{\pgfxy(12.7,1.3)}{\pgfxy(12.8,1.1)}{\pgfxy(12.8,1.1)}
\pgfcurveto{\pgfxy(8.6,.8)}{\pgfxy(8.45,.4)}{\pgfxy(8.35,.25)}
\pgfcurveto{\pgfxy(8.31,.2)}{\pgfxy(8.27,.1)}{\pgfxy(8.25,0)}
\pgffill
\color{red!50}
\pgfmoveto{\pgfxy(8.25,0)}
\pgfcurveto{\pgfxy(8.24,.1)}{\pgfxy(8.23,.2)}{\pgfxy(8.22,.25)}
\pgfcurveto{\pgfxy(8.25,.35)}{\pgfxy(8.28,.4)}{\pgfxy(8.45,.5)}
\pgfcurveto{\pgfxy(8.28,.3)}{\pgfxy(8.26,.1)}{\pgfxy(8.25,0)}
\pgffill
\end{pgfscope}
\begin{pgftranslate}{\pgfpoint{2cm}{0cm}}
\pgfmoveto{\pgfxy(5,3)}
\pgfcurveto{\pgfxy(6.2,4.2)}{\pgfxy(7,4.7)}{\pgfxy(7.5,5)}
\pgfstroke
\pgfmoveto{\pgfxy(7.5,5)}
\pgfcurveto{\pgfxy(7.8,5.2)}{\pgfxy(8.2,5.4)}{\pgfxy(8.6,5.6)}
\pgfstroke
\pgfmoveto{\pgfxy(5,3)}
\pgfcurveto{\pgfxy(3.8,4.2)}{\pgfxy(3,4.7)}{\pgfxy(2.5,5)}
\pgfstroke
\pgfmoveto{\pgfxy(2.5,5)}
\pgfcurveto{\pgfxy(2.2,5.2)}{\pgfxy(1.8,5.4)}{\pgfxy(1.4,5.6)}
\pgfstroke
\end{pgftranslate}
\pgfmoveto{\pgfxy(12.5,1.5)}
\pgfcurveto{\pgfxy(15.2,4.2)}{\pgfxy(16,4.7)}{\pgfxy(16.5,5)}
\pgfstroke
\pgfmoveto{\pgfxy(12.5,1.5)}
\pgfcurveto{\pgfxy(9.8,4.2)}{\pgfxy(9,4.7)}{\pgfxy(8.5,5)}
\pgfstroke
\pgfmoveto{\pgfxy(8.5,5)}
\pgfcurveto{\pgfxy(8.2,5.2)}{\pgfxy(7.8,5.4)}{\pgfxy(7.4,5.6)}
\pgfstroke
\pgfmoveto{\pgfxy(8.25,0)}
\pgfcurveto{\pgfxy(8.23,.1)}{\pgfxy(8.2,.2)}{\pgfxy(8.15,.25)}
\pgfstroke
\pgfmoveto{\pgfxy(8.15,.25)}
\pgfcurveto{\pgfxy(7.9,.4)}{\pgfxy(6.8,.8)}{\pgfxy(5.6,1.1)}
\pgfstroke
\pgfmoveto{\pgfxy(5.6,1.1)}
\pgfcurveto{\pgfxy(5.5,1.1)}{\pgfxy(6,2)}{\pgfxy(7,3)}
\pgfstroke
\pgfmoveto{\pgfxy(8.25,0)}
\pgfcurveto{\pgfxy(8.26,.1)}{\pgfxy(8.28,.2)}{\pgfxy(8.3,.25)}
\pgfstroke
\pgfmoveto{\pgfxy(8.3,.25)}
\pgfcurveto{\pgfxy(8.35,.4)}{\pgfxy(8.7,.8)}{\pgfxy(8.7,1.1)}
\pgfstroke
\pgfmoveto{\pgfxy(8.7,1.1)}
\pgfcurveto{\pgfxy(8.7,1.1)}{\pgfxy(8,2)}{\pgfxy(7,3)}
\pgfstroke
\pgfmoveto{\pgfxy(8.25,0)}
\pgfcurveto{\pgfxy(8.24,.1)}{\pgfxy(8.23,.2)}{\pgfxy(8.22,.25)}
\pgfstroke
\pgfmoveto{\pgfxy(8.22,.25)}
\pgfcurveto{\pgfxy(8.25,.4)}{\pgfxy(8.3,.8)}{\pgfxy(12.2,1.1)}
\pgfstroke
\pgfmoveto{\pgfxy(12.2,1.1)}
\pgfcurveto{\pgfxy(12.2,1.2)}{\pgfxy(12.4,1.4)}{\pgfxy(12.5,1.5)}
\pgfstroke
\pgfmoveto{\pgfxy(8.25,0)}
\pgfcurveto{\pgfxy(8.27,.1)}{\pgfxy(8.31,.2)}{\pgfxy(8.35,.25)}
\pgfstroke
\pgfmoveto{\pgfxy(8.35,.25)}
\pgfcurveto{\pgfxy(8.45,.4)}{\pgfxy(8.6,.8)}{\pgfxy(12.8,1.1)}
\pgfstroke
\pgfmoveto{\pgfxy(12.8,1.1)}
\pgfcurveto{\pgfxy(12.8,1.1)}{\pgfxy(12.7,1.3)}{\pgfxy(12.5,1.5)}
\pgfstroke
\pgfmoveto{\pgfxy(8.25,0)}
\pgfcurveto{\pgfxy(8.3,.1)}{\pgfxy(8.35,.2)}{\pgfxy(8.45,.25)}
\pgfstroke
\pgfmoveto{\pgfxy(8.45,.25)}
\pgfcurveto{\pgfxy(8.85,.4)}{\pgfxy(12.5,.8)}{\pgfxy(14.5,1.1)}
\pgfstroke
\pgfmoveto{\pgfxy(14.5,1.1)}
\pgfcurveto{\pgfxy(14.7,1.1)}{\pgfxy(14.9,2.2)}{\pgfxy(15.6,3.4)}
\pgfstroke
\pgfmoveto{\pgfxy(15.6,3.4)}
\pgfcurveto{\pgfxy(15.8,3.8)}{\pgfxy(16.1,4.2)}{\pgfxy(16.5,4.6)}
\pgfstroke
\pgfmoveto{\pgfxy(8.25,0)}
\pgfcurveto{\pgfxy(8.2,.1)}{\pgfxy(8.15,.2)}{\pgfxy(8.05,.25)}
\pgfstroke
\pgfmoveto{\pgfxy(8.05,.25)}
\pgfcurveto{\pgfxy(7.65,.4)}{\pgfxy(4,.8)}{\pgfxy(2,1.1)}
\pgfstroke
\pgfmoveto{\pgfxy(2,1.1)}
\pgfcurveto{\pgfxy(1.8,1.1)}{\pgfxy(1.5,2.2)}{\pgfxy(.9,3.4)}
\pgfstroke
\pgfmoveto{\pgfxy(.9,3.4)}
\pgfcurveto{\pgfxy(.7,3.8)}{\pgfxy(.4,4.2)}{\pgfxy(0,4.6)}
\pgfstroke
\begin{pgfscope}
\pgfsetlinewidth{1pt}
\pgfmoveto{\pgfxy(1.9,1.05)}
\pgfcurveto{\pgfxy(1.6,1.05)}{\pgfxy(2,.5)}{\pgfxy(1.7,.5)}
\pgfcurveto{\pgfxy(2,.5)}{\pgfxy(1.6,-.05)}{\pgfxy(1.9,-.05)}
\pgfstroke
\end{pgfscope}
\pgfnodecircle{Node00}[fill]{\pgfxy(8.25,-.15)}{0.14cm}
\pgfnodecircle{Node1}[fill]{\pgfxy(7,3)}{0.14cm}
\pgfnodecircle{Node2}[fill]{\pgfxy(12.5,1.5)}{0.14cm}
\pgfnodecircle{Node3}[fill]{\pgfxy(9,4.68)}{0.14cm}
\begin{pgfscope}
\pgfsetdash{{3pt}{3pt}}{0pt}
\pgfxyline(9,4.68)(6.48,3)
\pgfxyline(6.48,3)(4.83,1.9)
\pgfxyline(4.38,1.6)(1.98,0)
\pgfxyline(6.48,3)(10.98,0)
\pgfxyline(9,4.68)(13.77,1.5)
\pgfxyline(13.77,1.5)(16.02,0)
\pgfxyline(13.77,1.5)(11.52,0)
\end{pgfscope}
\begin{pgfscope}
\pgfsetlinewidth{1pt}
\pgfsetdash{{1pt}{3pt}}{0pt}
\pgfmoveto{\pgfxy(5.6,1.2)}
\pgfcurveto{\pgfxy(5.44,.9)}{\pgfxy(5.24,.7)}{\pgfxy(4.97,.1)}
\pgfstroke
\pgfmoveto{\pgfxy(8.62,1.2)}
\pgfcurveto{\pgfxy(8.85,.9)}{\pgfxy(9.15,.5)}{\pgfxy(9.47,0)}
\pgfstroke
\pgfmoveto{\pgfxy(12.24,1.2)}
\pgfcurveto{\pgfxy(11.96,.9)}{\pgfxy(11.64,.5)}{\pgfxy(11.4,0)}
\pgfstroke
\pgfmoveto{\pgfxy(12.74,1.2)}
\pgfcurveto{\pgfxy(12.99,.9)}{\pgfxy(13.25,.5)}{\pgfxy(13.5,0)}
\pgfstroke
\end{pgfscope}
\color{white}
\pgfnodecircle{Node10}[fill]{\pgfxy(6.48,3)}{0.14cm}
\pgfnodecircle{Node20}[fill]{\pgfxy(13.77,1.5)}{0.14cm}
\color{black}
\pgfnodecircle{Node10}[stroke]{\pgfxy(6.48,3)}{0.14cm}
\pgfnodecircle{Node20}[stroke]{\pgfxy(13.77,1.5)}{0.14cm}
\pgfsetarrowsend{Triangle[scale=1.7pt]}
\pgfmoveto{\pgfxy(7,4.35)}
\pgfcurveto{\pgfxy(7,3.6)}{\pgfxy(6.1,3.95)}{\pgfxy(6.35,3.2)}
\pgfstroke
\pgfmoveto{\pgfxy(13,3.35)}
\pgfcurveto{\pgfxy(13,2.7)}{\pgfxy(14,2.7)}{\pgfxy(13.8,1.73)}
\pgfstroke
\pgfmoveto{\pgfxy(3.5,1.55)}
\pgfcurveto{\pgfxy(3.3,.9)}{\pgfxy(2.5,1.8)}{\pgfxy(2,1.2)}
\pgfstroke
\pgfmoveto{\pgfxy(9.5,1.5)}
\pgfcurveto{\pgfxy(9.2,1.3)}{\pgfxy(9.1,1.1)}{\pgfxy(9,.9)}
\pgfstroke
\pgfmoveto{\pgfxy(11.4,1.5)}
\pgfcurveto{\pgfxy(12,1.1)}{\pgfxy(11.9,.7)}{\pgfxy(12.4,.15)}
\pgfstroke
\pgfmoveto{\pgfxy(8.9,2.6)}
\pgfcurveto{\pgfxy(8.9,1.7)}{\pgfxy(7.5,1.2)}{\pgfxy(8.2,.4)}
\pgfstroke
\pgfmoveto{\pgfxy(7.0,-.15)}
\pgfcurveto{\pgfxy(7.4,-.15)}{\pgfxy(7.7,-.15)}{\pgfxy(8,-.15)}
\pgfstroke
\pgfxyline(.5,4.3)(.5,5.3)
\pgfmoveto{\pgfxy(1.9,3.4)}
\pgfcurveto{\pgfxy(1.5,3.1)}{\pgfxy(1.5,3.7)}{\pgfxy(.75,3.8)}
\pgfstroke
\end{pgfmagnify}
\pgfputat{\pgfxy(5.7,2.3)}{\pgfbox[center,center]{$y$}}
\pgfputat{\pgfxy(9.9,1.17)}{\pgfbox[center,center]{$z$}}
\pgfputat{\pgfxy(7.25,3.63)}{\pgfbox[center,center]{$x$}}
\pgfputat{\pgfxy(5.4,3.55)}{\pgfbox[center,center]{apparent $y$}}
\pgfputat{\pgfxy(10.1,2.8)}{\pgfbox[center,center]{apparent $z$}}
\pgfputat{\pgfxy(6.8,3)}{\pgfbox[center,center]{actual}}
\pgfputat{\pgfxy(6.8,2.6)}{\pgfbox[center,center]{horizons}}
\pgfputat{\pgfxy(6.8,2.2)}{\pgfbox[center,center]{overlap}}
\pgfputat{\pgfxy(8.1,1.76)}{\pgfbox[center,center]{apparent}}
\pgfputat{\pgfxy(8.1,1.4)}{\pgfbox[center,center]{horizons}}
\pgfputat{\pgfxy(8.1,1)}{\pgfbox[center,center]{disjoint}}
\pgfputat{\pgfxy(2.8,1.8)}{\pgfbox[center,center]{slower}}
\pgfputat{\pgfxy(2.8,1.4)}{\pgfbox[center,center]{expansion}}
\pgfputat{\pgfxy(.9,3.7)}{\pgfbox[center,center]{time}}
\pgfputat{\pgfxy(2.5,3.2)}{\pgfbox[center,center]{resuming}}
\pgfputat{\pgfxy(2.5,2.83)}{\pgfbox[center,center]{accelerated}}
\pgfputat{\pgfxy(2.5,2.43)}{\pgfbox[center,center]{expansion}}
\pgfputat{\pgfxy(.55,.45)}{\pgfbox[center,center]{inflation}}
\pgfputat{\pgfxy(3.8,-.1)}{\pgfbox[center,center]{beginning of time}}
\end{pgftranslate}
\end{pgfpicture}
\caption{Inflationary approach to the horizon problem.}
\label{fighorizoninf}
\end{figure}

{\bf Inflationary hypothesis.}  Horizon stretching in the schematic diagram of Figure \hyperref[fighorizonGR]{\ref{fighorizonGR}} suggests a profound early change in the speed of light, a dark-horse mechanism for resolving the horizon problem \cite{Petit88,Moffat93}.  However, the dynamical Big-Bang model of spacetime genesis admits a different mechanism involving a brief, drastic period of spacetime expansion.  This is the {\it inflationary hypothesis} of Guth \cite{GuthPaper81} and Linde \cite{LindePaper82}, illustrated in Figure \hyperref[fighorizoninf]{\ref{fighorizoninf}}, which has dominated mainstream cosmology since 1980.  It posits that expansion spiked in an explosive {\it inflationary epoch} lasting roughly $10^{-32}$ seconds, caused by a mysterious {\it inflaton field,} which increased local spatial volumes by a factor of $10^{25}$ or more.  Afterward, inflating spacetime decayed into ordinary spacetime, with a much slower expansion rate.  Looking back in time from $x$ under the inflationary paradigm, the actual horizons of $y$ and $z$ are much smaller than their apparent horizons, yet overlap in the crowded environment near the beginning of time.  Looking in the opposite direction, adjacent regions inside this overlap are blasted far apart during the inflationary epoch, producing large but widely-separated futures. {\it Adjusted apparent pasts} of $y$ and $z$ (dotted curves), based on ordinary spacetime expansion inferred from red-shift data, are also disjoint, illustrating that the horizon problem is not merely a failure to account for obvious processes.   Horizon stretching near the top of the diagram represents the standard view that spacetime expansion resumed gradual acceleration around 5 billion years ago. 

Among other successes, inflation also addresses the {\it flatness problem,} which is the question of why spacetime seems fine-tuned to exhibit so little large-scale curvature, and offers a possible mechanism for {\it structure formation} in the early universe via quantum fluctuations. However, it also faces serious objections.  The hypothetical inflaton field requires such fine-tuning that reasonable inflationary scenarios may be less likely than coincidental homogeneity and flatness.  Recent astronomical data disfavors simple inflationary models \cite{Steinhardt13,Steinhardt17,SteinhardtReply17,AdeInflation18}.  Perhaps most troubling is the lack of motivation and context for the hypothesis.  It serves as an {\it ex post facto} patch for relativistic cosmology, extrapolating familiar spacetime geometry into a regime involving energies and interactions beyond present knowledge.  The identity, origin, and overall behavior of the inflaton field remain mysterious.  As described by Ellis, Maartens, and MacCallum \cite{Ellis12}, inflation {\it ``remains a phenomenological scenario that is yet to be rooted in a fundamental theory."}


{\bf Network-theoretic small-world alternative:} By contrast, spacetime emergence from network structure naturally favors horizon scaling of the type needed to resolve the horizon problem.  Assuming a {\it pre-geometric epoch} dominated by random network structure, exponential horizon scaling follows automatically by the reasoning described in Section \hyperref[subsecconnprob]{\ref{subsecconnprob}}. Feynman's path summation approach to quantum theory \cite{FeynmanSOH48} motivates such a pre-geometric epoch, as we now describe.  In ordinary quantum mechanics, Feynman's approach predicts the probabilistic behavior of a particle by integrating the complex exponential $\exp\left(\frac{i}{\hbar}\ms{S}(\gamma)\right)$ over all permissible particle trajectories $\gamma$, where $\ms{S}$ is the {\it classical action}, $\hbar$ is {\it Planck's reduced constant,} and $i$ is the imaginary unit $\sqrt{-1}$. Most trajectories deviate wildly from the classical trajectory $\gamma_{\tn{CL}}$, but these destructively interfere, effectively reproducing $\gamma_{\tn{CL}}$ for macroscopic particles.  However, random effects are significant for small particles.  In quantum gravity, the path integral is replaced by a sum over evolutionary pathways for spacetimes, which may be modeled via geometries, spin foams, causal dynamical triangulations, or in the present case, networks.  Every permissible process of network evolution contributes to this sum.  Networks closely approximating classical relativistic geometry play the role of $\gamma_{\tn{CL}}$, but these are vastly outnumbered by random networks.  Hence, random effects such as exponential horizon scaling tend to dominate a small network-theoretic universe near the beginning of time.

\begin{figure}[h]
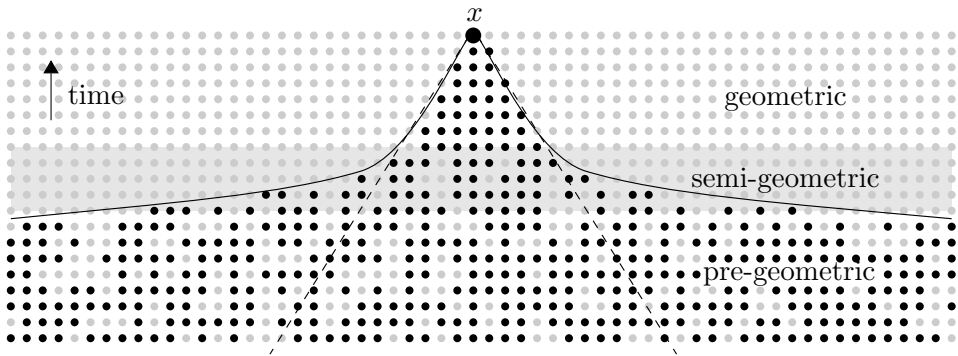


\caption{Pre-geometric epoch induces exponential past horizon growth.}
\label{figrandomcos}
\end{figure}

Figure \hyperref[figrandomcos]{\ref{figrandomcos}} illustrates the effect of a pre-geometric epoch on past horizon growth in a network toy model with  $20$ generations of $60$ nodes each and an average of $120$ edges between neighboring pairs of generations.  Edges are suppressed for clarity.  The shaded horizontal band indicates a brief {\it semi-geometric epoch} during which network structure changes from random to geometric.  The discussion of hybrid local/random structure in Section \hyperref[sectionNN]{\ref{sectionNN}} suggests that semi-geometric horizon growth might exceed even random horizon growth, but the figure does not show this possibility. For illustrative purposes, only two dimensions are shown in the geometric epoch.  Each node in this epoch is connected to exactly two previous-generation nodes.  This represents the chosen rate of information exchange, analogous to the speed of light, and independent of the dimension.  In the pre-geometric epoch, a total of $8\times 120=960$ non-generation-skipping edges are distributed randomly.  Black nodes indicate past horizon growth for a final-generation node $x$. Growth is linear with temporal depth in the geometric epoch because the spatial dimension is $1$, but increases drastically to exponential growth as one moves back in time through the semi-geometric epoch to the pre-geometric epoch.  Here, the network becomes a small world, and further horizon growth is arrested by its overall size.  In the pre-geometric epoch, thorough mixing occurs within a few generations, resolving the horizon problem.

\begin{figure}[h]
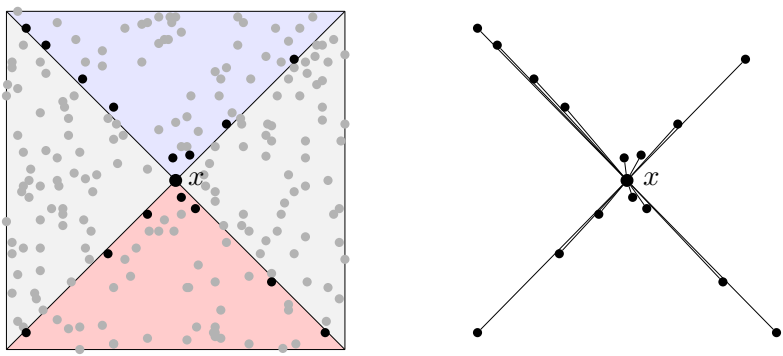

\begin{pgfpicture}{0cm}{0cm}{17cm}{4.8cm}
\begin{pgfmagnify}{1}{1}
\begin{pgfmagnify}{.75}{.75}
\begin{pgftranslate}{\pgfpoint{1cm}{0cm}}
\begin{pgftranslate}{\pgfpoint{-.5cm}{-.9cm}}
\begin{pgfmagnify}{1}{1}
\begin{pgfscope}
\color{black!5}
\pgfmoveto{\pgfxy(1,1)}
\pgflineto{\pgfxy(7,1)}
\pgflineto{\pgfxy(7,7)}
\pgflineto{\pgfxy(1,7)}
\pgflineto{\pgfxy(1,1)}
\pgffill
\color{red!20}
\pgfmoveto{\pgfxy(1,1)}
\pgflineto{\pgfxy(4,4)}
\pgflineto{\pgfxy(7,1)}
\pgflineto{\pgfxy(1,1)}
\pgffill
\color{blue!10}
\pgfmoveto{\pgfxy(1,7)}
\pgflineto{\pgfxy(4,4)}
\pgflineto{\pgfxy(7,7)}
\pgflineto{\pgfxy(1,7)}
\pgffill
\end{pgfscope}

\begin{pgfscope}
\color{black}
\pgfxyline(1,1)(7,7)
\pgfxyline(1,7)(7,1)

\end{pgfscope}

\begin{colormixin}{100!white}
\pgfxyline(1,1)(1,7)
\pgfxyline(7,1)(7,7)
\pgfxyline(1,1)(7,1)
\pgfxyline(1,7)(7,7)
\end{colormixin}
\begin{colormixin}{30!white}
\begin{pgftranslate}{\pgfpoint{0.03cm}{0cm}}
\pgfnodecircle{Node2}[fill]{\pgfxy(3.9,6.8)}{0.081cm}
\pgfnodecircle{Node3}[fill]{\pgfxy(6.9,4.7)}{0.081cm}
\pgfnodecircle{Node4}[fill]{\pgfxy(6.3,5.9)}{0.081cm}
\pgfnodecircle{Node6}[fill]{\pgfxy(5.6,5.0)}{0.081cm}
\pgfnodecircle{Node7}[fill]{\pgfxy(6.1,1.1)}{0.081cm}
\pgfnodecircle{Node8}[fill]{\pgfxy(2.5,3.8)}{0.081cm}
\pgfnodecircle{Node9}[fill]{\pgfxy(4.2,2.1)}{0.081cm}
\pgfnodecircle{Node10}[fill]{\pgfxy(1.4,4.2)}{0.081cm}
\end{pgftranslate}
\begin{pgftranslate}{\pgfpoint{0cm}{0.03cm}}
\pgfnodecircle{Node11}[fill]{\pgfxy(2.9,4.7)}{0.081cm}
\pgfnodecircle{Node12}[fill]{\pgfxy(5.6,4.4)}{0.081cm}
\pgfnodecircle{Node13}[fill]{\pgfxy(2.7,6.0)}{0.081cm}
\pgfnodecircle{Node14}[fill]{\pgfxy(4.7,5.9)}{0.081cm}
\pgfnodecircle{Node15}[fill]{\pgfxy(1.7,4.5)}{0.081cm}
\pgfnodecircle{Node16}[fill]{\pgfxy(2.7,1.4)}{0.081cm}
\pgfnodecircle{Node17}[fill]{\pgfxy(4.2,5.1)}{0.081cm}
\pgfnodecircle{Node18}[fill]{\pgfxy(4.0,5.0)}{0.081cm}
\pgfnodecircle{Node19}[fill]{\pgfxy(5.6,2.9)}{0.081cm}
\pgfnodecircle{Node20}[fill]{\pgfxy(2.4,4.0)}{0.081cm}
\end{pgftranslate}
\pgfnodecircle{Node21}[fill]{\pgfxy(1.1,6.1)}{0.081cm}
\pgfnodecircle{Node22}[fill]{\pgfxy(4.7,2.2)}{0.081cm}
\pgfnodecircle{Node23}[fill]{\pgfxy(4.5,4.1)}{0.081cm}
\pgfnodecircle{Node24}[fill]{\pgfxy(5.6,4.8)}{0.081cm}
\pgfnodecircle{Node25}[fill]{\pgfxy(7.0,5.0)}{0.081cm}
\pgfnodecircle{Node27}[fill]{\pgfxy(6.7,2.1)}{0.081cm}
\pgfnodecircle{Node28}[fill]{\pgfxy(1.2,4.8)}{0.081cm}
\pgfnodecircle{Node29}[fill]{\pgfxy(2.0,3.2)}{0.081cm}
\begin{pgftranslate}{\pgfpoint{-0.03cm}{0cm}}
\pgfnodecircle{Node31}[fill]{\pgfxy(2.5,2.6)}{0.081cm}
\pgfnodecircle{Node32}[fill]{\pgfxy(2.4,3.3)}{0.081cm}
\pgfnodecircle{Node33}[fill]{\pgfxy(6.7,1.5)}{0.081cm}
\pgfnodecircle{Node34}[fill]{\pgfxy(3.9,6.5)}{0.081cm}
\pgfnodecircle{Node35}[fill]{\pgfxy(4.3,5.5)}{0.081cm}
\pgfnodecircle{Node36}[fill]{\pgfxy(5.2,3.2)}{0.081cm}
\pgfnodecircle{Node37}[fill]{\pgfxy(6.5,6.2)}{0.081cm}
\pgfnodecircle{Node38}[fill]{\pgfxy(5.8,6.7)}{0.081cm}
\pgfnodecircle{Node39}[fill]{\pgfxy(1.05,5.7)}{0.081cm}
\pgfnodecircle{Node40}[fill]{\pgfxy(5.9,3.4)}{0.081cm}
\end{pgftranslate}
\begin{pgftranslate}{\pgfpoint{0cm}{-0.03cm}}
\pgfnodecircle{Node41}[fill]{\pgfxy(6.0,6.5)}{0.081cm}
\pgfnodecircle{Node42}[fill]{\pgfxy(6.6,5.1)}{0.081cm}
\pgfnodecircle{Node43}[fill]{\pgfxy(5.7,5.0)}{0.081cm}
\pgfnodecircle{Node44}[fill]{\pgfxy(4.2,1.2)}{0.081cm}
\pgfnodecircle{Node45}[fill]{\pgfxy(4.4,4.9)}{0.081cm}
\pgfnodecircle{Node46}[fill]{\pgfxy(3.4,6.8)}{0.081cm}
\pgfnodecircle{Node47}[fill]{\pgfxy(1.0,3.3)}{0.081cm}
\pgfnodecircle{Node48}[fill]{\pgfxy(6.6,5.8)}{0.081cm}
\pgfnodecircle{Node49}[fill]{\pgfxy(2.1,1.4)}{0.081cm}
\pgfnodecircle{Node50}[fill]{\pgfxy(5.9,1.05)}{0.081cm}
\end{pgftranslate}
\begin{pgftranslate}{\pgfpoint{0.03cm}{0cm}}
\pgfnodecircle{Node41}[fill]{\pgfxy(6.4,6.9)}{0.081cm}
\pgfnodecircle{Node42}[fill]{\pgfxy(4.5,3.9)}{0.081cm}
\pgfnodecircle{Node43}[fill]{\pgfxy(1.9,3.9)}{0.081cm}
\pgfnodecircle{Node44}[fill]{\pgfxy(6.3,2.2)}{0.081cm}
\pgfnodecircle{Node45}[fill]{\pgfxy(2.0,5.3)}{0.081cm}
\pgfnodecircle{Node46}[fill]{\pgfxy(2.2,6.5)}{0.081cm}
\pgfnodecircle{Node47}[fill]{\pgfxy(1.5,1.9)}{0.081cm}
\pgfnodecircle{Node48}[fill]{\pgfxy(3.5,3.1)}{0.081cm}
\pgfnodecircle{Node49}[fill]{\pgfxy(5.3,5.8)}{0.081cm}
\pgfnodecircle{Node50}[fill]{\pgfxy(5.9,3.6)}{0.081cm}
\end{pgftranslate}
\pgfnodecircle{Node51}[fill]{\pgfxy(4.7,3.8)}{0.081cm}
\pgfnodecircle{Node52}[fill]{\pgfxy(2.6,4.1)}{0.081cm}
\pgfnodecircle{Node53}[fill]{\pgfxy(6.1,2.7)}{0.081cm}
\pgfnodecircle{Node54}[fill]{\pgfxy(2.9,1.1)}{0.081cm}
\pgfnodecircle{Node55}[fill]{\pgfxy(6.3,3.0)}{0.081cm}
\pgfnodecircle{Node56}[fill]{\pgfxy(4.8,5.8)}{0.081cm}
\pgfnodecircle{Node57}[fill]{\pgfxy(2.8,5.1)}{0.081cm}
\pgfnodecircle{Node58}[fill]{\pgfxy(2.3,1.4)}{0.081cm}
\pgfnodecircle{Node59}[fill]{\pgfxy(5.5,2.7)}{0.081cm}
\pgfnodecircle{Node60}[fill]{\pgfxy(1.1,2.9)}{0.081cm}
\begin{pgftranslate}{\pgfpoint{0cm}{0.03cm}}
\pgfnodecircle{Node61}[fill]{\pgfxy(4.7,1.2)}{0.081cm}
\pgfnodecircle{Node62}[fill]{\pgfxy(1.2,5.6)}{0.081cm}
\pgfnodecircle{Node63}[fill]{\pgfxy(1.9,1.5)}{0.081cm}
\pgfnodecircle{Node64}[fill]{\pgfxy(4.1,6.6)}{0.081cm}
\pgfnodecircle{Node65}[fill]{\pgfxy(7.0,6.2)}{0.081cm}
\pgfnodecircle{Node66}[fill]{\pgfxy(1.2,2.3)}{0.081cm}
\pgfnodecircle{Node67}[fill]{\pgfxy(4.7,3.6)}{0.081cm}
\pgfnodecircle{Node69}[fill]{\pgfxy(2.2,4.1)}{0.081cm}
\pgfnodecircle{Node70}[fill]{\pgfxy(3.7,6.4)}{0.081cm}
\end{pgftranslate}
\pgfnodecircle{Node71}[fill]{\pgfxy(1.6,2.8)}{0.081cm}
\pgfnodecircle{Node72}[fill]{\pgfxy(5.2,4.6)}{0.081cm}
\pgfnodecircle{Node73}[fill]{\pgfxy(2.0,3.5)}{0.081cm}
\pgfnodecircle{Node74}[fill]{\pgfxy(2.4,6.1)}{0.081cm}
\pgfnodecircle{Node75}[fill]{\pgfxy(5.3,2.3)}{0.081cm}
\pgfnodecircle{Node77}[fill]{\pgfxy(4.8,1.7)}{0.081cm}
\pgfnodecircle{Node78}[fill]{\pgfxy(4.7,1.4)}{0.081cm}
\pgfnodecircle{Node79}[fill]{\pgfxy(6.8,3.4)}{0.081cm}
\pgfnodecircle{Node80}[fill]{\pgfxy(6.2,5.5)}{0.081cm}
\pgfnodecircle{Node82}[fill]{\pgfxy(6.6,5.3)}{0.081cm}
\pgfnodecircle{Node83}[fill]{\pgfxy(3.1,2.6)}{0.081cm}
\pgfnodecircle{Node84}[fill]{\pgfxy(1.3,2.8)}{0.081cm}
\pgfnodecircle{Node85}[fill]{\pgfxy(6.1,3.8)}{0.081cm}
\pgfnodecircle{Node86}[fill]{\pgfxy(6.5,5.7)}{0.081cm}
\pgfnodecircle{Node87}[fill]{\pgfxy(5.1,6.9)}{0.081cm}
\pgfnodecircle{Node88}[fill]{\pgfxy(6.1,6.0)}{0.081cm}
\pgfnodecircle{Node89}[fill]{\pgfxy(5.3,3.1)}{0.081cm}
\pgfnodecircle{Node91}[fill]{\pgfxy(5.8,3.1)}{0.081cm}
\pgfnodecircle{Node92}[fill]{\pgfxy(4.2,4.6)}{0.081cm}
\pgfnodecircle{Node93}[fill]{\pgfxy(4.7,1.3)}{0.081cm}
\pgfnodecircle{Node95}[fill]{\pgfxy(6.3,3.9)}{0.081cm}
\pgfnodecircle{Node96}[fill]{\pgfxy(1.2,7.0)}{0.081cm}
\pgfnodecircle{Node97}[fill]{\pgfxy(1.9,4.0)}{0.081cm}
\pgfnodecircle{Node99}[fill]{\pgfxy(2.4,2.8)}{0.081cm}
\pgfnodecircle{Node100}[fill]{\pgfxy(6.0,5.8)}{0.081cm}
\pgfnodecircle{Node101}[fill]{\pgfxy(3.5,6.8)}{0.081cm}
\pgfnodecircle{Node103}[fill]{\pgfxy(6.7,6.1)}{0.081cm}
\pgfnodecircle{Node105}[fill]{\pgfxy(6.3,6.1)}{0.081cm}
\pgfnodecircle{Node106}[fill]{\pgfxy(3.9,6.9)}{0.081cm}
\pgfnodecircle{Node107}[fill]{\pgfxy(2.9,2.1)}{0.081cm}
\pgfnodecircle{Node108}[fill]{\pgfxy(2.5,3.1)}{0.081cm}
\pgfnodecircle{Node109}[fill]{\pgfxy(4.8,1.2)}{0.081cm}
\begin{pgftranslate}{\pgfpoint{-0.03cm}{0cm}}
\pgfnodecircle{Node111}[fill]{\pgfxy(4.0,3.1)}{0.081cm}
\pgfnodecircle{Node112}[fill]{\pgfxy(1.3,3.8)}{0.081cm}
\pgfnodecircle{Node113}[fill]{\pgfxy(3.0,4.3)}{0.081cm}
\pgfnodecircle{Node114}[fill]{\pgfxy(4.8,5.5)}{0.081cm}
\pgfnodecircle{Node115}[fill]{\pgfxy(3.1,4.8)}{0.081cm}
\pgfnodecircle{Node116}[fill]{\pgfxy(5.9,6.1)}{0.081cm}
\pgfnodecircle{Node117}[fill]{\pgfxy(5.2,5.0)}{0.081cm}
\pgfnodecircle{Node118}[fill]{\pgfxy(4.7,1.3)}{0.081cm}
\pgfnodecircle{Node120}[fill]{\pgfxy(4.4,1.4)}{0.081cm}
\end{pgftranslate}
\pgfnodecircle{Node121}[fill]{\pgfxy(1.4,3.7)}{0.081cm}
\pgfnodecircle{Node122}[fill]{\pgfxy(4.0,6.9)}{0.081cm}
\pgfnodecircle{Node123}[fill]{\pgfxy(4.2,2.1)}{0.081cm}
\pgfnodecircle{Node124}[fill]{\pgfxy(5.6,6.2)}{0.081cm}
\pgfnodecircle{Node125}[fill]{\pgfxy(3.7,6.9)}{0.081cm}
\pgfnodecircle{Node126}[fill]{\pgfxy(3.5,1.2)}{0.081cm}
\pgfnodecircle{Node127}[fill]{\pgfxy(5.2,3.4)}{0.081cm}
\pgfnodecircle{Node128}[fill]{\pgfxy(6.3,4.4)}{0.081cm}
\pgfnodecircle{Node129}[fill]{\pgfxy(1.8,3.5)}{0.081cm}
\pgfnodecircle{Node130}[fill]{\pgfxy(6.8,6.5)}{0.081cm}
\pgfnodecircle{Node131}[fill]{\pgfxy(2.7,6.3)}{0.081cm}
\pgfnodecircle{Node132}[fill]{\pgfxy(5.3,1.1)}{0.081cm}
\pgfnodecircle{Node133}[fill]{\pgfxy(1.3,3.9)}{0.081cm}
\pgfnodecircle{Node134}[fill]{\pgfxy(3.8,6.5)}{0.081cm}
\pgfnodecircle{Node135}[fill]{\pgfxy(1.1,6.0)}{0.081cm}
\pgfnodecircle{Node136}[fill]{\pgfxy(4.1,3.4)}{0.081cm}
\pgfnodecircle{Node137}[fill]{\pgfxy(4.9,6.0)}{0.081cm}
\pgfnodecircle{Node138}[fill]{\pgfxy(6.9,1.9)}{0.081cm}
\pgfnodecircle{Node140}[fill]{\pgfxy(3.4,6.1)}{0.081cm}
\pgfnodecircle{Node142}[fill]{\pgfxy(6.0,4.1)}{0.081cm}
\pgfnodecircle{Node144}[fill]{\pgfxy(2.0,3.4)}{0.081cm}
\pgfnodecircle{Node145}[fill]{\pgfxy(1.6,5.2)}{0.081cm}
\pgfnodecircle{Node146}[fill]{\pgfxy(3.2,2.3)}{0.081cm}
\pgfnodecircle{Node147}[fill]{\pgfxy(1.3,4.5)}{0.081cm}
\pgfnodecircle{Node148}[fill]{\pgfxy(1.4,4.0)}{0.081cm}
\pgfnodecircle{Node149}[fill]{\pgfxy(5.0,5.2)}{0.081cm}
\pgfnodecircle{Node150}[fill]{\pgfxy(1.6,6.1)}{0.081cm}
\pgfnodecircle{Node152}[fill]{\pgfxy(6.9,2.9)}{0.081cm}
\pgfnodecircle{Node153}[fill]{\pgfxy(3.5,4.2)}{0.081cm}
\pgfnodecircle{Node154}[fill]{\pgfxy(5.2,6.8)}{0.081cm}
\pgfnodecircle{Node155}[fill]{\pgfxy(1.2,1.5)}{0.081cm}
\pgfnodecircle{Node156}[fill]{\pgfxy(1.3,1.4)}{0.081cm}
\pgfnodecircle{Node157}[fill]{\pgfxy(1.3,6.4)}{0.081cm}
\pgfnodecircle{Node158}[fill]{\pgfxy(6.9,3.5)}{0.081cm}
\pgfnodecircle{Node159}[fill]{\pgfxy(2.3,4.5)}{0.081cm}
\pgfnodecircle{Node160}[fill]{\pgfxy(4.9,2.9)}{0.081cm}
\pgfnodecircle{Node161}[fill]{\pgfxy(4.6,4.3)}{0.081cm}
\pgfnodecircle{Node163}[fill]{\pgfxy(4.7,4.6)}{0.081cm}
\pgfnodecircle{Node164}[fill]{\pgfxy(2.0,5.4)}{0.081cm}
\pgfnodecircle{Node165}[fill]{\pgfxy(6.7,4.0)}{0.081cm}
\pgfnodecircle{Node167}[fill]{\pgfxy(5.1,6.4)}{0.081cm}
\pgfnodecircle{Node168}[fill]{\pgfxy(2.3,1.0)}{0.081cm}
\pgfnodecircle{Node169}[fill]{\pgfxy(3.8,3.3)}{0.081cm}
\pgfnodecircle{Node170}[fill]{\pgfxy(5.7,4.7)}{0.081cm}
\pgfnodecircle{Node171}[fill]{\pgfxy(6.6,6.4)}{0.081cm}
\pgfnodecircle{Node172}[fill]{\pgfxy(7.0,2.6)}{0.081cm}
\pgfnodecircle{Node173}[fill]{\pgfxy(1.1,2.0)}{0.081cm}
\pgfnodecircle{Node174}[fill]{\pgfxy(6.0,1.8)}{0.081cm}
\pgfnodecircle{Node175}[fill]{\pgfxy(3.7,3.1)}{0.081cm}
\pgfnodecircle{Node176}[fill]{\pgfxy(1.6,3.3)}{0.081cm}
\pgfnodecircle{Node177}[fill]{\pgfxy(3.1,6.7)}{0.081cm}
\pgfnodecircle{Node178}[fill]{\pgfxy(2.1,5.1)}{0.081cm}
\pgfnodecircle{Node179}[fill]{\pgfxy(1.1,2.7)}{0.081cm}
\pgfnodecircle{Node180}[fill]{\pgfxy(1.9,6.8)}{0.081cm}
\pgfnodecircle{Node181}[fill]{\pgfxy(1.7,5.9)}{0.081cm}
\pgfnodecircle{Node183}[fill]{\pgfxy(6.0,1.7)}{0.081cm}
\pgfnodecircle{Node184}[fill]{\pgfxy(4.4,2.5)}{0.081cm}
\pgfnodecircle{Node185}[fill]{\pgfxy(2.1,4.8)}{0.081cm}
\pgfnodecircle{Node186}[fill]{\pgfxy(2.9,1.1)}{0.081cm}
\pgfnodecircle{Node187}[fill]{\pgfxy(1.9,4.4)}{0.081cm}
\pgfnodecircle{Node188}[fill]{\pgfxy(3.7,4.9)}{0.081cm}
\pgfnodecircle{Node189}[fill]{\pgfxy(5.8,5.5)}{0.081cm}
\pgfnodecircle{Node191}[fill]{\pgfxy(1.0,5.5)}{0.081cm}
\pgfnodecircle{Node192}[fill]{\pgfxy(1.5,2.0)}{0.081cm}
\pgfnodecircle{Node193}[fill]{\pgfxy(1.6,2.4)}{0.081cm}
\pgfnodecircle{Node194}[fill]{\pgfxy(2.2,5.5)}{0.081cm}
\pgfnodecircle{Node195}[fill]{\pgfxy(4.1,4.6)}{0.081cm}
\pgfnodecircle{Node196}[fill]{\pgfxy(1.6,3.9)}{0.081cm}
\pgfnodecircle{Node197}[fill]{\pgfxy(6.3,4.7)}{0.081cm}
\pgfnodecircle{Node199}[fill]{\pgfxy(6.2,6.7)}{0.081cm}
\pgfnodecircle{Node200}[fill]{\pgfxy(3.3,2.8)}{0.081cm}
\pgfnodecircle{Node76}[fill]{\pgfxy(2.95,5.0)}{0.081cm}
\pgfnodecircle{Node94}[fill]{\pgfxy(4.85,3.2)}{0.081cm}
\pgfnodecircle{Node98}[fill]{\pgfxy(5.15,5.1)}{0.081cm}
\pgfnodecircle{Node139}[fill]{\pgfxy(2.15,2.2)}{0.081cm}
\pgfnodecircle{Node151}[fill]{\pgfxy(6.5,6.4)}{0.081cm}
\pgfnodecircle{Node166}[fill]{\pgfxy(1.65,1.7)}{0.081cm}
\pgfnodecircle{Node182}[fill]{\pgfxy(5.7,2.35)}{0.081cm}
\end{colormixin}
\pgfnodecircle{Node1}[fill]{\pgfxy(4,4)}{0.115cm}
\pgfnodecircle{Node5}[fill]{\pgfxy(2.9,5.3)}{0.081cm}
\pgfnodecircle{Node26}[fill]{\pgfxy(4.35,3.5)}{0.081cm}
\pgfnodecircle{Node30}[fill]{\pgfxy(3.95,4.4)}{0.081cm}
\pgfnodecircle{Node68}[fill]{\pgfxy(3.5,3.4)}{0.081cm}
\pgfnodecircle{Node81}[fill]{\pgfxy(4.25,4.45)}{0.081cm}
\pgfnodecircle{Node90}[fill]{\pgfxy(2.35,5.8)}{0.081cm}
\pgfnodecircle{Node102}[fill]{\pgfxy(6.1,6.15)}{0.081cm}
\pgfnodecircle{Node104}[fill]{\pgfxy(5.7,2.2)}{0.081cm}
\pgfnodecircle{Node110}[fill]{\pgfxy(6.65,1.3)}{0.081cm}
\pgfnodecircle{Node119}[fill]{\pgfxy(2.8,2.7)}{0.081cm}
\pgfnodecircle{Node141}[fill]{\pgfxy(4.1,3.7)}{0.081cm}
\pgfnodecircle{Node143}[fill]{\pgfxy(1.7,6.4)}{0.081cm}
\pgfnodecircle{Node162}[fill]{\pgfxy(1.35,6.7)}{0.081cm}
\pgfnodecircle{Node190}[fill]{\pgfxy(1.35,1.3)}{0.081cm}
\pgfnodecircle{Node198}[fill]{\pgfxy(4.9,5.0)}{0.081cm}
\end{pgfmagnify}
\end{pgftranslate}
\end{pgftranslate}
\begin{pgftranslate}{\pgfpoint{9cm}{0cm}}
\begin{pgftranslate}{\pgfpoint{-.5cm}{-.9cm}}
\begin{pgfmagnify}{1}{1}
\pgfnodecircle{Node1}[fill]{\pgfxy(4,4)}{0.115cm}
\pgfnodecircle{Node5}[fill]{\pgfxy(2.9,5.3)}{0.081cm}
\pgfnodecircle{Node26}[fill]{\pgfxy(4.35,3.5)}{0.081cm}
\pgfnodecircle{Node30}[fill]{\pgfxy(3.95,4.4)}{0.081cm}
\pgfnodecircle{Node68}[fill]{\pgfxy(3.5,3.4)}{0.081cm}
\pgfnodecircle{Node81}[fill]{\pgfxy(4.25,4.45)}{0.081cm}
\pgfnodecircle{Node90}[fill]{\pgfxy(2.35,5.8)}{0.081cm}
\pgfnodecircle{Node102}[fill]{\pgfxy(6.1,6.15)}{0.081cm}
\pgfnodecircle{Node104}[fill]{\pgfxy(5.7,2.2)}{0.081cm}
\pgfnodecircle{Node110}[fill]{\pgfxy(6.65,1.3)}{0.081cm}
\pgfnodecircle{Node119}[fill]{\pgfxy(2.8,2.7)}{0.081cm}
\pgfnodecircle{Node141}[fill]{\pgfxy(4.1,3.7)}{0.081cm}
\pgfnodecircle{Node143}[fill]{\pgfxy(1.7,6.4)}{0.081cm}
\pgfnodecircle{Node162}[fill]{\pgfxy(1.35,6.7)}{0.081cm}
\pgfnodecircle{Node190}[fill]{\pgfxy(1.35,1.3)}{0.081cm}
\pgfnodecircle{Node198}[fill]{\pgfxy(4.9,5.0)}{0.081cm}
\pgfnodeconnline{Node1}{Node5}
\pgfnodeconnline{Node1}{Node26}
\pgfnodeconnline{Node1}{Node30}
\pgfnodeconnline{Node1}{Node68}
\pgfnodeconnline{Node1}{Node81}
\pgfnodeconnline{Node1}{Node90}
\pgfnodeconnline{Node1}{Node102}
\pgfnodeconnline{Node1}{Node104}
\pgfnodeconnline{Node1}{Node110}
\pgfnodeconnline{Node1}{Node119}
\pgfnodeconnline{Node1}{Node141}
\pgfnodeconnline{Node1}{Node143}
\pgfnodeconnline{Node1}{Node162}
\pgfnodeconnline{Node1}{Node190}
\pgfnodeconnline{Node1}{Node198}
\end{pgfmagnify}
\end{pgftranslate}
\end{pgftranslate}
\end{pgfmagnify}
\end{pgfmagnify}
\pgfputat{\pgfxy(3.65,2.35)}{\pgfbox[center,center]{$x$}}
\begin{pgftranslate}{\pgfpoint{6.2cm}{0cm}}
\pgfputat{\pgfxy(3.5,2.35)}{\pgfbox[center,center]{$x$}}
\end{pgftranslate}
\end{pgfpicture}
\caption{Approximately Lorentzian network; neighbors of a node $x$.}
\label{figsprinklingzoomedout}
\end{figure}

{\bf Lorentzian networks.} The toy model in Figure \hyperref[figrandomcos]{\ref{figrandomcos}} differs importantly from actual spacetime due to the {\it preferred frame of reference} defined by its generational structure.  Its geometric epoch is therefore approximately Euclidean rather than Lorentzian.   Even the network in Figure \hyperref[figdifferentgen]{\ref{figdifferentgen}}, which has no preferred frame, is still nearly Euclidean, with some frames ``more equal than others."   Lorentzian networks are not used in these figures for the prosaic reason that they make for awkward illustrations, but their qualitative horizon growth properties are the same as for other geometric networks.  The left-hand diagram in Figure \hyperref[figsprinklingzoomedout]{\ref{figsprinklingzoomedout}} shows the shape of an approximately Lorentzian network, with edges suppressed for clarity.  Edges attach each node $x$ to a family of nodes located just inside an imaginary ``light cone" with $x$ at its apex, and therefore appear to nearly coincide in drawings, as shown in the right-hand diagram.   As explained in Section \hyperref[Genacyclic]{\ref{Genacyclic}}, such a network may be partitioned into generations noncanonically, which amounts to choosing a frame of reference.  For any such choice, a large proportion of edges skip generations.  Since nature favors this type of network architecture, it is interesting to transport it into other settings, for example, artificial neural networks. Unfortunately, the term {\it Lorentzian neural network} has already been coined to mean something entirely different \cite{Giraud95}.

{\bf Emergence: causal metric hypothesis.} Small-world resolution of the horizon problem, as we have proposed it, depends on the hypothesis that spacetime {\it can} emerge from network microstructure. This is far from obvious; for example, strictly graded graphs such as $G_N^M(K)$ and $\ms{G}_N^M(p)$ are almost certainly ruled out due to experimental constraints on {\it Lorentz invariance violation} \cite{Sanner19}.  More promising are {\it causal network} models, in which nodes represent events, and directed edges represent causal relationships between pairs of events.   After foreshadowing by Finkelstein \cite{Finkelstein69, Finkelstein88}, Hawking \cite{Hawking76} and Malament \cite{Malament77} proved in the late 1970s that many Lorentzian geometries may be described simply in terms of {\it causal structure} and {\it scale data.}  It was soon understood that such scale data, represented by the {\it conformal factor} in the Lorentzian {\it metric,} comes for free in the discrete context, via enumeration of fundamental spacetime elements \cite{Sorkinetal87,Myrheim78,tHooft78}.  This suggests that spacetime geometry can emerge from causal network structure.  We refer to this idea as the {\it causal metric hypothesis}  \cite{DribusDCT, DribusFQXi15, DribusAxioms13}.  Popular among causal network models are {\it causal sets,} developed by Sorkin et al.  \cite{Sorkinetal87}, defined by transferring relativistic notions to the discrete context with as few changes as possible.  Physically-relevant causal sets, and other types of causal networks, may be constructed by randomly {\it sprinkling} elements into a pre-existing Lorentzian manifold $X$, defining edges via its causal structure, then erasing $X$.  This method is used in Figure \hyperref[figsprinklingzoomedout]{\ref{figsprinklingzoomedout}}, where $X$ is flat Minkowski spacetime.  A technical distinction is that causal sets are defined to be {\it transitive,} which inconveniently obscures the distinction between direct and indirect causation.  Our causal networks are {\it not} assumed to be transitive.

\begin{figure}[h]
\centering
\begin{subfigure}{.5\textwidth}
  \centering
  \includegraphics[width=1.1\linewidth]{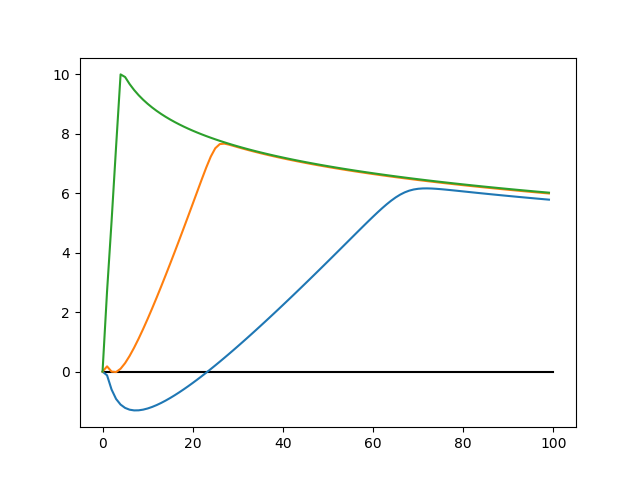}
\end{subfigure}%
\begin{subfigure}{.5\textwidth}
  \centering
\includegraphics[width=1.1\linewidth]{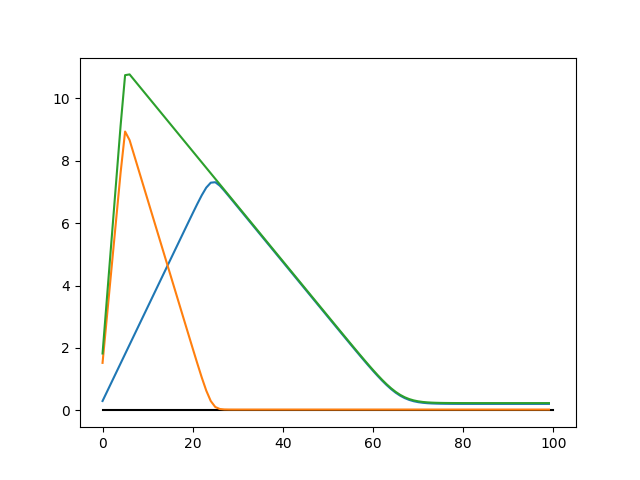}
\end{subfigure}
\caption{Geometric/non-geometric horizon growth; degree dependence.}
\label{figexpvspower}
\end{figure}

{\bf Example: trillion-node generations.}  The left-hand diagram in Figure \hyperref[figexpvspower]{\ref{figexpvspower}} illustrates the difference in horizon growth between large four-dimensional sprinkled causal networks and pre-geometric networks, using \eqref{nexthorizon} as an approximation.  Curves correspond to networks with $10^2$ generations of $10^{12}$ nodes each, and average node degrees $1.5$ (blue), $3$ (yellow), and $1000$ (green).  The abscissa is the generation number, and the ordinate is the base-$10$ logarithm of the ratio of horizon sizes for the pre-geometric and geometric cases.   Even for average degree $1.5$, the horizon size predicted by \eqref{nexthorizon} is nearly $10^6$ times larger than the geometric horizon size after about $75$ generations, while average degree $1000$ yields a horizon size nearly $10^{10}$ times larger after just a few generations. The eventual decay of each curve is due to small-world saturation, i.e., halting of exponential horizon growth due to limited network size. This corresponds to thorough mixing in the context of the horizon problem.

\vspace*{-.3cm} 

\begin{figure}[h]
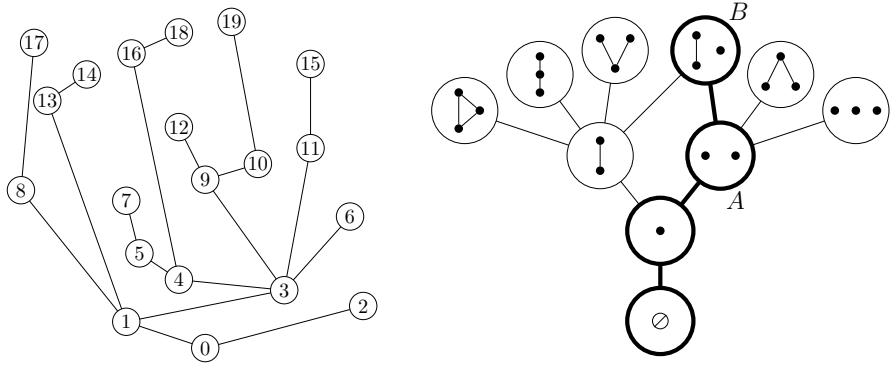

\begin{pgfpicture}{0cm}{0cm}{17cm}{4.6cm}
\begin{pgftranslate}{\pgfpoint{.6cm}{.1cm}}
\begin{pgfmagnify}{.7}{.7}
\pgfnodecircle{Node0}[stroke]{\pgfxy(3.5,0)}{0.26cm}
\pgfputat{\pgfxy(3.5,0)}{\pgfbox[center,center]{$0$}}

\pgfnodecircle{Node1}[stroke]{\pgfxy(2,.5)}{0.26cm}
\pgfputat{\pgfxy(2,.5)}{\pgfbox[center,center]{$1$}}

\pgfnodecircle{Node2}[stroke]{\pgfxy(6.5,.8)}{0.26cm}
\pgfputat{\pgfxy(6.5,.8)}{\pgfbox[center,center]{$2$}}

\pgfnodecircle{Node3}[stroke]{\pgfxy(5,1.1)}{0.26cm}
\pgfputat{\pgfxy(5,1.1)}{\pgfbox[center,center]{$3$}}

\pgfnodecircle{Node4}[stroke]{\pgfxy(3,1.3)}{0.26cm}
\pgfputat{\pgfxy(3,1.3)}{\pgfbox[center,center]{$4$}}
\pgfnodecircle{Node5}[stroke]{\pgfxy(2.25,1.8)}{0.26cm}
\pgfputat{\pgfxy(2.25,1.8)}{\pgfbox[center,center]{$5$}}
\pgfnodecircle{Node6}[stroke]{\pgfxy(6.25,2.5)}{0.26cm}
\pgfputat{\pgfxy(6.25,2.5)}{\pgfbox[center,center]{$6$}}
\pgfnodecircle{Node7}[stroke]{\pgfxy(2,2.8)}{0.26cm}
\pgfputat{\pgfxy(2,2.8)}{\pgfbox[center,center]{$7$}}
\pgfnodecircle{Node8}[stroke]{\pgfxy(0,3)}{0.26cm}
\pgfputat{\pgfxy(0,3)}{\pgfbox[center,center]{$8$}}
\pgfnodecircle{Node9}[stroke]{\pgfxy(3.5,3.2)}{0.26cm}
\pgfputat{\pgfxy(3.5,3.2)}{\pgfbox[center,center]{$9$}}
\pgfnodecircle{Node10}[stroke]{\pgfxy(4.5,3.5)}{0.26cm}
\pgfputat{\pgfxy(4.5,3.5)}{\pgfbox[center,center]{$10$}}
\pgfnodecircle{Node11}[stroke]{\pgfxy(5.5,3.8)}{0.26cm}
\pgfputat{\pgfxy(5.5,3.8)}{\pgfbox[center,center]{$11$}}
\pgfnodecircle{Node12}[stroke]{\pgfxy(3,4.2)}{0.26cm}
\pgfputat{\pgfxy(3,4.2)}{\pgfbox[center,center]{$12$}}
\pgfnodecircle{Node13}[stroke]{\pgfxy(.5,4.7)}{0.26cm}
\pgfputat{\pgfxy(.5,4.7)}{\pgfbox[center,center]{$13$}}
\pgfnodecircle{Node14}[stroke]{\pgfxy(1.25,5.2)}{0.26cm}
\pgfputat{\pgfxy(1.25,5.2)}{\pgfbox[center,center]{$14$}}
\pgfnodecircle{Node15}[stroke]{\pgfxy(5.5,5.4)}{0.26cm}
\pgfputat{\pgfxy(5.5,5.4)}{\pgfbox[center,center]{$15$}}
\pgfnodecircle{Node16}[stroke]{\pgfxy(2.1,5.6)}{0.26cm}
\pgfputat{\pgfxy(2.1,5.6)}{\pgfbox[center,center]{$16$}}
\pgfnodecircle{Node17}[stroke]{\pgfxy(.25,5.8)}{0.26cm}
\pgfputat{\pgfxy(.25,5.8)}{\pgfbox[center,center]{$17$}}
\pgfnodecircle{Node18}[stroke]{\pgfxy(3,6)}{0.26cm}
\pgfputat{\pgfxy(3,6)}{\pgfbox[center,center]{$18$}}
\pgfnodecircle{Node19}[stroke]{\pgfxy(4,6.2)}{0.26cm}
\pgfputat{\pgfxy(4,6.2)}{\pgfbox[center,center]{$19$}}
\pgfnodeconnline{Node1}{Node0}
\pgfnodeconnline{Node2}{Node0}
\pgfnodeconnline{Node3}{Node1}
\pgfnodeconnline{Node4}{Node3}
\pgfnodeconnline{Node5}{Node4}
\pgfnodeconnline{Node6}{Node3}
\pgfnodeconnline{Node7}{Node5}
\pgfnodeconnline{Node8}{Node1}
\pgfnodeconnline{Node9}{Node3}
\pgfnodeconnline{Node10}{Node9}
\pgfnodeconnline{Node11}{Node3}
\pgfnodeconnline{Node12}{Node9}
\pgfnodeconnline{Node13}{Node1}
\pgfnodeconnline{Node14}{Node13}
\pgfnodeconnline{Node15}{Node11}
\pgfnodeconnline{Node16}{Node4}
\pgfnodeconnline{Node17}{Node8}
\pgfnodeconnline{Node18}{Node16}
\pgfnodeconnline{Node19}{Node10}
\end{pgfmagnify}
\end{pgftranslate}
\begin{pgftranslate}{\pgfpoint{-1.8cm}{.15cm}}

\begin{pgftranslate}{\pgfpoint{4.3cm}{-.25cm}}
\begin{pgfmagnify}{.8}{.8}
\begin{pgftranslate}{\pgfpoint{0cm}{.7cm}}
\pgfnodecircle{Node1}[stroke]{\pgfxy(8.25,0)}{.55cm}%
\pgfnodecircle{Node2}[stroke]{\pgfxy(8.25,1.5)}{.55cm}%
\pgfnodecircle{Node3}[stroke]{\pgfxy(7.25,2.75)}{.55cm}%
\pgfnodecircle{Node4}[stroke]{\pgfxy(9.25,2.75)}{.55cm}%
\pgfputat{\pgfxy(9.5,2)}{\pgfbox[center,center]{{\large$A$}}}
\pgfnodecircle{Node5}[stroke]{\pgfxy(5,3.5)}{.55cm}%
\pgfnodecircle{Node6}[stroke]{\pgfxy(6.25,4.1)}{.55cm}%
\pgfnodecircle{Node7}[stroke]{\pgfxy(7.5,4.5)}{.55cm}%
\pgfnodecircle{Node8}[stroke]{\pgfxy(9,4.5)}{.55cm}%
\pgfputat{\pgfxy(9.55,5.15)}{\pgfbox[center,center]{{\large$B$}}}
\pgfnodecircle{Node9}[stroke]{\pgfxy(10.25,4.1)}{.55cm}%
\pgfnodecircle{Node10}[stroke]{\pgfxy(11.5,3.5)}{.55cm}%
\pgfnodeconnline{Node1}{Node2}
\pgfnodeconnline{Node2}{Node3}
\pgfnodeconnline{Node2}{Node4}
\pgfnodeconnline{Node3}{Node5}
\pgfnodeconnline{Node3}{Node6}
\pgfnodeconnline{Node3}{Node7}
\pgfnodeconnline{Node3}{Node8}
\pgfnodeconnline{Node4}{Node8}
\pgfnodeconnline{Node4}{Node9}
\pgfnodeconnline{Node4}{Node10}
\begin{pgfscope}
\pgfsetlinewidth{2pt}
\pgfnodecircle{Node1}[stroke]{\pgfxy(8.25,0)}{.55cm}
\pgfnodecircle{Node2}[stroke]{\pgfxy(8.25,1.5)}{.55cm}%
\pgfnodecircle{Node4}[stroke]{\pgfxy(9.25,2.75)}{.55cm}%
\pgfnodecircle{Node8}[stroke]{\pgfxy(9,4.5)}{.55cm}%
\pgfnodeconnline{Node1}{Node2}
\pgfnodeconnline{Node2}{Node4}
\pgfnodeconnline{Node4}{Node8}
\end{pgfscope}
\end{pgftranslate}
\begin{pgftranslate}{\pgfpoint{0cm}{-.3cm}}
\begin{pgftranslate}{\pgfpoint{0cm}{0cm}}
\pgfnodecircle{Node1}[stroke]{\pgfxy(8.25,1)}{0.13cm}
\pgfxyline(8.15,.9)(8.35,1.1)
\end{pgftranslate}
\begin{pgftranslate}{\pgfpoint{0cm}{1.5cm}}
\pgfnodecircle{Node1}[fill]{\pgfxy(8.25,1)}{0.071cm}
\end{pgftranslate}
\begin{pgftranslate}{\pgfpoint{-1cm}{2.75cm}}
\pgfnodecircle{Node1}[fill]{\pgfxy(8.25,.75)}{0.071cm}
\pgfnodecircle{Node2}[fill]{\pgfxy(8.25,1.25)}{0.071cm}
\pgfnodeconnline{Node1}{Node2}
\end{pgftranslate}
\begin{pgftranslate}{\pgfpoint{1cm}{2.75cm}}
\pgfnodecircle{Node1}[fill]{\pgfxy(8,1)}{0.071cm}
\pgfnodecircle{Node2}[fill]{\pgfxy(8.5,1)}{0.071cm}
\end{pgftranslate}
\begin{pgftranslate}{\pgfpoint{-3.25cm}{3.5cm}}
\pgfnodecircle{Node1}[fill]{\pgfxy(8.15,.7)}{0.071cm}
\pgfnodecircle{Node2}[fill]{\pgfxy(8.5,1)}{0.071cm}
\pgfnodecircle{Node3}[fill]{\pgfxy(8.15,1.3)}{0.071cm}
\pgfnodeconnline{Node1}{Node2}
\pgfnodeconnline{Node1}{Node3}
\pgfnodeconnline{Node2}{Node3}
\end{pgftranslate}
\begin{pgftranslate}{\pgfpoint{-2cm}{4.1cm}}
\pgfnodecircle{Node1}[fill]{\pgfxy(8.25,.7)}{0.071cm}
\pgfnodecircle{Node2}[fill]{\pgfxy(8.25,1)}{0.071cm}
\pgfnodecircle{Node3}[fill]{\pgfxy(8.25,1.3)}{0.071cm}
\pgfnodeconnline{Node1}{Node2}
\pgfnodeconnline{Node2}{Node3}
\end{pgftranslate}
\begin{pgftranslate}{\pgfpoint{-.75cm}{4.5cm}}
\pgfnodecircle{Node1}[fill]{\pgfxy(8.25,.7)}{0.071cm}
\pgfnodecircle{Node2}[fill]{\pgfxy(8,1.2)}{0.071cm}
\pgfnodecircle{Node3}[fill]{\pgfxy(8.5,1.2)}{0.071cm}
\pgfnodeconnline{Node1}{Node2}
\pgfnodeconnline{Node1}{Node3}
\end{pgftranslate}
\begin{pgftranslate}{\pgfpoint{.75cm}{4.5cm}}
\pgfnodecircle{Node1}[fill]{\pgfxy(8.1,.75)}{0.071cm}
\pgfnodecircle{Node2}[fill]{\pgfxy(8.1,1.25)}{0.071cm}
\pgfnodecircle{Node3}[fill]{\pgfxy(8.5,1)}{0.071cm}
\pgfnodeconnline{Node1}{Node2}
\end{pgftranslate}
\begin{pgftranslate}{\pgfpoint{2cm}{4.1cm}}
\pgfnodecircle{Node1}[fill]{\pgfxy(8.25,1.3)}{0.071cm}
\pgfnodecircle{Node2}[fill]{\pgfxy(8,.8)}{0.071cm}
\pgfnodecircle{Node3}[fill]{\pgfxy(8.5,.8)}{0.071cm}
\pgfnodeconnline{Node1}{Node2}
\pgfnodeconnline{Node1}{Node3}
\end{pgftranslate}
\begin{pgftranslate}{\pgfpoint{3.25cm}{3.5cm}}
\pgfnodecircle{Node1}[fill]{\pgfxy(7.9,1)}{0.071cm}
\pgfnodecircle{Node2}[fill]{\pgfxy(8.25,1)}{0.071cm}
\pgfnodecircle{Node3}[fill]{\pgfxy(8.6,1)}{0.071cm}
\end{pgftranslate}
\end{pgftranslate}
\begin{colormixin}{15!white}
\end{colormixin}
\end{pgfmagnify}
\end{pgftranslate}
\end{pgftranslate}
\end{pgfpicture}
\caption{Preferential attachment; configuration space structure.}
\label{figpreferential}
\end{figure}

{\bf Preferential attachment.} Rapid horizon growth in the early universe may be further enhanced by a ``rich-get-richer" mechanism called {\it preferential attachment} or {\it accumulated advantage} \cite{NewmanNetworks18}, whereby nodes that appear early in a network growth process, or with many incident edges, are likely to possess the most incident edges late in the process.  For example, early-written papers are more likely to accumulate many citations than late-written papers, because there are more opportunities for them to be cited, and more existing citations to advertise them.  The left-hand diagram  in Figure \hyperref[figpreferential]{\ref{figpreferential}} illustrates a simple process leading to a weak form of preferential attachment.  Nodes are added sequentially, each attached by a single edge.  The resulting network is a {\it random tree.} Early nodes such as $1$ and $3$ tend to acquire more edges than later nodes.  Many other network growth processes exhibit similar behavior, often more marked; for example, individuals with many social connections are likely to acquire more connections rapidly, regardless of chronology.   Preferential attachment is possible but not automatic in a causal network.  If the random tree in Figure \hyperref[figpreferential]{\ref{figpreferential}} is interpreted causally, then each event has a unique cause but may produce multiple effects, which seems unrealistic.  Allowing multiple causes introduces a balancing effect, since later nodes have more possible attachment sites.  However, there may not be an absolute balance, on average, between the number of causes and effects of events in the real world.  Idealized Lorentzian geometry is locally time-symmetric, but we know that real physics discriminates between cause and effect due to the second law of thermodynamics.  Even a small asymmetry of this type leads to significantly different early horizon growth.  The right-hand diagram in Figure \hyperref[figexpvspower]{\ref{figexpvspower}} compares horizon growth for different average node degrees, using the same random networks involved in the left-hand diagram.   For example, the blue curve shows that doubling the average degree from $1.5$ to $3$ produces a horizon more than $10^7$ times larger after about $25$ generations.  An analogous change in a geometric network merely doubles horizon sizes, which demonstrates that the potential effect of preferential attachment is vastly amplified in a pre-geometric epoch.

{\bf Problem of time; sequential growth.}  The significance of preferential attachment in resolving the horizon problem depends on the role of network growth processes in spacetime dynamics.  This  is distinct from the more basic question of emergence, and involves the so-called {\it problem of time.} Though relativity has standardized geometric unification of space and time, there remains deep confusion about how the two differ, and whether or not past, present, and future should really be accorded equal ontological status.  Reducing the distinction to a sign difference in the metric signature trivializes the issue by simply canonizing the time-symmetric relativistic viewpoint, leaving no obvious role for time-asymmetric growth-related mechanisms such as preferential attachment.  Quantum gravity then involves summing over a configuration space of fixed networks.  The right-hand diagram in Figure \hyperref[figpreferential]{\ref{figpreferential}} shows a small part of such a space, with member networks appearing inside large circles.  Connections among these circles are for future reference.  This summation procedure admits no intrinsic distinction between past and future directions, since each network in the configuration space has a {\it dual network} given by reversing its edge directions. 

By contrast, one may choose to regard network models of spacetime as evolving, rather than fixed, objects.  This is how Sorkin and Rideout approach the problem of time in their {\it sequential growth dynamics} for causal sets \cite{SorkinSequentialGrowthDynamics99}.  In sequential growth dynamics, a causal set develops one node at a time, much like the random tree in the left-hand diagram in Figure \hyperref[figpreferential]{\ref{figpreferential}}.   Similar processes apply to causal network models.  A {\it higher-level edge} may be defined between a pair of networks $A$ and $B$ whenever $A$ can evolve into $B$ by acquiring a single additional node and its accompanying edges.  The right-hand diagram in Figure \hyperref[figpreferential]{\ref{figpreferential}} illustrates such higher-level edges as connections between large circles.  Recognition of such relationships transforms the configuration space into a {\it network of networks,} in which past and future directions are intrinsically distinct \cite{DribusDCT}. Quantum gravity then involves summing over {\it network growth processes} instead of individual networks.  Bold circles and edges illustrate such a process.  One may then attempt to specify how networks are likely to develop, rather than interpreting each network as a fixed unit.  The resulting evolutionary rules may or may not favor preferential attachment, depending on the details of the model.


{\bf Spacetime $\leftrightarrow$ DNN analogies.} We close with a brief discussion of analogies between our two application topics: deep learning, and the cosmological horizon problem.  Throughout the paper, we have used spacetime-inspired conventions and heuristics to describe general network structure, and the fact that popular DNN architectures such as CNNs incorporate geometric properties inherited from applications strengthens the conceptual connections between the two.  At the end of Section \hyperref[sectionNN]{\ref{sectionNN}}, we explicitly compared a trained dense DNN to spacetime,  anticipating our analysis of the horizon problem.  Conversely, one may {\it compare spacetime to a DNN,} and this leads to a variety of interesting, though speculative, ideas.  Unsurprisingly, there has been a recent explosion of interest in connections between fundamental physics and machine learning, involving such topics as the {\it renormalization group} \cite{MehtaDLRenorm14}, {\it quantum field theory} \cite{LeeDLQFT17}, and the intersection of machine learning and {\it quantum computing} \cite{LloydQuantumDL13}.  The more radical idea of comparing spacetime itself to a DNN is not without historical precedent in the ``physics as computation" tradition \cite{Finkelstein69,Wheeler89,Wolfram02}. 

Beginning from a network-theoretic spacetime model such as causal set theory, it is not difficult to incorporate the remaining characteristics of an individual DNN: processing is similar to path summation, edge weights to a Lagrangian, classical action, or phase map, and activation functions to an operator field.  More tenuous, but still interesting, are comprehensive cosmological analogies attempting to identify universal counterparts to other aspects of machine learning.  For example, the discussion at the end of Section \hyperref[sectionNN]{\ref{sectionNN}} suggests comparing spacetime to a {\it trained} DNN, since early small-world behavior reflects broad distribution of information near the input layer before localization due to sorting.  An obvious question is how to reconcile this sorting viewpoint with the second law of thermodynamics in a network-theoretic setting \cite{DribusEntropy17}, but we cannot analyze this topic here.  Biological synaptic pruning offers an interesting analogy related to the horizon problem: high random connectivity in the early universe may be compared to excess connectivity in an immature brain, and structural changes through the ``inflationary epoch" and up to the present era may be interpreted as {\it cosmological synaptic pruning.}  Even more speculative is to compare the idea of cyclic cosmological models \cite{Steinhardt13,Steinhardt17,PenroseCCC10} to repeated training cycles of a vast neural network.  This could ``explain," for instance, why spacetime appears to be ``trained," in the sense that late layers are more local than early layers under our approach to the horizon problem: previous aeons adjusted the weights.  While such analogies have no particular claim to credence, they are impossible to ignore once more modest connections between the two topics have been broached.  An environment populated by such ideas might offer further contributions to network design and/or theoretical physics.

\subsection*{Acknowledgements}

We thank Jalynn Roberts, Jessica Garriga, Thomas Naugle, Haley Dozier, Joshua Deaton, Lillie Blackmon, Madeline Leboeuf, and Stephanie Dribus for stimulating discussions and technical assistance.



\end{document}